\documentclass[preprint,notoc]{JHEP3}
\usepackage{graphicx}
\usepackage{amsmath}
\usepackage{psfrag}
\usepackage[sort&compress,numbers]{natbib}

\def\be{\begin{equation}}
\def\ee{\end{equation}}
\def\bea{\begin{eqnarray}}
\def\eea{\end{eqnarray}}

\def\e{\epsilon}

\def\Psl{\not{\hbox{\kern-2.3pt $P$}}}
\def\psl{\not{\hbox{\kern-2.3pt $p$}}}
\def\qsl{\not{\hbox{\kern-2.3pt $q$}}}
\def\Ksl{\not{\hbox{\kern-2.3pt $K$}}}
\def\ksl{\not{\hbox{\kern-2.3pt $k$}}}
\def\esl{\not{\hbox{\kern-2.3pt $\pol$}}}

\def\pol{\varepsilon}

\def\spa#1.#2{\left\langle#1\,#2\right\rangle}
\def\spb#1.#2{\left[#1\,#2\right]}
\def\spba#1#2#3{\left[#1\,P_#2\,#3\right]}
\def\lor#1.#2{\left(#1\,#2\right)}
\def\sand#1.#2.#3{%
\left\langle\smash{#1}{\vphantom1}^{-}\right|{#2}%
\left|\smash{#3}{\vphantom1}^{-}\right\rangle}
\def\sandp#1.#2.#3{%
\left\langle\smash{#1}{\vphantom1}^{-}\right|{#2}%
\left|\smash{#3}{\vphantom1}^{+}\right\rangle}
\def\sandpp#1.#2.#3{%
\left\langle\smash{#1}{\vphantom1}^{+}\right|{#2}%
\left|\smash{#3}{\vphantom1}^{+}\right\rangle}
\def\sandpm#1.#2.#3{%
\left\langle\smash{#1}{\vphantom1}^{+}\right|{#2}%
\left|\smash{#3}{\vphantom1}^{-}\right\rangle}
\def\sandmp#1.#2.#3{%
\left\langle\smash{#1}{\vphantom1}^{-}\right|{#2}%
\left|\smash{#3}{\vphantom1}^{+}\right\rangle}
\def\sandmm#1.#2.#3{%
\left\langle\smash{#1}{\vphantom1}^{-}\right|{\slash\!\!\! #2}%
\left|\smash{#3}{\vphantom1}^{-}\right\rangle}
\def\spab#1.#2.#3{\sandmm#1.#2.#3}
\def\spbb#1.#2.#3.#4{\sandpm#1.{\slash\!\!\! #2\slash\!\!\! #3}.#4}
\newbox\charbox
\newbox\slabox
\def\s#1{{      
        \setbox\charbox=\hbox{$#1$}
        \setbox\slabox=\hbox{$/$}
        \dimen\charbox=\ht\slabox
        \advance\dimen\charbox by -\dp\slabox
        \advance\dimen\charbox by -\ht\charbox
        \advance\dimen\charbox by \dp\charbox
        \divide\dimen\charbox by 2
        \raise-\dimen\charbox\hbox to \wd\charbox{\hss/\hss}
        \llap{$#1$}
}}
\def\ksl{\s{k}}

\def\beqa{\begin{eqnarray}}
\def\eeqa{\end{eqnarray}}
\def\beq{\begin{equation}}
\def\eeq{\end{equation}}
\def\hf{{\textstyle{\frac{1}{2}}}}

\newcommand{\wh}[1]{\widehat{#1}}

\def\A#1#2{\langle#1#2\rangle}
\def\B#1#2{[#1#2]}

\def\spa#1.#2{\left\langle#1\,#2\right\rangle}
\def\spb#1.#2{\left[#1\,#2\right]}
\def\spab#1.#2.#3{\left\langle#1\,#2\,#3\right]}
\def\spba#1.#2.#3{\left[#1\,#2\,#3\right\rangle}
\def\spaa#1.#2.#3.#4{\left\langle#1\,#2\,#3\,#4\right\rangle}
\def\spahr#1.#2{\langle#1\,\hat{#2}\rangle}
\def\spaah#1.#2.#3.#4{\langle#1\,#2\,#3\,\hat{#4}\rangle}
\def\spaahl#1.#2.#3.#4{\langle\hat{#1}\,#2\,#3\,#4\rangle}
\def\spabh#1.#2.#3{\langle#1\,\widehat{#2}\,#3]}
\def\spahl#1.#2{\langle\hat{#1}\,#2\rangle}
\def\spahh#1.#2{\langle\hat{#1}\,\hat{#2}\rangle}
\def\spaas#1.#2.#3{\left\langle#1\,#2\,#3\right\rangle}
\def\spbhl#1.#2{\left[\hat{#1}\,#2\right]}
\def\spbhr#1.#2{\left[#1\,\hat{#2}\right]}
\def\spabhh#1.#2.#3{\langle\hat{#1}\,\hat{#2}\,#3]}
\def\spbah#1.#2.#3{[#1\,\widehat{#2}\,#3\rangle}
\def\lo{\ell_1}

\def\lt{\ell_2}

\def\ap{\alpha+1}
\def\mpl{(m+1)}

\def\mm{(m-1)}

\def\NQ{\left(1-\frac{\NF}{4N}\right)}
\def\NP{\left(1-\frac{\NF}{N}\right)}
\def\NF{N_{F}}
\def\NN{$\mathcal{N}=4$}
\DeclareMathOperator{\F}{\mathit{F}}

\DeclareMathOperator{\tr}{ {\rm tr}}
\def\trm{\tr_-}

\DeclareMathOperator{\Tri}{ {\rm F}^{1m}_3}
\DeclareMathOperator{\Ftme}{ {\rm F}^{2me}_4}
\DeclareMathOperator{\Fftme}{ {\rm F}^{2me}_{4F}}
\DeclareMathOperator{\Fom}{ {\rm F}^{1m}_4}
\DeclareMathOperator{\FFom}{ {\rm F}^{1m}_{4F}}
\def\FF#1{\sideset{_2}{_1}\F\bigg(#1\bigg)}

\def\mc#1{\mathcal{#1}}
\def\sl#1{\slash\!\!\!{#1}}
\def\hl#1.#2{\langle\hat{#1}#2\rangle}
\def\hr#1.#2{\langle#1\hat{#2}\rangle}
\def\bhl#1.#2.#3.#4{\langle #1\widehat{#2}#3\hat{#4}\rangle}
\def\thl#1.#2.#3{\langle#1\widehat{#2}#3]}
\def\spabt#1.#2.#3{$\begin{tiny}$\langle#1\,#2\,#3]$\end{tiny}$}

\def\dea{\langle \ell \ d \ell \rangle}
\def\deb{[\ell \ d \ell]}
\def\dex{\int_0^1 dx}
\def\dedea{\langle d \ell \ \partial_\ell \rangle}
\def\dedeb{[d \ell \ \partial_{\tilde{\ell}}]}
\def\nn{{\nonumber}}

\preprint{
  IPPP/08/17\\
  CERN-PH-TH/2008-082\\
  \today}

\title{One-loop $\phi$-MHV amplitudes using the unitarity 
bootstrap: the general helicity case}

\author{E. W. Nigel Glover$^*$,
    \ Pierpaolo Mastrolia$^\dagger$,
    \ Ciaran Williams$^*$
	\\
	$^*$Department of Physics, University of Durham, Durham, DH1 3LE, UK
	\\
        $^\dagger$Theory Division, CERN CH-1211 Geneva 23, Switzerland
	\\
	E-mails: 
        {\tt e.w.n.glover@durham.ac.uk}, 
        {\tt Pierpaolo.Mastrolia@cern.ch}, 
	{\tt Ciaran.Williams@durham.ac.uk}.
}

\abstract{ We consider a Higgs boson coupled to gluons via the  five-dimensional
effective operator $H {\rm tr} G_{\mu\nu}G^{\mu\nu}$. We treat $H$ as the real
part of a complex field $\phi$ that couples to the selfdual gluon field
strengths and compute the one-loop corrections to the $\phi$-MHV amplitudes
involving $\phi$, two negative helicity gluons and an arbitrary number of
positive helicity gluons. Our results generalise earlier work where the two
negative helicity gluons were constrained to be colour adjacent.
We use four-dimensional unitarity to construct the
cut-containing contributions and the recently developed recursion relations to
obtain the rational contribution for an arbitrary number of external gluons. We
solve the recursion relations and give explicit results for up to four external
gluons. These amplitudes are relevant for Higgs plus jet production via gluon
fusion in the  limit where the top quark mass is large compared to all other
scales in the problem. \\  \today }

\begin{document}
\section{Introduction}
The startup of the LHC anticipated for the autumn of 2008  heralds the arrival
of a new arena for the exploration of particle physics.  The large centre of
mass energy is expected to produce complex multiparticle final states both as
decay products of putative new physics Beyond the Standard Model and through the
Standard Model itself.   Extracting the signals of new phenomena and
discriminating between different models of new physics is only possible if the
predictions  for the Standard Model, and its prominent extensions, have
sufficient accuracy. The precision which can be achieved using calculations at
leading order in perturbation theory is, in most cases, not sufficient for
detailed studies of signals and especially backgrounds at the LHC.  In many
cases, the calculation of multi-particle final states at next to leading order
(NLO) will be essential to the successful interpretation of the data. Over the
past few years vast leaps in our understanding of the structure of one-loop
amplitudes in gauge theories has lead to the widespread belief that soon
predictions for many multi-jet final states will soon become available. 

The use of four-dimensional on-shell techniques, originally pioneered by Bern et
al~\cite{BDDK:uni1,BDDK:uni2}  in the mid-90's has lead to a vast
reduction in the complexity of one-loop calculations. The use of gauge-invariant
physical amplitudes (at tree level) as building blocks means that
simplifications due to the large cancellation of Feynman diagrams occur in the
preliminary stages of the calculation, rather than the latter.  The unitarity
method sews together four-dimensional tree-level amplitudes and, using unitarity
to reconstruct the (poly)logarithmic cut constructible part of the amplitude,
successfully reproduces the coefficients of the cut-constructible pieces of a
one-loop amplitude. This has extensive uses in supersymmetric Yang-Mills
theories, which are cut-constructible i.e. the whole amplitude can be
reconstructed from knowledge of its discontinuities. 

The more modern applications of unitarity were kick-started by the discovery of
the MHV rules by Witten and collaborators in 2004~\cite{Cachazo:MHVtree}.  The
realisation that MHV tree amplitudes could be promoted to vertices which could
be used to create amplitudes with any number of negative helicity gluons sparked
a revolution in the field of on-shell QCD. In a series of remarkable papers,
Brandhuber, Spence and Travaglini (BST)~\cite{Brandhuber:n4} showed how the MHV
rules can be used at one-loop for the calculation of $n$-point gluonic MHV
amplitudes.  Around the same time, the quadruple cut~\cite{Britto:genuni}  using
complex momenta was  introduced to reduce the determination of the coefficients
of box integrals to simple algebraic manipulation of four tree level
amplitudes.  
Double and triple unitarity cuts 
have led to direct 
techniques for extracting triangle and bubble integral 
coefficients analytically~\cite{Britto:sqcd,Britto:ccqcd,Mastrolia:2006ki,Forde:intcoeffs}.
In cases where fewer than four denominators are cut, the loop momentum
is not frozen, so the explicit integration over the phase space is
still required.
In the BBCFM-approach ~\cite{Britto:sqcd,Britto:ccqcd,Mastrolia:2006ki}, 
double or
triple cut phase-space integration has been reduced to  
extraction of residues in spinor variables, and, in the case of a
triple cut, residues in a Feynman parameter.
This method has been recently used for the 
evaluation of the complete six-photon 
amplitudes~\cite{Binoth:rational,Binoth:2007ca}.

Despite its success, the four-dimensional unitarity method does not give the
complete result for  non-supersymmetric theories such as QCD, since there are
missing rational functions which are cut-free and as result do not possess 
discontinuities in physical channels.  
The missing rational parts have only simple poles and
are therefore tree-like.  Since the rational pieces of one-loop amplitudes are
tree like in their discontinunity structure they can be calculated using a
straightforward generalization of the tree level recursion relations. One can
then use  the tree-level on-shell recursion
relations~\cite{Britto:rec,Britto:proof}  to compute the rational pieces of
one-loop amplitudes recursively.  The ability to calculate the rational pieces
of amplitudes independently of the cut-constructible terms lead to the
development of the unitarity bootstrap
approach~\cite{BDK:1lonshell,BDK:1lrecfin,Bern:bootstrap,Forde:MHVqcd,Berger:genhels,Berger:allmhv,Badger:2007si}.
Recently, an automated package {\tt BlackHat}
has been developed to compute these rational terms for pure QCD amplitudes~\cite{Berger:2008sj}.

Another approach is to extend use of unitarity to
$D=4-2\epsilon$
dimensions~\cite{Brandhuber:ddimgu,Anastasiou:DuniI,Mastrolia:2006ki,Britto:Duni,Anastasiou:DuniII,Britto:2008sw,Britto:2008vq}
and to take the cut particles into $D=4-2\epsilon$ dimensions. This approach has
the great advantage of calculating both the cut containing and the elusive
rational terms at once, but care must be taken with application of the
four-dimensional spinor helicity formalism in $D$ dimensions. 

It has also been observed that the rational parts are related to the ultraviolet
behaviour of the amplitude, and can be directly obtained from the traditional
Feynman diagram approach~\cite{Xiao:rationalI,
Su:rationalII,Xiao:rationalIII,Binoth:rational}. In a very interesting work,
Ossola, Papadopoulos and  Pittau~\cite{Ossola:2006us} have applied the
unitarity ideas directly to the integrand of the Feynman amplitude, without
necessarily appealing to the simplified forms of the cut diagrams.  They find
algebraic identities which can be automatically solved to give the coefficients
of the master integrals as well as the rational part.  This approach is being
further
developed~\cite{Ellis:2007br,Ossola:2007ax,Giele:2008ve,Ossola:2008xq,Ellis:2008kd,Mastrolia:2008jb,Binoth:2008kt}
with a view to providing automated computations of both cut-constructible {\it
and} rational parts of one-loop scattering amplitudes.
A summary of the current state of the art is given in Ref.~\cite{Bern:2008ef}.

In this paper, we exploit the unitarity  bootstrap
approach~\cite{BDK:1lonshell,BDK:1lrecfin,Bern:bootstrap,Forde:MHVqcd,Berger:genhels,Berger:allmhv} 
which meshes together
the calculation of the cut-constructible parts of an amplitude (via generalised
unitarity, one-loop MHV rules etc.) with the ability of the BCFW recursion
relations to calculate the rational pieces. As a result of the splitting the
total amplitude is given by the combination

\begin{equation}
A^{(1)}_n=C_n+R_n.
\end{equation}

Here the $C_n$ are the purely cut-constructible pieces which arise from box,
triangle and bubble (and in massive theories tadpole) loop integrals, the
functions in $C_n$ are those which contain discontinuities, in general
poly-logarithims (and associated $\pi^2$ terms).  $C_n$ may contain
unphysical singularities which are produced by tensor loop integrals
and must be cancelled by rational contributions. To make this cancellation
explicit, 
we add the cut-completion terms $CR_n$, 
so that the ``full'' cut-constructible pieces are
defined as,
\begin{equation}
\hat{C}_n=C_n+CR_n.
\end{equation}
These additional rational terms would be double counted if we
naively calculated the rational terms with the BCFW recursion relations, so we
redefine the rational pieces as
\begin{equation}
\hat{R}_n=R_n-CR_n.
\end{equation}
The rational part now contains only simple poles, and can, in principle, be
constructed recursively using the multiparticle factorisation properties of
amplitudes. We label this direct recursive term by $R_n^D$.  
By construction, the recursive approach generates the 
complete residues of physical poles. 
However, the cut-completion term $CR_n$ may also produce a contribution at the
residue of the physical poles, and may lead to double counting.   These
potential unwanted contributions are removed by the overlap terms, $O_n$. 

To generate the recursive contribution, one generally shifts two of the external
momenta by an amount proportional to $z$.  Complex analysis~\cite{Britto:proof} 
then generates the correct amplitude provided that  
\begin{equation}
\label{eq:zcond}
A_n(z) \to 0\qquad {\rm as} \qquad z \to \infty.
\end{equation}
For a generic tree-level process it is
frequently possible to shift two momenta such that \eqref{eq:zcond} is obeyed.
Similarly, for one-loop processes, one can often make a similar shift.
However, because the choice of $CR_n$ is not unique, the shift 
may introduce a  ``spurious" large $z$ behaviour in $CR_n$, 
labelled by $\mathrm{Inf}\, CR_n$,
which should be explicitly removed~\cite{Berger:genhels,Berger:allmhv}.
The rational part (provided that $A_n(z) \to 0$ as $z \to
\infty$) is given by,
\begin{equation}
\hat{R}_n=R_n^D+O_n-\mathrm{Inf}\, CR_n,
\end{equation}
while the physical one-loop amplitude   
is given by~\cite{Berger:genhels,Berger:allmhv},
\begin{equation}
A^{(1)}_n=C_n+CR_n+R_n^D+O_n-\mathrm{Inf}\, CR_n.
\end{equation}

In this paper, we focus on the 
$\phi$-MHV amplitudes involving $\phi$, two negative helicity gluons and an arbitrary number of
positive helicity gluons. Our results generalise earlier work~\cite{Badger:2007si}
where the two negative helicity gluons were constrained to be colour adjacent.
The paper proceeds as follows. In section~\ref{sec:higgs}, 
we give a brief overview of
the Higgs couples to gluons, and how this is related to $\phi$-amplitudes.
Section~\ref{sec:cutconstructible}
reviews the four-dimensional unitarity methods
for constructing the
cut-containing contribution $C_n$.   There are many similarities with the
pure-gluon case, and we develop the derivation of the cut-constructible parts 
of pure-glue MHV amplitudes 
and $\phi$-MHV amplitudes in sections~\ref{subsec:pureglue} and \ref{subsec:phi}.
Section~\ref{sec:rational} deals with computation of the three separate
rational pieces, the cut-reconstructible part $CR_n$, 
the on-shell recursive part $R_n^D$ and the overlap term $O_n$.
As an example,  we derive the
four-point amplitudes $A^{(1)}_4(\phi,1^-,2^+,3^-,4^+)$ and 
$A^{(1)}_4(H,1^-,2^+,3^-,4^+)$ in section~\ref{sec:fourpoint}, 
while section~\ref{sec:checks} describes 
the checks we have
performed on our result. Finally, in section~\ref{sec:conclusions}, 
we present our conclusions.
Two
appendices detailing the explicit construction of the cut-completion terms and
the forms of the one-loop basis functions are enclosed.

\section{The Higgs Model}
\label{sec:higgs}

The coupling of the Higgs to gluons in the Standard Model is produced via 
a fermion loop. Since the Yukawa coupling depends on the mass of the 
fermion, the interaction is dominated by the top quark loop. 
For large $m_t$ this can be integrated out, leading to an effective interaction, 
\begin{equation}
\mathcal{L}^{int}_{H}=\frac{C}{2}H\,\mathrm{tr}\,G_{\mu\nu}G^{\mu\nu}.
\end{equation}
This approximation works very well when the kinematic
scales involved are smaller
than twice the top quark mass~\cite{Kramer:1996iq,Baur:ptdist,Ellis:ptdist}.
For the interesting $pp \to H$ plus two jet process, the approximation is valid
when $m_H,~p_{Tj} < m_t$~\cite{DelDuca:2001fn}. 
The strength of the interaction $C$ has been calculated through to 
order $\mathcal{O}(\alpha_s^4)$ in the 
standard model~\cite{Chetyrkin:heffalpha3}. To order $\mathcal{O}(\alpha_s^2)$~\cite{Inami:Heff2l},
this is
\begin{equation}
\label{eq:C}
C=\frac{\alpha_s}{6\pi v}\bigg(1+\frac{11}{4}\frac{\alpha_s}{\pi}+\dots\bigg)
\end{equation}

The MHV-structure of Higgs-plus-gluons is best 
understood~\cite{Dixon:MHVhiggs} by defining the Higgs to be the real part of a complex scalar $\phi=\frac{1}{2}(H+iA)$ so that 
\begin{eqnarray}
\mathcal{L}^{int}_{\phi,\phi^{\dagger}}=C\bigg[\phi \mathrm{tr}G_{SD\,\mu\nu}G^{\mu,\nu}_{SD}+\phi^{\dagger}\mathrm{tr}G_{ASD\,\mu\nu}G^{\mu,\nu}_{ASD}\bigg]
\end{eqnarray}
where the purely selfdual (SD) and purely anti-selfdual gluon field strength tensors are given as
\begin{eqnarray}
G^{\mu\nu}_{SD}=\frac{1}{2}(G^{\mu\nu}+ \,^*G^{\mu\nu}) \quad G^{\mu\nu}_{ASD}=\frac{1}{2}(G^{\mu\nu}-\, ^*G^{\mu\nu}),
\end{eqnarray}
with
\begin{equation}
^*G^{\mu\nu}=\frac{i}{2}\epsilon^{\mu\nu\rho\sigma}G_{\rho\sigma}.
\end{equation}
Because of selfduality, the amplitudes for $\phi$ and $\phi^{\dagger}$ have a simpler  structure than those for the Higgs field~\cite{Dixon:MHVhiggs}. 
The following relations allow for the construction of Higgs amplitudes from those involving $\phi$ and $\phi^{\dagger}$.
\begin{eqnarray}
\label{eq:H}
A^{(m)}_n(H,g^{\lambda_1}_1,\dots,g^{\lambda_n}_n)=A^{(m)}_n(\phi,g^{\lambda_1}_1,\dots,g^{\lambda_n}_n)+A^{(m)}_n(\phi^{\dagger},g^{\lambda_1}_1,\dots,g^{\lambda_n}_n),\\
\label{eq:A}
A^{(m)}_n(A,g^{\lambda_1}_1,\dots,g^{\lambda_n}_n)=
\frac{1}{i}\left(
A^{(m)}_n(\phi,g^{\lambda_1}_1,\dots,g^{\lambda_n}_n)-A^{(m)}_n(\phi^{\dagger},g^{\lambda_1}_1,\dots,g^{\lambda_n}_n)
\right).
\end{eqnarray}
Furthermore parity relates $\phi$ and $\phi^{\dagger}$ amplitudes,
\begin{equation}
\label{eq:phidagger}
A^{(m)}_n(\phi^{\dagger},g^{\lambda_1}_1,\dots,g^{\lambda_n}_n)=\bigg(A^{(m)}_n(\phi,g^{-\lambda_1}_1,\dots,g^{-\lambda_n}_n)\bigg)^*.
\end{equation}
From now on, we will only consider $\phi$-amplitudes, knowing that all others can be obtained
using eqs.~(\ref{eq:H})--(\ref{eq:phidagger}).

The tree level amplitudes linking a $\phi$ with $n$ gluons 
can be decomposed into colour ordered amplitudes as~\cite{Dawson:Htomultijet,DelDuca:Hto3jets},
\begin{align}
	{\cal A}^{(0)}_n(\phi,\{k_i,\lambda_i,a_i\}) = 
	i C g^{n-2}
	\sum_{\sigma \in S_n/Z_n}
	\tr(T^{a_{\sigma(1)}}\cdots T^{a_{\sigma(n)}})\,
	A^{(0)}_n(\phi,\sigma(1^{\lambda_1},..,n^{\lambda_n})).
	\label{TreeColorDecompositionQ}
\end{align}
Here $S_n/Z_n$ is the group of non-cyclic permutations on $n$
symbols, and $j^{\lambda_j}$ labels the momentum $p_j$ and helicity
$\lambda_j$ of the $j^{\rm th}$ gluon, which carries the adjoint
representation index $a_i$.  The $T^{a_i}$ are fundamental
representation SU$(N_c)$ color matrices, normalized so that
${\rm Tr}(T^a T^b) = \delta^{ab}$.  The strong coupling constant is
$\alpha_s=g^2/(4\pi)$.

Tree-level amplitudes with a single quark-antiquark pair 
can be decomposed into colour-ordered amplitudes as follows,
\begin{eqnarray}
\lefteqn{
{\cal A}^{(0)}_n(\phi,\{p_i,\lambda_i,a_i\},\{p_j,\lambda_j,i_j\}) }\\
&&= 
i C g^{n-2}
\sum_{\sigma \in S_{n-2}} (T^{a_{\sigma(2)}}\cdots T^{a_{\sigma(n-1)}})_{i_1i_n}\,
A_n(\phi,1^{\lambda},\sigma(2^{\lambda_2},\ldots,{(n-1)}^{\lambda_{n-1}}),
n^{-\lambda})\,.\nonumber 
\label{TreeColorDecomposition}
\end{eqnarray}
where $S_{n-2}$ is the set of permutations of $(n-2)$ gluons.
Quarks are characterised with fundamental colour 
label $i_j$  and
helicity $\lambda_j$ for $j=1,n$.
By current conservation, the quark and antiquark helicities are  related such
that $\lambda_1 = -\lambda_n \equiv \lambda$ where $\lambda = \pm \hf$.

The one-loop amplitudes which are the main subject of this paper
follow the same colour ordering as the pure QCD amplitudes \cite{BDDK:uni1,Bern:1990ux}
and can be decomposed as \cite{Berger:higgsrecfinite,Badger:1lhiggsallm,Badger:2007si},
\begin{align}
	\mc A^{(1)}_n(\phi,\{k_i,\lambda_i,a_i\}) &= i C g^{n}
	\sum_{c=1}^{[n/2]+1}\sum_{\sigma \in S_n/S_{n;c}} G_{n;c}(\sigma)
	A^{(1)}_n(\phi,\sigma(1^{\lambda_1},\ldots,n^{\lambda_n}))
	\label{eq:1lhtogcolour}
\end{align}
where
\begin{align}
	G_{n;1}(1) &= N \tr( T^{a_1}\cdots T^{a_n} ) \\
	G_{n;c}(1) &=   \tr( T^{a_1}\cdots T^{a_{c-1} } )
			\tr( T^{a_c}\cdots T^{a_n} )
	\,\, , \, c>2.
	\label{eq:colourfactors}
\end{align}
The sub-leading terms can be computed by summing over various permutations of the leading colour
amplitudes~\cite{BDDK:uni1}

The tree level $\phi$-MHV amplitude has the same form as the pure-glue MHV amplitude,
\begin{equation}
A^{(0)}_n(\phi,1^-,2^+,\dots,m^-,\dots,n^+)=\frac{\spa1.m^4}{\spa1.2\dots\spa n.1}.
\end{equation}
The only difference between the gluon only and the $\phi$-MHV amplitude being momentum conservation, here the sum of all the
gluon momenta equals $-p_{\phi}$. 
Since we will encounter MHV diagrams in which a fermion circulates in the loop we will also need the amplitudes involving a $\phi$ with a quark anti-quark
pair~\cite{Badger:MHVhiggs2},
\begin{eqnarray} 
A^{(0)}_n(\phi,1_{q}^-,2^+,\dots,m^-,\dots,n_{\overline{q}}^+)=\frac{\spa1.m^3\spa n.m}{\spa1.2\dots\spa n.1},\nonumber\\
A^{(0)}_n(\phi,1_{q}^+,2^+,\dots,m^-,\dots,n_{\overline{q}}^-)=\frac{\spa n.m^3\spa 1.m}{\spa1.2\dots\spa n.1}.
\end{eqnarray}
Also as a consequence of the 1-loop nature of the $\phi$-gluon vertex the following all minus amplitude is non-zero at tree-level;
\begin{equation}
A^{(0)}_n(\phi,1^-,2^-,\dots,n^-)=(-1)^n\frac{m_{\phi}^4}{\spb1.2\dots\spb n.1}.
\end{equation}

Amplitudes with fewer (but more than two) negative helicities have been 
computed with Feynman diagrams (up to 4 partons) in Ref.~\cite{DelDuca:Hto3jets}
and using MHV rules and on-shell recursion relations in Refs.~\cite{Dixon:MHVhiggs,Badger:MHVhiggs2}.
The MHV amplitude for an arbitrary number of gluons but with two adjacent negative helicity
gluons was computed in Refs.~\cite{Badger:2007si,Badger:2007gy}.

In this paper we concentrate on the general helicity case for the one-loop $\phi$-MHV amplitude.
For definiteness, we focus on the specific helicity configuration  $(1^-,\ldots,m^-,\cdots,n^+)$.
Throughout,  we will use the notation,
\begin{eqnarray}
&&s_{i,j}=(p_i+p_{i+1}+\dots+p_{j-1}+p_j)^2 = P^{2}_{(i,j)} \nonumber\\
&&s_{ij}=2(p_i.p_j)=\spa i.j\spb j.i,
\end{eqnarray}
with the exception of section~\ref{sec:fourpoint} where we use the notation $P_{abc}$ to represent $p_a+p_b+p_c$.

\section{The cut-constructible parts}
\label{sec:cutconstructible}

The calculation of the cut-constructible terms has been performed within both the BST
approach~\cite{Brandhuber:n4} and the BBCFM approach~\cite{Britto:sqcd,Britto:ccqcd}. Both
methods rely on reconstructing the amplitude using four-dimensional  unitarity with a double
cut.  Compared to conventional methods,  one is attempting to compute the (four-dimensional)
coefficients of the loop integrals as efficiently as possible.  The methods  differ in how the
integration over the phase space of the cut particles is carried out.    The BST method uses
Passarino-Veltman techniques  to eliminate any remaining tensor integrals, and aims to cast the
integrand into the form of well-known phase space integrals.   It has been shown to work well
for MHV amplitudes.

On the other hand, in the BBCFM method, the use of spinor variables yields an alternative to
the Passarino-Veltman reduction of tensor integrals,  based on spinor algebraic manipulation
and  integration of complex analytic functions. It has been applied successfully to non-MHV
amplitudes. Here, we use both methods as a check of our results.

\begin{figure}[t]
	\psfrag{L1}{$L_1$}
	\psfrag{L2}{$L_2$}
	\psfrag{A1}{$A_L$}
	\psfrag{A2}{$A_R$}
	\begin{center}
		\includegraphics[width=6cm]{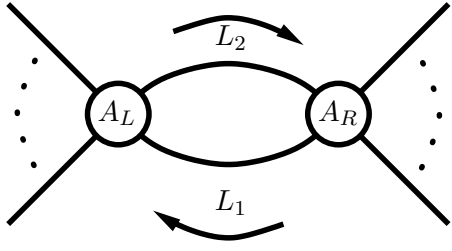}
	\end{center}
	\caption{A generic one-loop MHV diagram or unitarity cut.}
	\label{fig:1lmhvdiag}
\end{figure}

\subsection{The BST approach}

In the BST approach~\cite{Brandhuber:n4}
a generic diagram can be written:
\begin{align}
	\mc D = \frac{1}{(2\pi)^4}\int \frac{d^4L_1}{L_1^2}\frac{d^4L_2}{L_2^2}
	\delta^{(4)}(L_1-L_2-P)A_L(l_1,-P,-l_2) A_R(l_2,P,-l_1)
\end{align}
where $A_{L(R)}$ are the amplitudes for the left(right) vertices and $P$ is the sum of momenta
incoming to the right hand amplitude. The key step in the
evaluation of this expression is to re-write the integration measure as an integral over the
on-shell degrees of freedom and a separate integral over the complex variable $z$
\cite{Brandhuber:n4}:
\begin{align}
	\frac{d^4L_1}{L_1^2}\frac{d^4L_2}{L_2^2} &=
	(4i)^2\frac{dz_1}{z_1}\frac{dz_2}{z_2}d^4l_1d^4l_2\delta^{(+)}(l_1^2)\delta^{(+)}(l_2^2)\nonumber\\
		&=(4i)^2 \frac{2 dzdz'}{(z-z')(z+z')}d^4l_1d^4l_2\delta^{(+)}(l_1^2)\delta^{(+)}(l_2^2),
	\label{eq:intmeasure}
\end{align}
where $z=z_1-z_2$ and $z'=z_1+z_2$. The integrand can only depend on $z,z'$ through the momentum
conserving delta function,
\begin{equation}
	\delta^{(4)}(L_1-L_2-P) = \delta^{(4)}(l_1-l_2-P+z\eta) = \delta^{(4)}(l_1-l_2-\wh{P}),
\end{equation}
where $\wh{P} = P-z\eta$. This means that the integral over $z'$ can be performed so that,
\begin{align}
	\mc D &=\frac{(4i)^2 2\pi i}{(2\pi)^4}\int\frac{dz}{z}\int d^4l_1 d^4l_2\delta^{(+)}(l_1^2)\delta^{(+)}(l_2^2)
	\delta^{(4)}(l_1-l_2-\wh{P}) A_L(l_1,-P,-l_2)  A_R(l_2,P,-l_1) \nonumber\\
	&=(4i)^2 2\pi i\int\frac{dz}{z}\int d{\rm LIPS}^{(4)}(-l_1,l_2,\wh{P}) A_L(l_1,-P,-l_2)
	 A_R(l_2,P,-l_1),
\end{align}
where,
\begin{equation}
	d{\rm LIPS}^{(4)}(-l_1,l_2,\wh{P}) = \frac{1}{(2\pi)^4} d^4l_1 d^4l_2\delta^{(+)}(l_1^2)\delta^{(+)}(l_2^2)\delta^{(4)}(l_1-l_2-\wh{P})	
\end{equation}
The phase space integral is regulated using dimensional regularisation. 
Tensor integrals arising from the product of tree amplitudes can be reduced 
to scalar integrals either by using spinor algebra
or standard Passarino-Veltman reduction. The remaining scalar integrals have been evaluated
previously by van Neerven~\cite{vanNeerven:dimreg}. 

At this point, one has obtained the discontinuity, or imaginary part,
of the amplitude.   However, by making
a change of variables
the final integration over the $z$ variable
can be cast as a dispersion integral
\begin{equation}
	\frac{dz}{z} = \frac{d(\wh{P})^2}{\wh{P}^2-{P}^2}
\end{equation}
that re-constructs the full (cut-constructible part of the) amplitude.
 
\subsection{Spinorial Integration}

\def\spab#1.#2.#3{\left\langle#1|\,#2\,|#3\right]}
\def\spba#1.#2.#3{\left[#1|\,#2\,|#3\right\rangle}
\def\spaa#1.#2.#3.#4{\left\langle#1|\,#2\,#3\,|#4\right\rangle}

In the BBCFM approach~\cite{Britto:sqcd,Britto:ccqcd}, we make a conventional
double cut, so that
a generic diagram can be written:
\begin{align}
	\mc D = \frac{1}{(2\pi)^4}\int \frac{d^4l_1}{l_1^2}\frac{d^4l_2}{l_2^2}
	\delta^{(4)}(l_1-l_2-P)A_L(l_1,-P,-l_2) A_R(l_2,P,-l_1),
\end{align}
with $l_1^2=l_2^2=0$.

The double-cut can be written as,
\bea
&& {\cal D} = 
\int d{\rm LIPS}^{(4)} \ A_L(l_1,-P,-l_2) A_R(l_2,P,-l_1), 
\eea
where
 the $d{\rm LIPS}^{(4)}$ can be parametrised
in spinorial variables, as follows \cite{Cachazo:MHVtree},
\bea
\int d{\rm LIPS}^{(4)}  
&=&
\frac{1}{(2\pi)^4}
\int d^4 l_1d^4 l_2 \ \delta^{(+)}{(l_1^2)} \ \delta^{(+)}(l_2^2)\delta^{(4)}(l_1-l_2-P)
\nonumber \\
&=&
\frac{1}{(2\pi)^4}
\int {\dea \deb \over \spab \ell.P.\ell}
\int t \ dt \ 
\delta \Bigg(t - {P^2 \over \spab \ell.P.\ell}\Bigg) \ ,
\label{pm:phi4massless}
\eea
where the delta function eliminates the integration over $l_2$, and
the remaining $l_1$ integration variable has been rescaled, $l_1 \equiv t \ \ell$,
corresponding to,
\bea
|l_1\rangle \equiv \sqrt{t} \ |\ell \rangle \ , \qquad 
|l_1] \equiv \sqrt{t} \ |\ell]
\label{pm:rescaling}
\eea
with $l_1^2 = \ell^2 = 0$.
Accordingly, the double-cut can be written as,
\bea
&& {\cal D} =
\frac{1}{(2\pi)^4}
\int {\dea \deb \over \spab \ell.P.\ell}
\int t \ dt \ 
\delta \Bigg(t - {P^2 \over \spab \ell.P.\ell}\Bigg)  
\ A_L(t, |\ell\rangle, |\ell]) \ A_R(t, |\ell\rangle, |\ell])
\qquad \qquad
\eea
where we indicate only the dependence of the tree-level amplitudes
on the integration variables. 
By means of Schouten identities,
one can  disentangle the dependence on $|\ell\rangle$ and $|\ell]$, and
express the result of the $t$-integration (trivialised
by the presence of the $\delta$-function) as a combination 
of terms whose general form looks like,
\bea
&& {\cal D} = 
\frac{1}{(2\pi)^4}
\sum_i \int \dea \deb \ {\cal I}_i \ ,
\eea
with
\bea
{\cal I}_i &=& 
\rho_i\left(|\ell\rangle \right)
{ 
\spb \eta.\ell^{n}
\over 
\spab \ell.P_1.\ell^{n+1}
\spab \ell.P_2.\ell
}
\label{pm:eq:4Dgendeco}
\eea
where $P_1$ and $P_2$ can either be equal to the cut-momentum $P$, 
or be a linear combination of external vectors;
and where the $\rho_i$'s depend solely on one spinor flavour, say 
$|\ell\rangle$ 
(and not on $|\ell]$), and may contain poles in $|\ell\rangle$ through 
factors like $1/\spa \ell.\Omega$ 
(with $|\Omega\rangle$ being a massless spinor, either associated to 
any of the external legs, say $|k_i\rangle$, 
or to the action of a vector on it, like $\s{P} |k_i]$). \\
The explicit form of the vectors $P_1$ and $P_2$ 
in eq.~(\ref{pm:eq:4Dgendeco}) is determining 
the nature of the double-cut, logarithmic or not,
and correspondingly the topology of the 
diagram which is associated to.
Let us distinguish among the two possibilities one encounters,
in carrying on the spinor integration of ${\cal I}_i$:
\begin{enumerate}
\item $P_1 = P_2 =P$ (momentum across the cut).
In this case, the result is rational, hence containing only  
the cut of the 2-point function with external momentum $P$
(or degenerate 3-point functions which can be expressed as combination
of 2-point ones).
\item $P_1 = P$, $P_2 \ne P$, or $P_1 \ne P_2 \ne P$. 
In this case, the result is logarithmic, hence containing 
the cut of a linear combination of 
$n$-point functions with $n\ge3$.
\end{enumerate}

\noindent
If $P_1 = P_2 =P$,
\bea
{\cal I}_i &=& 
\rho_i\left( |\ell\rangle \right)
{ 
\spb \eta.\ell^{n}
\over 
\spab \ell.P.\ell^{n+2} 
} 
\ . \quad
\label{pm:eq:InoFpar}
\eea
If, however,  $P_1 = P$, $P_2 \ne P$ or $P_1 \ne P_2 \ne P$. 
one proceeds by introducing
a Feynman parameter, to write ${\cal I}_i$ as,
\bea
{\cal I}_i &=& 
(n+1) \dex \ 
(1-x)^n \ 
\rho_i\left( |\ell\rangle \right)
{ 
\spb \eta.\ell^{n}
\over 
\spab \ell.R.\ell^{n+2} 
} 
\ , \quad
\label{pm:eq:IwithFpar}
\eea
with 
\bea
\s{R} = x \s{P}_1 + (1-x) \s{P}_2 \ .
\label{pm:eq:Rdef}
\eea
The spinorial structure of eq.~(\ref{pm:eq:InoFpar}) and 
eq.~(\ref{pm:eq:IwithFpar}) is the same. Therefore,
we proceed with the spinor integration of eq.~(\ref{pm:eq:IwithFpar})
because it is more general than the eq.~(\ref{pm:eq:InoFpar}), 
because of the presence of the
Feynman parameter. 

First, the order of the integrations over the spinor variables and over 
the Feynman parameter is exchanged and we
perform the integration over the $|\ell]$-variable
by parts, using \cite{Britto:sqcd}
\be
\deb {\spb \eta.\ell^n \over \spab \ell.P.\ell^{n+2}} 
=
{\dedeb \over (n+1)}
{
\spb \eta.\ell^{n+1} 
\over \spab \ell.P.\ell^{n+1} \spab \ell.P.\eta} \ ,
\label{pm:ibp}
\ee
obtaining,
\bea
{\cal D}_i && = 
\frac{1}{(2\pi)^4}
\int \dea \deb \ {\cal I}_i = 
\nn \\
&& = 
\frac{1}{(2\pi)^4}
\dex \ (1-x)^n
\int \dea \dedeb
{
\rho_i(|\ell\rangle) \ 
\spb \eta.\ell^{n+1} 
\over 
\spab \ell.R.\ell^{n+1} 
\spab \ell.R.\eta
} \ .
\eea
Afterwards, the integration over the $|\ell\rangle$-variable is achieved  
using Cauchy's residues theorem, in the fashion of the holomorphic anomaly
\cite{Cachazo:2004by,Cachazo:2004dr,Britto:holo}, 
by taking the residues at $|\ell\rangle = \s{R}|\eta]$ and
at the simple poles of $\rho_i$, say $|\ell\rangle = |\ell_{ij} \rangle $,
\bea
{\cal D}_i &=& 
\frac{1}{(2\pi)^4}
\int \dea \deb \ {\cal I}_i = 
\nn \\
&=&
\frac{(2 \pi i)}{(2\pi)^4}
\dex \ (1-x)^n \ 
\bigg\{
{
\rho_i(\s{R}|\eta])
\over 
(R^2)^{n+1}
}
+
\sum_j \lim_{\ell \to \ell_{ij}} 
\spa \ell.{\ell_{ij}}
{
\rho_i(|\ell\rangle) \ 
\spb \eta.\ell^{n+1} 
\over 
\spab \ell.R.\ell^{n+1} 
\spab \ell.R.\eta
}
\bigg\} 
\ . \quad 
\label{pm:ImplicitDoubleCut}
\eea
To complete the integration of eq.~(\ref{pm:ImplicitDoubleCut}),
one has to perform the parametric integration which is finally
responsible for the appearence of logarithmic terms in the double-cut. 
Alternatively, the spinorial integration of eq.~(\ref{pm:eq:InoFpar})
would generate a pure rational contribution. 
We remark that the role of $|\ell\rangle$ and $|\ell]$ in the integration
could be interchanged.

\def\spab#1.#2.#3{\left\langle#1\,#2\,#3\right]}
\def\spba#1.#2.#3{\left[#1\,#2\,#3\right\rangle}
\def\spaa#1.#2.#3.#4{\left\langle#1\,#2\,#3\,#4\right\rangle}

\subsection{Gluonic amplitudes}
\label{subsec:pureglue}

We note that there are many similarities between $\phi$-amplitudes and pure glue
amplitudes, and we will exploit this by first rederiving the cut-constructible
contribution to pure glue MHV amplitudes with the same helicity configuration.

\begin{figure}[t]
	\psfrag{phi}{$\phi$}
	\psfrag{A}{$(a)$}
	\psfrag{B}{$(b)$}
	\psfrag{pm}{$\pm$}
	\psfrag{mp}{$\mp$}
	\psfrag{a}{$(j+1)^+$}
	\psfrag{b}{$1^-$}
	\psfrag{c}{$m^-$}
	\psfrag{d}{$i^+$}
	\psfrag{e}{$(i+1)^+$}
	\psfrag{f}{$j^+$}
	\psfrag{p}{$+$}
	\psfrag{m}{$-$}
	\begin{center}
		\includegraphics[width=12cm]{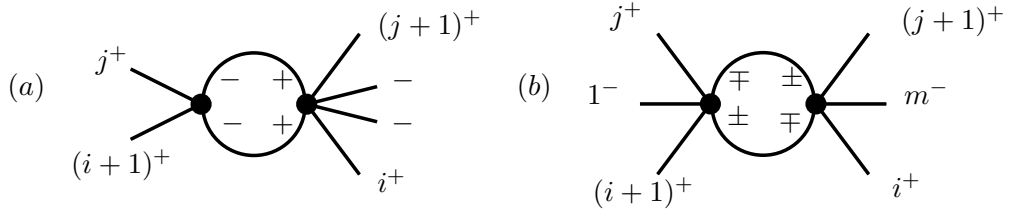}
	\end{center}
	\caption{The MHV diagrams contributing to one-loop gluonic MHV amplitudes}
	\label{fig:pureqcd}
\end{figure}

The graphs contributing to the one-loop gluonic amplitude
$A_n^{(1)}(1^-,\ldots,m^-,\ldots,n^+)$ 
are shown in Fig.~\ref{fig:pureqcd}.  There are two distinct
types of diagram, labelled (a) and (b).  In type (a), only gluons circulate in
the loop, while in type (b) gluons, fermions (and scalars) may circulate.
They can be characterised by the following sums

\begin{equation}
\label{eq:qcdsums}
(a)\sum_{i=m}^{n-2} \sum_{j=i+2}^{n}+\sum_{i=1}^{m-3}\sum_{j=i+2}^{m-1}
\qquad\qquad{\rm and}\qquad\qquad
(b)
\sum_{i=n}^{n}\sum_{j=2}^{m-1} 
+\sum_{i=m+1}^{n-1}\sum_{j=1}^{m-1} 
+\sum_{i=m}^{m}\sum_{j=1}^{m-2}.
\end{equation}

The various contributions have been computed using the MHV rules in 
Refs.~\cite{Brandhuber:n4,Bedford:n1,Bedford:nonsusy}.
We note that contributions of type (a) associated with a cut in the
$s_{(j+1),i}$ channel have an integrand of the form,
\begin{eqnarray}
\label{eq:typea}
\left(A_L A_R\right)_{(j+1),i}
&=&
\frac{\A{\ell_1}{\ell_2}^4}{\A{\ell_1}{(i+1)}\cdots\A{j}{\ell_2}\A{\ell_2}{\ell_1}}
\frac{\A{1}{m}^4}{\A{\ell_2}{(j+1)}\cdots\A{i}{\ell_1}\A{\ell_1}{\ell_2}}\nonumber
\\
&=& \frac{\A{1}{m}^4}{\A{1}{2}\cdots\A{n}{1}}
~\widehat{\cal G}(i,i+1,j,j+1)\nonumber \\
&\equiv&
A_n^{(0)}
~\widehat{\cal G} (i,i+1,j,j+1) 
\end{eqnarray}
where
\begin{equation}
\label{eq:gdef}
\widehat{\cal G} (i,i+1,j,j+1)  
=  \frac{\A{\ell_2}{\ell_1}\A{i}{(i+1)}}{\A{i}{\ell_1}\A{\ell_1}{(i+1)}}
\frac{\A{\ell_1}{\ell_2}\A{j}{(j+1)}}{\A{j}{\ell_2}\A{\ell_2}{(j+1)}}. 
 \end{equation}

For diagrams of type (b), there are three possible contributions - depending on whether
gluons, fermions (or for supersymmetric theories scalars) are circulating in the loop.
It is convenient to consider both (b)-type diagrams in the $s_{(j+1),i}$ channel together.
Immediately, we write down
\begin{eqnarray}
\left(A_L A_R\right)_{(j+1),i}^{\rm gluons} &=&
\frac{\A{1}{\ell_2}^4\A{m}{\ell_1}^4+\A{1}{\ell_1}^4\A{m}{\ell_2}^4}
{\A{\ell_1}{(i+1)}\cdots\A{j}{\ell_2}\A{\ell_2}{\ell_1}\A{\ell_2}{(j+1)}\cdots\A{i}{\ell_1}\A{\ell_1}{\ell_2}}
\nonumber \\
\left(A_L A_R\right)_{(j+1),i}^{\rm fermions}  &=&
\frac{\A{1}{\ell_1}\A{1}{\ell_2}^3\A{m}{\ell_2}\A{m}{\ell_1}^3+\A{1}{\ell_2}\A{1}{\ell_1}^3\A{m}{\ell_1}\A{m}{\ell_2}^3}
{\A{\ell_1}{(i+1)}\cdots\A{j}{\ell_2}\A{\ell_2}{\ell_1}\A{\ell_2}{(j+1)}\cdots\A{i}{\ell_1}\A{\ell_1}{\ell_2}}
\nonumber \\
\label{eq:sloop}
\left(A_L A_R\right)_{(j+1),i}^{\rm scalars}  &=&
\frac{\A{1}{\ell_1}^2\A{1}{\ell_2}^2\A{m}{\ell_2}^2\A{m}{\ell_1}^2}
{\A{\ell_1}{(i+1)}\cdots\A{j}{\ell_2}\A{\ell_2}{\ell_1}\A{\ell_2}{(j+1)}\cdots\A{i}{\ell_1}\A{\ell_1}{\ell_2}}
\nonumber \\
\end{eqnarray} 
In each case, the denominator has the same structure as in the (a)-type diagrams and only the numerator changes
with the particle type.
We can exploit the Schouten identity
\begin{equation}
\A{1}{\ell_2} \A{m}{\ell_1} -\A{1}{\ell_1} \A{m}{\ell_2} +\A{1}{m}\A{\ell_1}{\ell_2} = 0
\end{equation}
to rewrite each of the numerators into a simpler form. 
\begin{eqnarray}
\label{eq:gloop}
\A{1}{\ell_2}^4\A{m}{\ell_1}^4&+&\A{1}{\ell_1}^4\A{m}{\ell_2}^4
= \A{1}{m}^4\A{\ell_1}{\ell_2}^4 \nonumber \\
&&+4 \A{1}{\ell_2} \A{m}{\ell_1} \A{1}{\ell_1} \A{m}{\ell_2} \A{1}{m}^2\A{\ell_1}{\ell_2}^2\nonumber \\
&&+2 \A{1}{\ell_2}^2 \A{m}{\ell_1}^2 \A{1}{\ell_1}^2 \A{m}{\ell_2}^2\nonumber \\
&&\\
\label{eq:floop}
\A{1}{\ell_1}\A{1}{\ell_2}^3\A{m}{\ell_2}\A{m}{\ell_1}^3&+&\A{1}{\ell_2}\A{1}{\ell_1}^3\A{m}{\ell_1}\A{m}{\ell_2}^3
= 
\A{1}{\ell_2} \A{m}{\ell_1} \A{1}{\ell_1} \A{m}{\ell_2} \A{1}{m}^2\A{\ell_1}{\ell_2}^2\nonumber \\
&&+2 \A{1}{\ell_2}^2 \A{m}{\ell_1}^2 \A{1}{\ell_1}^2 \A{m}{\ell_2}^2\nonumber \\
\end{eqnarray}
We see that the first term on the RHS of eq.~(\ref{eq:gloop}) corresponds to an (a)-type gluonic contribution which we 
label with $G$, while the third term looks like the scalar
contribution  of eq.~(\ref{eq:sloop}) which we label with $S$. 
Similarly, the fermion contribution can be separated into a fermionic 
piece $F$ and
a scalar contribution $S$.
We define the three contributions as,
\begin{eqnarray}
\left(A_L A_R\right)_{(j+1),i}^{\rm G}  &=&
\frac{ \A{1}{m}^4\A{\ell_1}{\ell_2}^4}
{\A{\ell_1}{(i+1)}\cdots\A{j}{\ell_2}\A{\ell_2}{\ell_1}\A{\ell_2}{(j+1)}\cdots\A{i}{\ell_1}\A{\ell_1}{\ell_2}}
 \nonumber \\
 &=& A_n^{(0)}~\widehat{\cal G} (i,i+1,j,j+1)\\
\left(A_L A_R\right)_{(j+1),i}^{\rm F}  &=&
\frac{\A{1}{\ell_2} \A{m}{\ell_1} \A{1}{\ell_1} \A{m}{\ell_2} \A{1}{m}^2\A{\ell_1}{\ell_2}^2}
{\A{\ell_1}{(i+1)}\cdots\A{j}{\ell_2}\A{\ell_2}{\ell_1}\A{\ell_2}{(j+1)}\cdots\A{i}{\ell_1}\A{\ell_1}{\ell_2}}
 \nonumber \\
&=& -A_n^{(0)}~\widehat{\cal F} (i,i+1,j,j+1)\\
\left(A_L A_R\right)_{(j+1),i}^{\rm S}  &=&
\frac{\A{1}{\ell_1}^2\A{1}{\ell_2}^2\A{m}{\ell_2}^2\A{m}{\ell_1}^2}
{\A{\ell_1}{(i+1)}\cdots\A{j}{\ell_2}\A{\ell_2}{\ell_1}\A{\ell_2}{(j+1)}\cdots\A{i}{\ell_1}\A{\ell_1}{\ell_2}}
\nonumber\\
&=& -A_n^{(0)}~\widehat{\cal S} (i,i+1,j,j+1)
\end{eqnarray}
where $\widehat{\cal G} (i,i+1,j,j+1)$ is defined in eq.~\eqref{eq:gdef} and,
\begin{eqnarray}
\label{eq:fdef}
\widehat{\cal F} (i,i+1,j,j+1)&=& 
\frac{\A{i}{(i+1)}\A{j}{(j+1)}\A{1}{\ell_1} \A{m}{\ell_1} \A{1}{\ell_2} \A{m}{\ell_2}  }
{\A{1}{m}^2 \A{i}{\ell_1}\A{\ell_1}{(i+1)}\A{j}{\ell_2}\A{\ell_2}{(j+1)}}
\\
\label{eq:sdef}
\widehat{\cal S} (i,i+1,j,j+1)&=&
\frac{\A{i}{(i+1)}\A{j}{(j+1)}\A{1}{\ell_1}^2\A{m}{\ell_1}^2\A{1}{\ell_2}^2\A{m}{\ell_2}^2}
{\A{1}{m}^4\A{\ell_1}{\ell_2}^2\A{i}{\ell_1}\A{\ell_1}{(i+1)}\A{j}{\ell_2}\A{\ell_2}{(j+1)}}
\end{eqnarray}

Restoring the particle multiplicities in supersymmetric theories, we see that
for ${\cal N} = 4$ SYM with four fermions and six scalars (in the adjoint representation), 
only the ``gluonic'' part remains
\begin{equation}
\left(A_L A_R\right)_{(j+1),i}^{\rm gluons}-4\left(A_L A_R\right)_{(j+1),i}^{\rm fermions}
+6\left(A_L A_R\right)_{(j+1),i}^{\rm scalars}
= \left(A_L A_R\right)_{(j+1),i}^{\rm G}.
\end{equation}
On the other hand,  for QCD with $\NF$ fermion flavours in the fundamental representation, 
the contribution from this graph is,
\begin{equation}
\left(A_L A_R\right)_{(j+1),i}^{\rm QCD}
= \left(A_L A_R\right)_{(j+1),i}^{\rm G} 
+4\NQ \left(A_L A_R\right)_{(j+1),i}^{\rm F}
+2\NP \left(A_L A_R\right)_{(j+1),i}^{\rm S}.
\end{equation}

The functions ${\widehat {\cal X}}$ for ${\cal X} = G,F,S$ represent contributions to the cut amplitude.
Performing the phase space and dispersion integrals generates the ``cut-constructible''
contribution to the full amplitude.
We define,
\begin{equation}
\widehat{X}(i,i+1,j,j+1) = \int \frac{dz}{z} ~\int d^D{\rm LIPS}(-l_1,l_2,P)  ~\widehat{\cal X}(i,i+1,j,j+1).
\end{equation}
Explicit expressions for $\widehat{X}(i,i+1,j,j+1)$ are written down in Appendix~\ref{app:X}.
The one-loop gluonic MHV amplitude is thus obtained by summing combinations of 
the ``cut-constructible" contributions according to eq.~(\ref{eq:qcdsums}). 
As a result the one-loop gluonic MHV amplitude is given by,
\begin{eqnarray}
\lefteqn{C_{n;1}(1^-,2^+,\dots,m^-,\dots,n^+)}\nonumber \\
&=&
c_{\Gamma}A^{(0)}_{n}\bigg(A^{G}_{n;1}(m,n)-4\bigg(1-\frac{\NF}{4N}\bigg)A^{F}_{n;1}(m,n)-2\bigg(1-\frac{\NF}{N}\bigg)A^{S}_{n;1}(m,n)\bigg)
\label{eq:gluMHV}
\end{eqnarray}
where
\begin{eqnarray}
A^{G}_{n;1}(m,n)=-\frac{1}{2}\sum_{i=1}^n\Fom(s_{i,i+2};s_{i,i+1},s_{i+1,i+2})
-\frac{1}{4}\sum_{i=1}^{n}\sum_{j=i+3}^{n+i-3}
\Ftme(s_{i,j},s_{i+1,j-1};s_{i+1,j},s_{i,j-1}).\nonumber\\
\label{eq:Gqcdmhv}
\end{eqnarray}
The terms associated with the fermion loop have the following form:
\begin{eqnarray}
A^{F}_{n;1}(m,n)=
&&\sum_{i=m+1}^{n}\sum_{j=2}^{m-1} b^{ij}_{1m}\Fftme(s_{i,j},s_{i-1,j+1};s_{i-1,j},s_{i,j+1})\nonumber\\
&&-\sum_{i=2}^{m-1}\sum_{j=m}^{n} \frac{\trm(1,P_{(i,j)},i,m)}{s^2_{1m}}\mathcal{A}^{ij}_{1m}T_1(P_{(i+1,j)},P_{(i,j)})\nonumber\\
&&+\sum_{i=2}^{m}\sum_{j=m+1}^{n} \frac{\trm(1,P_{(i,j-1)},j,m)}{s^2_{1m}}\mathcal{A}^{j(i-1)}_{1m}T_1(P_{(i,j-1)},P_{(i,j)}).
\label{eq:Fqcdmhv}
\end{eqnarray}
Here we have introduced the shorthand notation
\bea
{\rm tr}_{-}(abcd) = \spa a.b \spb b.c \spa c.d \spb d.a
\eea
and the auxiliary functions,
\begin{eqnarray}
\label{eq:bdef}
&&b^{ij}_{1m}=\frac{\trm(m,i,j,1)\trm(m,j,i,1)}{s_{ij}^2s_{1m}^2}\\
\label{eq:Adef}
&&\mathcal{A}^{ij}_{1m}=\bigg(\frac{\trm(1,i,j,m)}{s_{ij}}-(j\rightarrow j+1)\bigg) \ ,
\end{eqnarray}
Note that $b^{ij}_{m1}$ is symmetric under both $i \leftrightarrow j$ and $1 \leftrightarrow m$,
while $\mathcal{A}^{ij}_{1m}$ is antisymmetric under $1 \leftrightarrow m$.
The function $\Fftme$ is the finite pieces of the two mass easy box function 
(or the finite pieces of the one mass box function in the limit where one of the massive legs becomes massless). 
We define the triangle function $T_i(P,Q)$ as,
\begin{equation}
T_i(P,Q)=L_i(P,Q)=\frac{\log{(P^2/Q^2)}}{(P^2-Q^2)^i} \qquad P^2 \neq 0, ~~Q^2 \neq 0.
\end{equation} 
If one of the invariants becomes massless then the triangle function becomes the divergent function,
\begin{equation}
T_i(P,Q)\rightarrow(-1)^i\frac{1}{\epsilon}\frac{(-P^2)^{-\epsilon}}{(P^2)^i}, ~~Q^2\rightarrow0.
\label{eq:Ti}
\end{equation}

The terms associated with a scalar circulating in the loop have the form,
\begin{eqnarray}
A^{S}_{n;1}(m,n)=
\sum_{i=m+1}^{n}\sum_{j=2}^{m-1}&&-(b^{ij}_{1m})^2\Fftme(s_{i,j},s_{i-1,j+1};s_{i-1,j},s_{i,j+1})\nonumber\\
+\sum_{i=2}^{m-1}\sum_{j=m}^{n}&\Bigg[ &
-\frac{\trm(1,P_{(i,j)},i,m)^3}{3s_{1m}^4} \mathcal{A}^{ij}_{1m} T_3(P_{(i+1,j)},P_{(i,j)}) \nonumber\\
&&-\frac{\trm(1,P_{(i,j)},i,m)^2}{2s_{1m}^4} \mathcal{K}^{ij}_{1m} T_2(P_{(i+1,j)},P_{(i,j)})\nonumber\\
&&+\frac{\trm(1,P_{(i,j)},i,m)}{s_{1m}^4} \mathcal{I}^{ij}_{1m}
T_1(P_{(i+1,j)},P_{(i,j)})\Bigg]
\nonumber\\
+\sum_{i=2}^{m}\sum_{j=m+1}^{n}&\Bigg[
 &+\frac{\trm(1,P_{(i,j-1)},j,m)^3}{3s_{1m}^4} \mathcal{A}^{j(i-1)}_{1m} T_3(P_{(i,j-1)},P_{(i,j)}) \nonumber\\
&&+\frac{\trm(1,P_{(i,j-1)},j,m)^2}{2s_{1m}^4} \mathcal{K}^{j(i-1)}_{1m} T_2(P_{(i,j-1)},P_{(i,j)})\nonumber\\
&&-\frac{\trm(1,P_{(i,j-1)},j,m)}{s_{1m}^4} \mathcal{I}^{j(i-1)}_{1m}    T_1(P_{(i,j-1)},P_{(i,j)})\Bigg].
\end{eqnarray}
Here we have introduced two further auxiliary functions which are defined as follows,
\begin{eqnarray}
\label{eq:Kdef}
&&\mathcal{K}^{ij}_{1m}=\bigg(\frac{\trm(1,i,j,m)^2}{s_{ij}^2}-(j\rightarrow j+1)\bigg),\\
\label{eq:Idef}
&&\mathcal{I}^{ij}_{1m}=\bigg(\frac{\trm(1,i,j,m)^2\trm(1,j,i,m)}{s_{ij}^3}-(j\rightarrow j+1)\bigg). 
\end{eqnarray}

\subsection{$\phi$-amplitudes}
\label{subsec:phi}

\begin{figure}[t]
	\psfrag{phi}{$\phi$}
	\psfrag{A}{$(a)$}
	\psfrag{B}{$(b)$}
	\psfrag{pm}{$\pm$}
	\psfrag{mp}{$\mp$}
	\psfrag{p}{$+$}
	\psfrag{m}{$-$}
	\psfrag{h}{$\phi$}
	\psfrag{a}{$(j+1)^+$}
	\psfrag{b}{$1^-$}
	\psfrag{c}{$m^-$}
	\psfrag{d}{$i^+$}
	\psfrag{e}{$(i+1)^+$}
	\psfrag{g}{$(i+2)^+$}
	\psfrag{f}{$j^+$}
	\begin{center}
		\includegraphics[width=12cm]{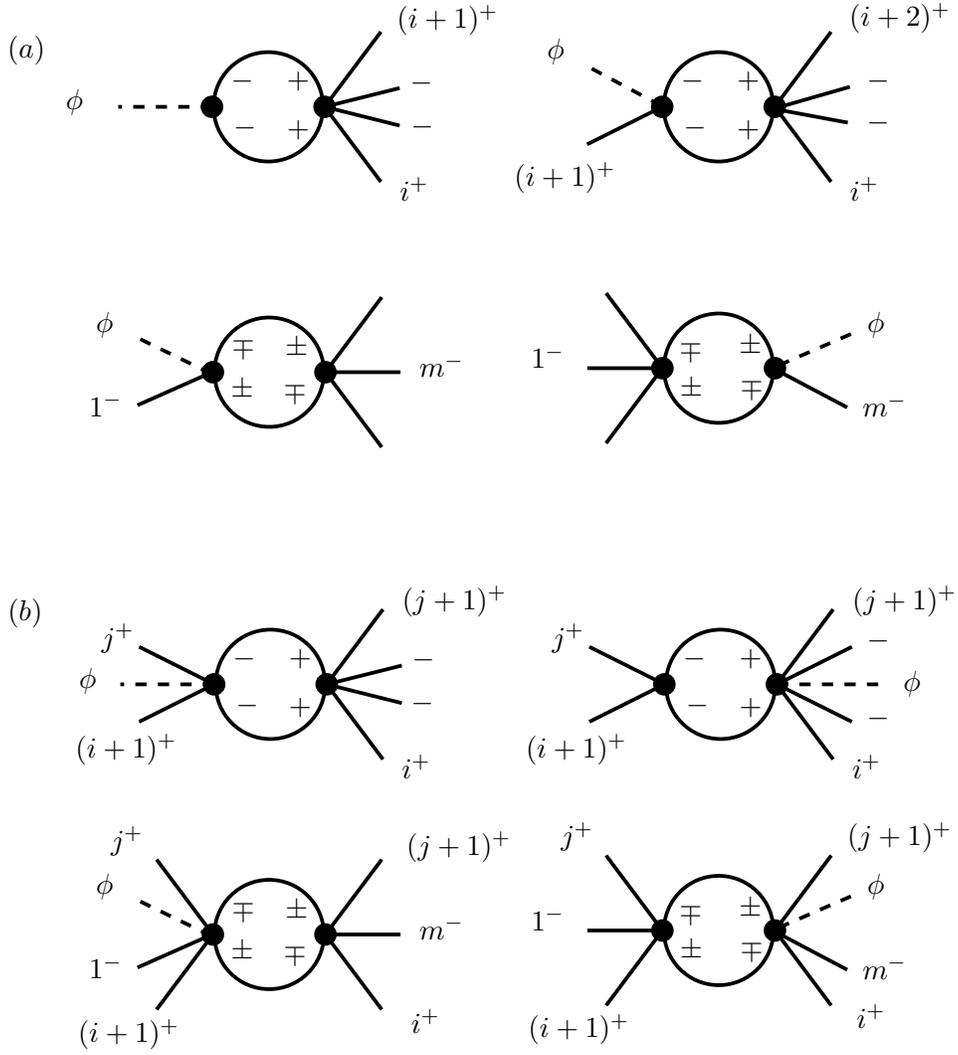}
	\end{center}
		\caption{The MHV diagrams contributing to one-loop $\phi$-MHV amplitudes. }
	\label{fig:phidiags}
\end{figure}

The graphs contributing to one-loop $\phi$-MHV amplitudes 
are shown in Fig.~\ref{fig:phidiags}.  Diagrams of type (b) are the QCD graphs dressed 
with an additional $\phi$, which may couple at either the left or right vertex.
The presence of the $\phi$ does not alter the spinor structure of the amplitudes, so these graphs
are exactly those for the pure-QCD amplitudes of the previous section, modified to
account for the momentum carried by the $\phi$.  The ranges of summations
correspond to those given in eq.~\eqref{eq:qcdsums}. 

On the other hand,  the diagrams shown in Fig.~\ref{fig:phidiags}(a) have no counterpart
in pure-QCD.  They all vanish in the limit where the four-momentum of the $\phi$ vanishes.
\begin{figure}[t]
	\psfrag{phi}{$\phi$}
	\psfrag{a}{$(i+1)^+$}
	\psfrag{m}{$-$}
	\psfrag{p}{$+$}
	\psfrag{b}{$i^+$}
	\psfrag{c}{$i^+$}
	\begin{center}
		\includegraphics[width=4cm]{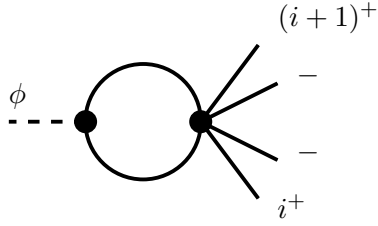}
	\end{center}
	\caption{A $\phi$ only diagram in the $s_{i+1,i}$ channel}
	\label{fig:dia1}
\end{figure}
The diagram contributing to a cut in the $s_{i+1,i}$ channel is shown in Fig.~\ref{fig:dia1}.
\begin{eqnarray}
\left(A_L A_R\right)_{1,n} &=& \frac{ \A{\ell_1}{\ell_2}^2 \A{1}{m}^4}{\A{\ell_2}{(i+1)}\cdots\A{i}{\ell_1}\A{\ell_1}{\ell_2}}
=
A_n^{(0)}~\frac{\A{i}{(i+1)}\A{\ell_1}{\ell_2}}{\A{\ell_2}{(i+1)}\A{i}{\ell_1}}\nonumber\\
&=&
A_n^{(0)}~\left(-1 + {\cal G}(i,i+1)\right)
\end{eqnarray}
with ${\cal G}(i,j)$ defined in eq.~(\ref{eq:GGij}).
\begin{figure}[t]
	\psfrag{phi}{$\phi$}
	\psfrag{a}{$(i+2)^+$}
	\psfrag{m}{$-$}
	\psfrag{p}{$+$}
	\psfrag{b}{$i^+$}
	\psfrag{c}{$(i+1)^+$}
	\begin{center}
		\includegraphics[width=4cm]{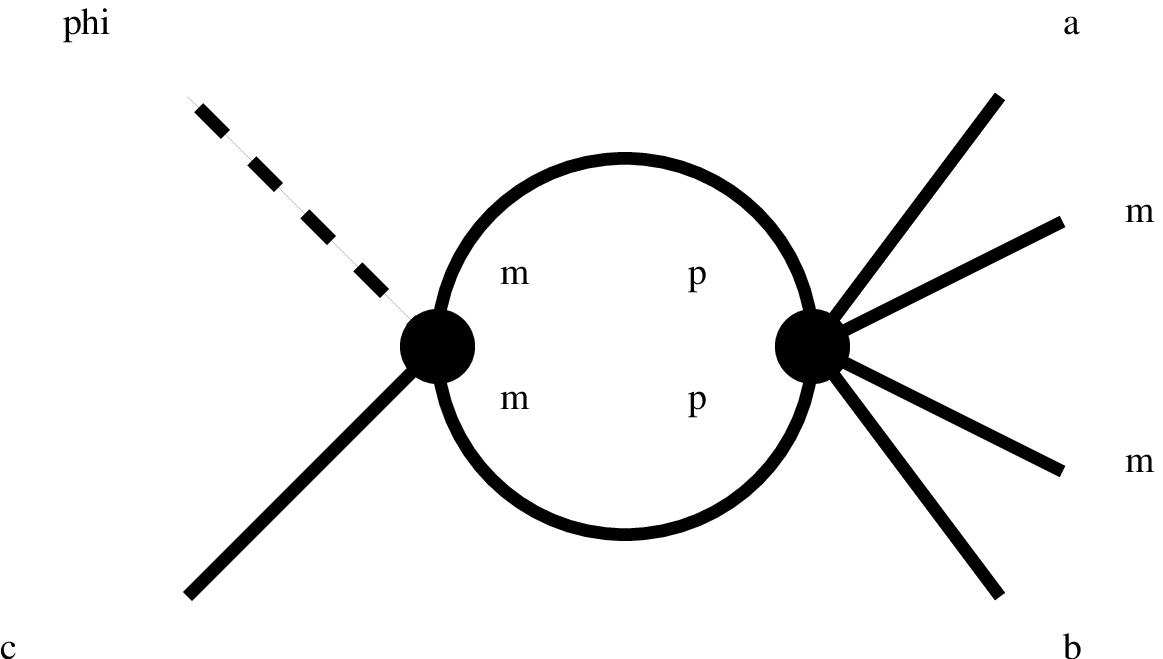}
	\end{center}
	\caption{A $\phi$ only diagram in the $s_{i+2,i}$ channel}
	\label{fig:dia2}
\end{figure}
The diagram contributing to a cut in the $s_{i+2,i}$ channel is shown in Fig.~\ref{fig:dia2}.
\begin{eqnarray}
\left(A_L A_R\right)_{i+2,i} &=& \frac{ \A{\ell_1}{\ell_2}^4 \A{1}{m}^4}{
\A{\ell_1}{(i+1)}\A{(i+1)}{\ell_2}\A{\ell_2}{\ell_1}\A{\ell_2}{(i+2)}\cdots\A{i}{\ell_1}\A{\ell_1}{\ell_2}}\nonumber\\
&=&
A_n^{(0)}~\widehat{\cal G}(i,i+1,i+1,i+2).
\end{eqnarray}
\begin{figure}[t]
	\psfrag{phi}{$\phi$}
	\psfrag{pm}{$\pm$}
	\psfrag{mp}{$\mp$}
	\psfrag{h}{$\phi$}
	\psfrag{b}{$1^-$}
	\psfrag{c}{$m^-$}
	\begin{center}
		\includegraphics[width=4cm]{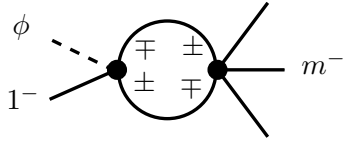}
	\end{center}
	\caption{A $\phi$ only diagram in the $s_{2,n}$ channel}
		\label{fig:dia3}
\end{figure}
The diagram contributing to a cut in the $s_{2,n}$ channel is shown in Fig.~\ref{fig:dia3}.
There are contributions from both gluon and fermion loops, and we find,
\begin{equation}
\left(A_L A_R\right)_{2,n}^{\rm QCD}
= A_n^{(0)} \left( \widehat{{\cal G}}(n,1,1,2)
-4\NQ \widehat{{\cal F}}(n,1,1,2)
-2\NP \widehat{{\cal S}}(n,1,1,2)\right).
\end{equation}
\begin{figure}[t]
	\psfrag{phi}{$\phi$}
	\psfrag{pm}{$\pm$}
	\psfrag{mp}{$\mp$}
	\psfrag{h}{$\phi$}
	\psfrag{b}{$1^-$}
	\psfrag{c}{$m^-$}
	\begin{center}
		\includegraphics[width=4cm]{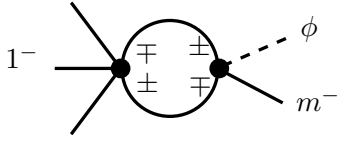}
	\end{center}
	\caption{A $\phi$ only diagram in the $s_{m+1,m-1}$ channel}
	\label{fig:dia4}
\end{figure}
The diagram contributing to a cut in the $s_{m+1,m-1}$ channel is shown in Fig.~\ref{fig:dia4}.
There are contributions from both gluon and fermion loops, and we find,
\begin{eqnarray}
\left(A_L A_R\right)_{m+1,m-1}^{\rm QCD}
&=& A_n^{(0)} \Biggl(\widehat{{\cal G}}(m,m+1,m-1,m)
-4\NQ \widehat{{\cal F}}(m,m+1,m-1,m)\nonumber\\
&&-2\NP \widehat{{\cal S}}(m,m+1,m-1,m)\Biggr).
\end{eqnarray}
Combining all of the diagrams together we find that the cut-constructible pieces of the general $\phi$-MHV amplitude is given by,
\begin{eqnarray}
\lefteqn{C_{n;1}(\phi,1^-,2^+,\dots,m^-,\dots,n^+)}\nonumber\\
&=&c_{\Gamma}A^{(0)}_{n}\bigg(A^{\phi G}_{n;1}(m,n)
-4\bigg(1-\frac{\NF}{4N}\bigg)A^{\phi F}_{n;1}(m,n)-2\bigg(1-\frac{\NF}{N}\bigg)A^{\phi S}_{n;1}(m,n)\bigg),
\label{eq:phimhv}
\end{eqnarray}
where
\begin{eqnarray}
A^{\phi G}_{n;1}(m,n)&=&-\frac{1}{2}\sum_{i=1}^{n}\sum_{j=i+3}^{n+i-1}\Ftme(s_{i,j},s_{i+1,j-1};s_{i+1,j},s_{i,j-1})-\frac{1}{2}\sum_{i=1}^{n}\Fom(s_{i,i+2};s_{i,i+1},s_{i+1,i+2})\nonumber\\
&&+\sum_{i=1}^{n}(\Tri(s_{i,n+i-2})-\Tri(s_{i,n+i-1})).
\label{eq:phiG}
\end{eqnarray}
We notice that $A^{\phi G}_{n;1}(m,n)$ is independent of the position of the 
two negative helicity gluons; this is exactly as one would expect from an $\mathcal{N}=4$ theory. 
Nevertheless, the presence of the colourless scalar has 
removed the supersymmetry and as a result we see the 
appearance of $\Tri$ functions which are not present in eq.~(\ref{eq:Gqcdmhv}). 
We can write the fermionic pieces as,
\begin{eqnarray}
A^{\phi F}_{n;1}(m,n)=
&&\phantom{+}\sum_{i=2}^{m-1}\sum_{j=m+1}^{n}b^{ij}_{1m}\Fftme(s_{i,j},s_{i+1,j-1};s_{i+1,j},s_{i,j-1})\nonumber\\
&&+\sum_{i=2}^{m-1}\sum_{j=m+1}^{n} b^{ij}_{1m}\Fftme(s_{j,i},s_{j+1,i-1};s_{j+1,i},s_{j,i-1})\nonumber\\
&&-\sum_{i=2}^{m-1}\sum_{j=m}^n\frac{\trm(m,P_{(i,j)},i,1)}{s_{1m}^2}\mathcal{A}^{ij}_{m1}L_1(P_{(i+1,j)},P_{(i,j)})\nonumber\\
&&+\sum_{i=2}^{m-1}\sum_{j=m+1}^{1}\frac{\trm(1,P_{(j,i)},i,m)}{s_{1m}^2}\mathcal{A}^{i(j-1)}_{1m}L_1(P_{(j,i-1)},P_{(j,i)})\nonumber\\
&&+\sum_{i=2}^{m}\sum_{j=m+1}^{n}\frac{\trm(m,P_{(i,j)},j,1)}{s_{1m}^2}\mathcal{A}^{j(i-1)}_{m1}L_1(P_{(i,j-1)},P_{(i,j)})\nonumber\\
&&-\sum_{i=1}^{m-1}\sum_{j=m+1}^{n}\frac{\trm(1,P_{(j,i)},j,m)}{s_{1m}^2}\mathcal{A}^{ji}_{1m}L_1(P_{(j+1,i)},P_{(j,i)})
\label{eq:phiFs}
\end{eqnarray}
where the functions $b^{ij}_{m1}$ and $\mathcal{A}^{ij}_{m1}$ are the same auxiliary functions
as in the pure-glue case and are given by eqs.~\eqref{eq:bdef} and \eqref{eq:Adef} respectively.

Finally the scalar pieces are given by,
\begin{eqnarray}
A^{\phi S}_{n;1}(m,n)=
-\sum_{i=2}^{m-1}\sum_{j=m+1}^{n}
&&(b^{ij}_{1m})^2\Fftme(s_{i,j},s_{i+1,j-1};s_{i+1,j},s_{i,j-1}) \nonumber\\
-\sum_{i=m+1}^{n}\sum_{j=2}^{m-1}
&&(b^{ij}_{1m})^2\Fftme(s_{i,j},s_{i+1,j-1};s_{i+1,j},s_{i,j-1}) \nonumber\\
+\sum_{i=2}^{m-1}\sum_{j=m}^n
&\Bigg[&-\frac{\trm(m,P_{(i,j)},i,1)^3}{3s_{1m}^4}\mathcal{A}^{ij}_{m1}L_3(P_{(i+1,j)},P_{(i,j)})\nonumber\\
&&-\frac{\trm(m,P_{(i,j)},i,1)^2}{2s_{1m}^4}\mathcal{K}^{ij}_{m1}L_2(P_{(i+1,j)},P_{(i,j)})\nonumber\\
&&+\frac{\trm(m,P_{(i,j)},i,1)}{s_{1m}^4}\mathcal{I}^{ij}_{m1}L_1(P_{(i+1,j)},P_{(i,j)})\Bigg]\nonumber\\
+\sum_{i=1}^{m-1}\sum_{j=m+1}^{n}
&\Bigg[&-\frac{\trm(1,P_{(j,i)},j,m)^3}{3s_{1m}^4}\mathcal{A}^{ji}_{1m}L_3(P_{(j+1,i)},P_{(j,i)})\nonumber\\
&&      -\frac{\trm(1,P_{(j,i)},j,m)^2}{2s_{1m}^4}\mathcal{K}^{ji}_{1m}L_2(P_{(j+1,i)},P_{(j,i)})\nonumber\\
&&      +\frac{\trm(1,P_{(j,i)},j,m)}{s_{1m}^4}   \mathcal{I}^{ji}_{1m}L_1(P_{(j+1,i)},P_{(j,i)})\Bigg]\nonumber\\
+\sum_{i=2}^{m}\sum_{j=m+1}^{n}
&\Bigg[&\frac{\trm(m,P_{(i,j)},j,1)^3}{3s_{1m}^4}\mathcal{A}^{j(i-1)}_{m1}L_3(P_{(i,j-1)},P_{(i,j)})\nonumber\\
&&     +\frac{\trm(m,P_{(i,j)},j,1)^2}{2s_{1m}^4}\mathcal{K}^{j(i-1)}_{m1}L_2(P_{(i,j-1)},P_{(i,j)})\nonumber\\
&&     -\frac{\trm(m,P_{(i,j)},j,1)}{s_{1m}^4}   \mathcal{I}^{j(i-1)}_{m1}L_1(P_{(i,j-1)},P_{(i,j)})\Bigg]\nonumber\\
+\sum_{i=2}^{m-1}\sum_{j=m+1}^{1}
&\Bigg[&\frac{\trm(1,P_{(j,i)},i,m)^3}{3s_{1m}^4}\mathcal{A}^{i(j-1)}_{1m}L_3(P_{(j,i-1)},P_{(j,i)})\nonumber\\
&&+\frac{\trm(1,P_{(j,i)},i,m)^2}{2s_{1m}^4}\mathcal{K}^{i(j-1)}_{1m}L_2(P_{(j,i-1)},P_{(j,i)})\nonumber\\
&&-\frac{\trm(1,P_{(j,i)},i,m)}{s_{1m}^4}\mathcal{I}^{i(j-1)}_{1m}L_1(P_{(j,i-1)},P_{(j,i)})\Bigg]
\label{eq:phiSs}
\end{eqnarray}
where the auxiliary
functions $\mathcal{K}^{ij}_{m1}$ and $\mathcal{I}^{ij}_{m1}$ are the same as in the pure-glue case and are given 
by eqs.~\eqref{eq:Kdef} and \eqref{eq:Idef} respectively.

The similarities and differences between the gluonic MHV and the $\phi$-MHV calculation are now most obvious. It is clear that
both have the same type of auxiliary functions multiplying the one-loop basis functions, however the presence of the scalar has 
introduced a second set of summations. One difference is that in the $\phi$-MHV result there are no
degenerate triangles. This is a consequence of the absence of $\mathcal{O}(\epsilon^{-1})$ terms as predicted by the infrared pole
structure. 

\subsection{Cross Check: The adjacent minus amplitude} 
The one-loop $(\phi, 1^-,2^-\dots n^+)$ amplitude has been calculated~\cite{Badger:2007si} and provides a check of our calculation.
As mentioned earlier $A^{\phi G}_{n;1}$ is independent of $m$ so we only need explicitly check the remaining two contributions, which collapse to,
\begin{eqnarray}
A^{\phi F}_{n;1}(2,n)=\sum_{i=3}^{n}\frac{\trm(1,P_{(i+1,n)},i,2)}{s_{12}}L_1(P_{(i+1,1)},P_{(i,1)})\nonumber\\
+\sum_{i=4}^{n}\frac{\trm(2,P_{(3,i-1)},i,1)}{s_{12}}L_1(P_{(2,i-1)},P_{(2,i)}),
\end{eqnarray}
and
\begin{eqnarray}
A^{\phi S}_{n;1}(2,n)=\sum_{i=4}^{n}\Bigg(\frac{\trm(2,P_{(3,i-1)},i,1)^3}{3s_{12}^3}L_3(P_{(2,i-1)},P_{(2,i)})\nonumber\\
+\frac{\trm(2,P_{(3,i-1)},i,1)^2}{2s_{12}^2}L_2(P_{(2,i-1)},P_{(2,i)})\Bigg)\nonumber\\
+\sum_{i=3}^{n-1}\Bigg(\frac{\trm(1,P_{(i+1,n)},i,2)^3}{3s_{12}^3}L_3(P_{(i+1,1)},P_{(i,1)})\nonumber\\
+\frac{\trm(1,P_{(i+1,n)},i,2)^2}{2s_{12}^2}L_2(P_{(i+1,1)},P_{(i,1)})\Bigg),
\end{eqnarray}
respectively, and which is in agreement with the result of~\cite{Badger:2007si}.

\subsection{Cut-completion Terms}

The basis-set of logarithmic functions in which the results are expressed 
contains unphysical singularities, which we remove by adding in rational pieces, the so-called cut completion terms. 
The new basis is given by the transformation,
\begin{eqnarray}
&L_1(s,t)&=\hat{L}_1(s,t),\nonumber\\
&L_2(s,t)&=\hat{L}_2(s,t)+\frac{1}{2(s-t)}\bigg(\frac{1}{t}+\frac{1}{s}\bigg),\nonumber\\
&L_3(s,t)&=\hat{L}_3(s,t)+\frac{1}{2(s-t)^2}\bigg(\frac{1}{t}+\frac{1}{s}\bigg). 
\end{eqnarray}
From the breakdown of our amplitude it is clear that only the scalar pieces contribute. 
When considering the overlap terms in the next section it proves most convenient to write the cut-completion terms in the following form,
\begin{eqnarray}
&&CR_n(\phi ,1^-,\dots,m^-,\dots,n^+)=\Gamma_n\bigg[\nonumber\\
&&\sum_{i=2}^{m}\sum_{j=m+1}^{n} \rho^{j,i-1}_{m1}(P_{(i,j-1)})\bigg(\frac{1}{s_{i,j-1}}+\frac{1}{s_{i,j}}\bigg)
-\sum_{i=2}^{m-1}\sum_{j=m}^{n}\rho^{i,j}_{m1}(P_{(i+1,j)})\bigg(\frac{1}{s_{i+1,j}}+\frac{1}{s_{i,j}}\bigg)\nonumber\\
+&&\sum_{i=2}^{m-1} \sum_{j=m+1}^{n+1}
\rho^{i,j-1}_{1m}(P_{(j,i-1)})\bigg(\frac{1}{s_{j,i-1}}+\frac{1}{s_{j,i}}\bigg)
-\sum_{i=1}^{m-1} \sum_{j=m+1}^{n}\rho^{j,i}_{1m}(P_{(j+1,i)})\bigg(\frac{1}{s_{j+1,i}}+\frac{1}{s_{j,i}}\bigg)
\bigg].\nonumber\\
\label{eq:CR}
\end{eqnarray}

The factor $\Gamma_n$ is given by,
\begin{equation}
\Gamma_n=\frac{c_{\Gamma}N_{P}}{2\Pi_{\alpha=1}^{n}\spa \alpha.\ap},
\end{equation}
and
\begin{eqnarray}
\label{eq:rhodef}
\rho^{a,b}_{m1}(P_{(i,j)})=\frac{\spaa m|.P_{(i,j)}.a|.1^3} {3\spab a.|P_{(i,j)}|.a^2}A^{ab}_{m1}+\frac{\spaa
m.|P_{(i,j)}.a|.1^2}{2\spab a.|P_{(i,j)}|.a}K^{ab}_{m1},
\end{eqnarray}
with
\begin{eqnarray}
A^{ab}_{m1}=\frac{\spa m.a\spa b.1}{\spa a.b}-(b\rightarrow b+1),\\
K^{ab}_{m1}=\frac{\spa m.a^2\spa b.1^2}{\spa a.b^2}-(b\rightarrow b+1).
\end{eqnarray}
We have also introduced the short-hand notation,
\begin{equation}
N_P=2\bigg(1-\frac{\NF}{N_c}\bigg).
\end{equation}

\section{The Rational Pieces}
\label{sec:rational}

In addition to the cut-constructible terms calculated in the previous section, one-loop amplitudes in non-supersymmetric theories
also contain rational terms with no discontinuities.  By definition this means that these terms can only contain simple poles in
physical invariants, which makes these terms amenable to the BCFW recursion relation techniques.  So far successful applications
have included amplitudes in QCD~\cite{BDK:1lonshell,BDK:1lrecfin,Bern:bootstrap} and the finite and adjacent minus $\phi$
amplitudes~\cite{Berger:higgsrecfinite}.

In an earlier section, we cancelled unphysical poles in $C_n$ by introducing the cut-completion terms $CR_n$. 
If we naively set up the recursion relations we would double count on these rational pieces. 
To avoid this, we define the recursion relation as a function of the physical poles $\hat{R}_n=R_n-CR_n$. 
We make a complex shift of the two negative gluons such that 
\begin{equation}
\label{eq:zshifts}
|\hat{1}\rangle=|1\rangle+z|m\rangle, \qquad |\hat{m}]=|m]-z|1],
\end{equation}
ensuring that overall momentum is conserved since
\begin{equation}
p_1^{\mu}(z)=p^{\mu}_1+\frac{z}{2}\spab m.|\gamma^{\mu}|.1,
\qquad p_m^{\mu}(z)=p^{\mu}_m-\frac{z}{2}\spab m.|\gamma^{\mu}|.1.
\end{equation}
The recursion relation on $\hat{R}_n$ is defined through the following integral,
\begin{equation}
\frac{1}{2\pi i}\oint_C\frac{dz}{z}\hat{R}_n=\frac{1}{2\pi i}\oint_C\frac{dz}{z}(R_n-CR_n).
\end{equation}
Provided that $z$ is chosen such that $A(z)\rightarrow 0$ as $z$ goes to infinity, the integral vanishes. The residues of the integrand are fixed by multiparticle factorisation so that the rational pieces are given by: 
\begin{eqnarray}
\hat{R}_n(0)&&=-\sum_{\mathrm{phys}\,\mathrm{poles}\, z_{i}}\mathrm{Res}_{z=z_i}\frac{(R_n(z)-CR_n(z))}{z}
\nonumber\\
&&=\sum_{i}\frac{A_{L}^{(0)}(z)R_R(z)+R_L(z)A^{(0)}_R(z)}{P^2_i}+\sum_{i}\mathrm{Res}_{z=z_i}\frac{CR_n(z)}{z}.
\end{eqnarray}
The final piece of this equation is called the overlap term. It's calculation is relatively simple if the poles are all first order. 

\begin{figure}[t]
	\psfrag{P}{$\phi$}
	\psfrag{a}{$1^-$}
	\psfrag{b}{$m^-$}
	\psfrag{bpt}{$(m+2)^+$}
	\psfrag{bp}{$(m+1)^+$}
	\psfrag{bm}{$(m-1)^+$}
	\psfrag{c}{$3^+$}
	\psfrag{ap}{$2^+$}
	\psfrag{bmt}{$(m-2)^+$}
	\psfrag{i}{$i^+$}
	\psfrag{ip}{$(i+1)^+$}
	\psfrag{j}{$j^+$}
	\psfrag{jp}{$(j+1)^+$}
	\psfrag{n}{$n^+$}
	\psfrag{nm}{$(n-1)^+$}
	\psfrag{p}{$+$}
	\psfrag{m}{$-$}
	\begin{center}
		\includegraphics[width=12cm]{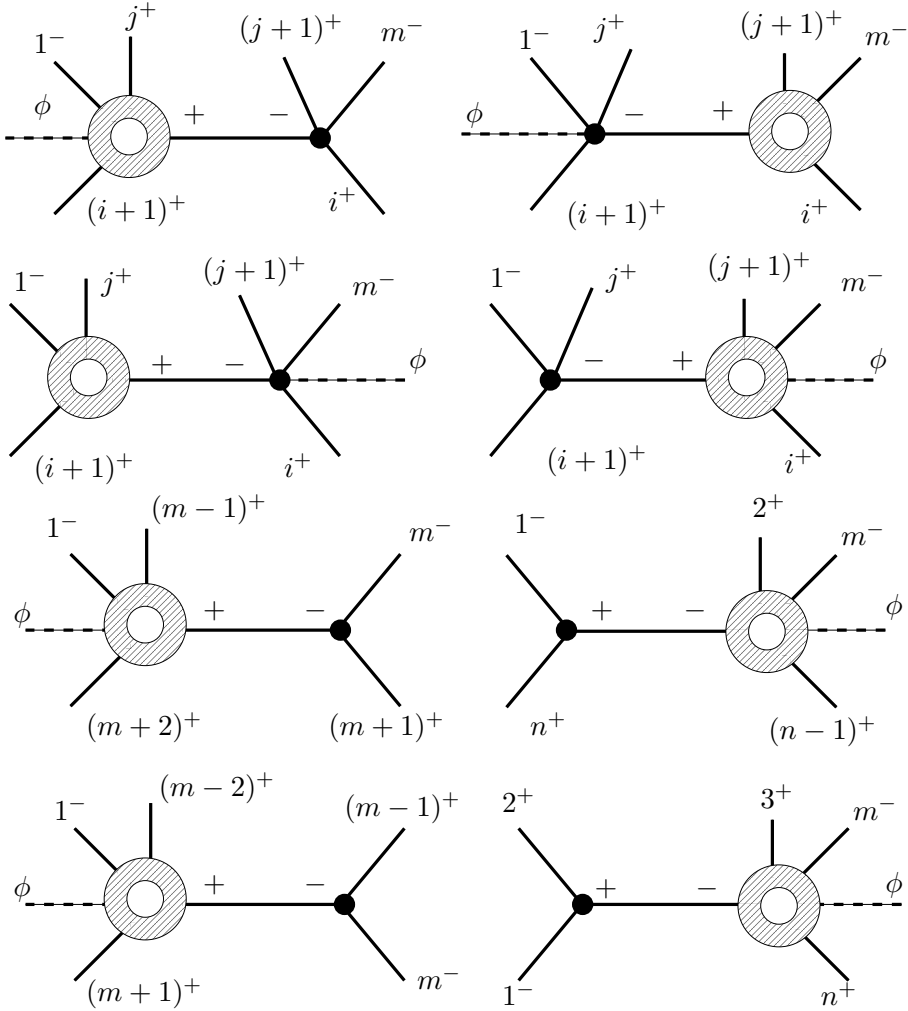}
	\end{center}
	\caption{The direct recursive terms contributing to $R_n(\phi,1^-,\dots,m^-,\dots,n^+)$}
	\label{fig:rat}
\end{figure}

\subsection{Recursive terms}
\label{rec}
The direct recursive terms are obtained by using the following formula
\begin{equation}
R^D_n=\sum_{i}\frac{A_{L}^{(0)}(z)R_R(z)+R_L(z)A^{(0)}_R(z)}{P^2_i}.
\end{equation}
For our chosen shift \eqref{eq:zshifts}, the allowed diagrams are 
shown in Fig.~\ref{fig:rat}, and the summation over these given by eq.~(\ref{eq:rats}). 
In the sum $R$ is defined as the full rational part of the 
amplitude with fewer than $n$ external legs. 
Due to our choice 
of shifts the tree amplitudes 
\begin{displaymath}
A^{(0)}(j^+,\hat{1}^-,-P_{(1,j)}^-), \qquad  A^{(0)}(j^+,\hat{m}^-,-P_{(m,j)}^+)
\end{displaymath}
are both zero, (here $j \in \{2,n,(m\pm1)\}$). 
These three point amplitudes are hence not included in either the diagram or the sum. 
Other terms that vanish are $R_2(\phi,-+)$ which is required to be zero by angular momentum conservation, 
and $R(j^+,\hat{m}^-,\hat{P}^{\pm})$ since the 
corresponding splitting function has no rational pieces.

Because the tree amplitudes with fewer than two negative
helicities vanish, the one requires the one-loop contributions  
with one negative helicity. These are finite one-loop amplitudes 
and are entirely rational.  The finite $\phi-+\ldots+$ amplitudes were computed 
for arbitrary numbers of positive helicity gluons in ref.~\cite{Berger:higgsrecfinite}.
As a concrete example, the three-gluon amplitude is given by,
\begin{align}
	{R}_3(\phi;1^-,2^+,3^+) &=
	\frac{N_P}{96\pi^2}\frac{\A{1}{2}\A{3}{1}\B{2}{3}}{\A{2}{3}^2}
	-\frac{1}{8\pi^2}A^{(0)}_3(\phi^\dagger;1^-,2^+,3^+).
	\label{eq:ncc3-mpp}
\end{align}
Similarly, the pure QCD $-+\ldots+$ amplitudes are given to all orders in ref.~\cite{Mahlon:oneminus,BDK:1lonshell}.
In the four gluon case, the result is,
\begin{align}
\label{eq:ncc4-mppp}
R_4(1^-,2^+,3^+,4^+) &= \frac{N_P}{96\pi^2}
\frac{\spa 2.4 \spb 2.4^3}{\spb 1.2 \spa 2.3 \spa 3.4 \spb 4.1}
\end{align}
 
Finally, there the ``homogenous'' terms in the recursion which depend on the
$\phi$-MHV amplitude with one gluon fewer.
The first few $\phi$-MHV amplitudes are known,
\begin{eqnarray}
\label{eq:ncc2-mm}
{R}_2(\phi;1^-,2^-) &=&  \frac{1}{8\pi^2}A^{(0)}(\phi,1^-,2^-),\\
\label{eq:ncc3-mmp}
{R}_3(\phi;1^-,2^-,3^+) &=& \frac{1}{8\pi^2}A^{(0)}(\phi,1^-,2^-,3^+),\\
\label{eq:ncc3-mpm}
{R}_3(\phi;1^-,2^+,3^-) &=& \frac{1}{8\pi^2}A^{(0)}(\phi,1^-,2^+,3^-).
\end{eqnarray}

Combining the various diagrams, we find that recursive terms obey the following relation,
\begin{eqnarray}
&&R^D_n(\phi,1^-,\dots,m^-,\dots,n^+)=\nonumber\\
&&+\sum_{i=m}^{n}\sum_{j=1}^{m-1}R(\phi,\hat{1}^-,\dots,j^+,\hat{P}^+_{(j+1,i)},(i+1)^+)\frac{1}{s_{j+1,i}}A^{(0)}(-\hat{P}^-_{(j+1,i)},(j+1)^+,\dots,\hat{m}^-,\dots,i^+)\nonumber\\
&&+\sum_{j=1}^{m-1}\sum_{i=m}^{n}A^{(0)}(\phi,\hat{1}^-,\dots,j^+,\hat{P}^-_{(j+1,i)},(i+1)^+)\frac{1}{s_{j+1,i}}R(-\hat{P}^+_{(j+1,i)},(j+1)^+,,\dots,\hat{m}^-,\dots,i^+)\nonumber\\
&&+\sum_{j=1}^{m-1}\sum_{i=m}^{n}R(\hat{1}^-,\dots,j^+,-\hat{P}^+_{(i+1,j)},(i+1)^+)\frac{1}{s_{i+1,j}}A^{(0)}(\phi,\hat{P}^-_{(i+1,j)},(j+1)^+,,\dots,\hat{m}^-,\dots,i^+)\nonumber\\
&&+\sum_{j=1}^{m-1}\sum_{i=m}^{n}A^{(0)}(\hat{1}^-,\dots,j^+,\hat{P}^-_{(i+1,j)},(i+1)^+)\frac{1}{s_{i+1,j}}R(\phi,-\hat{P}^+_{(i+1,j)},(j+1)^+,,\dots,\hat{m}^-,\dots,i^+)\nonumber\\
&&+R(\phi,\hat{1}^-,\dots,(m-1)^+,-\hat{P}^+_{(m,m+1)},(m+2)^+)\frac{1}{s_{m,m+1}}A^{(0)}(\hat{P}^-_{(m,m+1)},\hat{m}^-,(m+1)^+)\nonumber\\
&&+R(\phi,\hat{1}^-,\dots,(m-2)^+,-\hat{P}^+_{(m-1,m)},(m+1)^+)\frac{1}{s_{m-1,m}}A^{(0)}(\hat{P}^-_{(m-1,m)},(m-1)^+,\hat{m}^-)\nonumber\\
&&+A^{(0)}(\hat{1}^-,\hat{P}^+_{(n,1)},n^+)\frac{1}{s_{n,1}}R(\phi,-\hat{P}^-_{(n,1)},2^+,,\dots,\hat{m}^-,\dots,(n-1)^+)\nonumber\\
&&+A^{(0)}(\hat{1}^-,\hat{P}^+_{(1,2)},2^+)\frac{1}{s_{1,2}}R(\phi,-\hat{P}^-_{(1,2)},3^+,,\dots,\hat{m}^-,\dots,n^+).
\label{eq:rats}
\end{eqnarray}

The value that $z$ takes is obtained by requiring that the shifted momenta  
\begin{equation}
\wh{P}^\mu_{(i,j)} = P^\mu_{(i,j)} \pm \frac{z}{2}\langle m | \gamma^\mu | 1 ],
\end{equation}
is on-shell.  In this equation, the sign is positive when the momentum set $\{p_i,p_j\}$ includes 
$p_1$ and is negative when it includes $p_m$.
There are six independent channels, each one specified by a particular invariant mass, 
$s_{j+1,i}$, $s_{j,i+1}$, or by the double invariants, $s_{m,m+1}$,
$s_{m-1,m}$, $s_{n,1}$ and $s_{1,2}$.
In each channel, we find that the value of $z$ and the hatted variables 
are given by,
\begin{eqnarray}
&&s_{j+1,i}\quad\mathrm{channels}\qquad z_{j+1,i}=\frac{s_{j+1,i}}{\spab m.|P_{(j+1,i)}|.1}\nonumber\\
&&|\hat{1}\rangle=\frac{|(p_1+P_{(j+1,i)})P_{(j+1,i)}m\rangle}{\spab m.|P_{(j+1,i)}|.1}, \quad
|\hat{m}]=\frac{|(p_m-P_{(j+1,i)})P_{(j+1,i)}1]}{\spab m.|P_{(j+1,i)}|.1},\nonumber\\
&&\hat{P}_{(j+1,i)}=\frac{|P_{(j+1,i)}1]\langle m
P_{j+1,i}|}{\spab m.|P_{(j+1,i)}|.1}\nonumber\\
&&s_{j,i+1}\quad\mathrm{channels}\qquad z_{j,i+1}=-\frac{s_{j,i+1}}{\spab m.|P_{(j,i+1)}|.1}\nonumber\\
&&|\hat{1}\rangle=\frac{|(p_1-P_{(j,i+1)})P_{(j,i+1)}m\rangle}{\spab m.|P_{(j,i+1)}|.1}, \quad
|\hat{m}]=\frac{|(p_m+P_{(j,i+1)})P_{(j,i+1)}1]}{\spab m.|P_{(j,i+1)}|.1},\nonumber\\
&&\hat{P}_{(j,i+1)}=\frac{|P_{(j,i+1)}1]\langle m
P_{(j,i+1)}|}{\spab m.|P_{(j,i+1)}|.1}\nonumber\\
&&s_{m,m+1}\quad\mathrm{channel}\qquad z_{m,m+1}=\frac{\spb \mpl.m}{\spb \mpl.1}\nonumber\\
&&|\hat{1}\rangle=\frac{|(p_1+p_m),m+1]}{\spb 1.\mpl}, \quad |\hat{m}]=|m+1]\frac{\spb 1.m}{\spb
1.\mpl},\quad\hat{P}_{(m,m+1)}=\frac{|P_{(m,m+1)}1][m+1|}{\spb \mpl.1}\nonumber\\
&&s_{m-1,m}\quad\mathrm{channel}\qquad z_{m-1,m}=\frac{\spb \mm.m}{\spb \mm.1}\nonumber\\
&&|\hat{1}\rangle=\frac{|(p_1+p_m),m-1]}{\spb 1.\mm}, \quad |\hat{m}]=|m-1]\frac{\spb 1.m}{\spb
1.\mm},\quad\hat{P}_{(m-1,m)}=\frac{|P_{(m-1,m)}1][m-1|}{\spb \mm.1}\nonumber\\
&&s_{n,1}\quad\mathrm{channel}\qquad z_{n,1}=-\frac{\spa 1.n}{\spa m.n}\nonumber\\
&&|\hat{1}\rangle=|n\rangle\frac{\spa1.m}{\spa n.m}, \quad |\hat{m}]=\frac{|(p_1+p_m)n\rangle}{\spa
m.n},\quad\hat{P}_{(n,1)}=\frac{|n\rangle\langle m P_{(n,1)}|}{\spa m.n}\nonumber\\
&&s_{1,2}\quad\mathrm{channel}\qquad z_{1,2}=-\frac{\spa 1.2}{\spa m.2}\nonumber\\
&&|\hat{1}\rangle=|2\rangle\frac{\spa1.m}{\spa 2.m}, \quad |\hat{m}]=\frac{|(p_1+p_m)2\rangle}{\spa m.2},
\quad\hat{P}_{(1,2)}=\frac{|2\rangle\langle m P_{(1,2)}|}{\spa m.2}\nonumber
\end{eqnarray}

\subsection{The large $z$ behaviour of the completion terms}
\label{subsec:infcr}

In order for the direct recursive contribution to correctly generate the rational terms,  
the shifted amplitude $A^{(1)}_n(z)$ must vanish as $z \to \infty$.    
With the shift defined in eq.~\eqref{eq:zshifts} acting on two negative helicity 
gluons this is indeed the case.  However,  the cut-completion term $CR_n(z)$ introduced in eq.~\eqref{eq:CR}
to ensure that the cut constructible part does not have any spurious poles, 
does not vanish as $z \to \infty$.  We therefore have to explicitly remove the
contribution at infinity from the rational part, which now becomes
\cite{Berger:genhels,Berger:allmhv}, 
\begin{equation}
\hat{R}_n=R^D_n+O_n-\mathrm{Inf}\,CR_n,
\end{equation}
where 
\begin{equation}
\mathrm{Inf}\,CR_n = \lim_{z\to\infty} CR_n(z).
\end{equation} 
The calculation of $\mathrm{Inf}\,CR_n$ is straightforward. 
For the special case of adjacent negative helicities, corresponding to $m=2$, the
cut-completion terms behaves as $1/z$ as $z \to \infty$ so that,
\begin{equation}
\mathrm{Inf}\,CR_n(\phi ,1^-,2^-,\dots,n^+)=0.
\end{equation}
For the general, non-adjacent, case, there is a contribution as $z \to \infty$ and 
we find 
the contribution to be subtracted is, 
\begin{eqnarray}
\lefteqn{\mathrm{Inf} \,CR_n(\phi ,1^-,\dots,m^-,\dots,n^+)
=\frac{c_{\Gamma}N_P}{2\spa m.2\spa n.m\Pi_{\alpha=2}^{n-1}\spa \alpha.\ap}\bigg[}\nonumber \\
 &&\sum_{i=3}^{m}\sum_{j=m+1}^{n} \omega^{j,i-1} (P_{(i,j)})\bigg(\frac{1}{\spab m.|P_{(i,j-1)}|.1}+\frac{1}{\spab m.|P_{(i,j)}|.1}\bigg)\nonumber\\
-&&\sum_{i=2}^{m-1}\sum_{j=m+1}^{n} \omega^{i,j} (P_{(i,j)})\bigg(\frac{1}{\spab m.|P_{(i+1,j)}|.1}+\frac{1}{\spab m.|P_{(i,j)}|.1}\bigg)\nonumber\\
-&&\sum_{i=2}^{m-1}\sum_{j=m+1}^{n} \omega^{j,i} (P_{(i,j)})\bigg(\frac{1}{\spab m.|P_{(i,j-1)}|.1}+\frac{1}{\spab m.|P_{(i,j)}|.1}\bigg)\nonumber\\
+&&\sum_{i=2}^{m-1}\sum_{j=m}^{n-1}\omega^{i,j+1} (P_{(i,j)})\bigg(\frac{1}{\spab m.|P_{(i+1,j)}|.1}+\frac{1}{\spab m.|P_{(i,j)}|.1}\bigg)\nonumber\\
+&&\sum_{i=2}^{m-1} \sum_{j=m+2}^{n+1} \omega^{i,j-1} (\widetilde{P_{(j,i)}})\bigg(\frac{1}{\spab m.|P_{(j,i-1)}|.1}+\frac{1}{\spab m.|P_{(j,i)}|.1}\bigg)\nonumber\\
-&&\sum_{i=2}^{m-1} \sum_{j=m+1}^{n}\omega^{j,i} (\widetilde{P_{(j,i)}})\bigg(\frac{1}{\spab m.|P_{(j+1,i)}|.1}+\frac{1}{\spab m.|P_{(j,i)}|.1}\bigg)\nonumber \\
-&&\sum_{i=2}^{m-1} \sum_{j=m+1}^{n} \omega^{i,j} (\widetilde{P_{(j,i)}})\bigg(\frac{1}{\spab m.|P_{(j,i-1)}|.1}+\frac{1}{\spab m.|P_{(j,i)}|.1}\bigg)\nonumber\\
+&&\sum_{i=1}^{m-2} \sum_{j=m+1}^{n}\omega^{j,i+1}(\widetilde{P_{(j,i)}})\bigg(\frac{1}{\spab m.|P_{(j+1,i)}|.1}+\frac{1}{\spab m.|P_{(j,i)}|.1}\bigg)\bigg], 
\label{eq:infC}
\end{eqnarray}
with
\begin{equation}
\omega^{a,b}(P_{(i,j)}) = 
\frac{\spaa m.| P_{(i,j)}.a|.m^2 \spa a.m \spa b.m^2}{2 \spb 1.a \spa a.b^2},
\end{equation}
and $\widetilde{P_{(j,i)}} = P_{(j,i)} - p_1$. 

\subsection{Overlap Terms}
\label{subsec:overlap}

The overlap terms are defined as~\cite{Berger:genhels,Badger:2007si},
\begin{equation}
O_n=\sum_{i}\mathrm{Res}_{z=z_i}\frac{CR_n(z)}{z}.
\end{equation}
They can be obtained by evaluating the residue of the cut completion term $CR_n$ given
in eq.~(\ref{eq:CR}) in each of the physical channels.
To expose the coefficients of the poles most clearly the 
cut-completion terms are rewritten as follows,
\begin{eqnarray}
CR_n=&&\Gamma_{n}\bigg[
\sum_{i=3}^{m}\sum_{j=m}^{n-1}\frac{1}{s_{i,j}}\bigg(\rho^{j,i-1}_{m1}(P_{(i,j)})+\rho^{j+1,i-1}_{m1}(P_{(i,j)})-\rho^{i,j}_{m1}(P_{(i,j)})-\rho^{i-1,j}_{m1}(P_{(i,j)})\bigg)\nonumber\\
&&+\sum_{i=m+1}^{n}\sum_{j=2}^{m-1}\frac{1}{s_{i,j}}\bigg(\rho^{j,i-1}_{1m}(P_{(i,j)})+\rho^{j+1,i-1}_{1m}(P_{(i,j)})-\rho^{i,j}_{1m}(P_{(i,j)})-\rho^{i-1,j}_{1m}(P_{(i,j)})\bigg)\nonumber\\
&&+\sum_{i=3}^{m}\frac{1}{s_{i,n}}\bigg(\rho^{n,i-1}_{m1}(P_{(i,n)})-\rho^{i,n}_{m1}(P_{(i,n)})-\rho^{i-1,n}_{m1}(P_{(i,n)})\bigg)\nonumber\\
&&+\sum_{i=m+1}^{n}\frac{1}{s_{i,1}}\bigg(\rho^{2,i-1}_{1m}(P_{(i,1)})-\rho^{i,1}_{1m}(P_{(i,1)})-\rho^{i-1,1}_{1m}(P_{(i,1)})\bigg)\nonumber\\ 
&&+\sum_{j=m}^{n-1}\frac{1}{s_{2,j}}\bigg(\rho^{j,1}_{m1}(P_{(2,j)})+\rho^{j+1,1}_{m1}(P_{(2,j)})-\rho^{2,j}_{m1}(P_{(1,j)})\bigg)\nonumber\\
&&+\sum_{j=2}^{m-1}\frac{1}{s_{1,j}}\bigg(\rho^{j,n}_{1m}(P_{(1,j)})+\rho^{j+1,n}_{1m}(P_{(1,j)})-\rho^{n,j}_{1m}(P_{(1,j)})
\bigg)\nonumber\\
&&+\frac{1}{s_{2,n}}\bigg(\rho^{n,1}_{m1}(P_{(2,n)})-\rho^{2,n}_{m1}(P_{(2,n)})\bigg)\bigg],
\end{eqnarray}
with $\rho^{a,b}_{1m}$ defined in eq.~\eqref{eq:rhodef}.

The cut-completion terms contain many different simple poles in $s_{i,j}$ but only 
those invariants 
which contain either $p_1$ or $p_m$ (but not both) have non-trivial overlap terms. 
We observe that the cut completion term contain
only simple residues, so for the $P_{(i,j)}$ pole,
the overlap term is given by,
\begin{equation}
O^{i,j}_n = CR_n(z_{i,j}) \frac{\wh{s_{i,j}}}{s_{i,j}}
\end{equation}
where $z_{i,j}$ is the value of $z$ that puts $\wh{P}_{(i,j)}$ on-shell.
The multiplicative factor removes the $\wh{s_{i,j}}$ pole in $CR_n$ and replaces it with the
correct propagator ${s}_{i,j}$. 

The cut-completion terms also contribute to the overlap terms
because of singularities associated with the multiplicative tree factor in  eq.~(\ref{eq:CR}). 
The poles in
$\spahl1.2$ and $\spahr n.1$ must be treated carefully, but,
since the shift  leaves $\langle m|$
unaltered,
there are no overlap terms generated by $\spa m.\mpl$ or $\spa \mm.m$.

Splitting up the cut-completion terms in this way gives the overlap terms the following structure,
\begin{eqnarray}
\label{eq:otot}
O_n=&&\sum_{i=3}^{m}\sum_{j=m}^{n-1}O^{i,j}_{m,n}+\sum_{i=m+1}^{n}\sum_{j=2}^{m-1}O^{i,j}_{1,n}+\sum_{i=3}^{m}O^{i,n}_n+\sum_{i=m+1}^{n}O^{i,1}_n\nonumber\\
&&+\sum_{j=m}^{n-1}O^{2,j}_n+\sum_{j=2}^{m-1}O^{1,j}_n+O^{2,n}_n+O^{12}_n+O^{n1}_n.
\end{eqnarray}
We now describe in detail the derivation of each of these terms.
 
\subsubsection{The overlap term $O^{i,j}_{m,n}$}

The first overlap terms we
consider are those arising from the $s_{i,j}$ channel when $3\le i \le m$ and $m \le
j \le n-1$. Since it is always the case that $p_m \in P_{(i,j)}$, 
we use the shift $z_1=s_{i,j}/\spab
m.|P_{(i,j)}.1$. Under this shift the various functions become,
\begin{eqnarray}
\Gamma_{n}(z_1)=-\frac{c_{\Gamma}N_P}{2\Pi_{\alpha=2}^{n-1}\spa \alpha.\ap}
\frac{\spab m.|P_{(i,j)}|.1^2}{\spaa m.|P_{(i,j)}.(p_1+P_{(i,j)}|.2\spaa n.|(p_1+P_{(i,j)}).P_{(i,j)}|.m},
\end{eqnarray}
while,
\begin{eqnarray}
A^{ab}_{m1}(z_{1})=\bigg(\frac{\spa m.a\spaa b.|(p_1+P_{(i,j)}).P_{(i,j)}|.m}{\spa a.b\spab m.|P_{(i,j)}|.1}-(b\rightarrow b+1)\bigg)\nonumber\\
K^{ab}_{m1}(z_1)=\bigg(\frac{\spa m.a^2\spaa b.|(p_1+P_{(i,j)}).P_{(i,j)}|.m^2}{\spa a.b^2\spab m.|P_{(i,j)}|.1^2}-(b\rightarrow b+1)\bigg).
\end{eqnarray}
The prefactor multiplying the $A$ and $K$ functions is simplified since $P_{(i,j)}$ in the numerator is never shifted (as it is always adjacent to a $\langle m|$),  
\begin{eqnarray}
\frac{\spaah m.|P_{(i,j)}.a|.1^n}{\spabh a|.P_{(i,j)}.{|a}^{n-1}}=\frac{\spab m.|P_{(i,j)}|.a}{\spab m.|P_{(i,j)}|.1}
\bigg(\frac{\spaa a.|(p_1+P_{(i,j)}).P_{(i,j)}|.m^n}{\spab a.|P_{(i,j)}|.1^{n-1}}\bigg).
\end{eqnarray}
$O^{i,j}_{m,n}$ is thus given by,
\begin{eqnarray}
\label{eq:oijm}
O^{i,j}_{m,n}=&&\Gamma_n(z_1)\bigg[\frac{1}{s_{i,j}}\bigg\{\frac{\spab m.|P_{(i,j)}|.j}{\spab m.|P_{(i,j)}|.1}\bigg(\frac{\spaa j.|(p_1+P_{(i,j)}).P_{(i,j)}|.m^3}{3\spab j.|P_{(i,j)}|.1^2}A^{j(i-1)}_{m1}(z_1)\nonumber\\
&&+\frac{\spaa j.|(p_1+P_{(i,j)}).P_{(i,j)}|.m^2}{2\spab j.|P_{(i,j)}|.1}K^{j(i-1)}_{m1}(z_1)\bigg)+(j \rightarrow j+1,P_{(i,j)}\rightarrow P_{(i,j)})\nonumber\\&&+\frac{\spab m.|P_{(i,j)}|.i}{\spab m.|P_{(i,j)}|.1}\bigg(-\frac{\spaa i.|(p_1+P_{(i,j)}).P_{(i,j)}|.m^3}{3\spab i.|P_{(i,j)}|.1^2}A^{ij}_{m1}(z_1)\nonumber\\
&&-\frac{\spaa i.|(p_1+P_{(i,j)}).P_{(i,j)}|.m^2}{2\spab i.|P_{(i,j)}|.1}K^{ij}_{m1}(z_1)\bigg)+(i \rightarrow i-1,P_{(i,j)}\rightarrow P_{(i,j)})\bigg\}\bigg].\nonumber\\
\end{eqnarray}

\subsubsection{The overlap terms $O^{i,n}_n$, $O^{2,j}_n$ and $O^{2,n}_{n}$}
The contributions in the $s_{i,n}$, $s_{2,j}$ and $s_{2,n}$ channels are evaluated under the same shift as $O^{ij}_{m,n}$, such that, 
\begin{eqnarray}
\label{eq:oin}
O^{i,n}_{m,n}=&&
\Gamma_n(z_1)\bigg[\frac{1}{s_{i,n}}\bigg\{
\frac{\spab m.|P_{(i,n)}|.i}{\spab m.|P_{(i,n)}|.1}\bigg(-\frac{\spaa i|.|P_{(i,1)}.P_{(i,n)}|.m^3}{3\spab i.|P_{(i,n)}|.1^2}\frac{\spa m.i\spaa n.|P_{(i,1)}.P_{(i,n)}|.m}{\spa i.n\spab m.|P_{(i,n)}|.1}\nonumber\\
&&-\frac{\spaa i.|P_{(i,1)}.P_{(i,n)}|.m^2}{2\spab i.|P_{(i,n)}|.1}\frac{\spa m.i^2\spaa n.|P_{(i,1)}.P_{(i,n)}|.m^2}{\spa i.n^2\spab m.|P_{(i,n)}|.1^2}
\bigg)+(i \rightarrow i-1,P_{(i,j)}\rightarrow P_{(i,j)})\nonumber\\&&+\frac{\spab m.|P_{(i,n)}|.n}{\spab m.|P_{(i,n)}|.1}\bigg(\frac{\spaa n.|P_{(i,1)}.P_{(i,n)}|.m^3}{3\spab n.|P_{(i,n)}|.1^2}A^{n(i-1)}_{m1}(z_1)+
\frac{\spaa n.|P_{(i,1)}.P_{(i,n)}|.m^2}{2\spab n.|P_{(i,n)}|.1}K^{n(i-1)}_{m1}(z_1)\bigg)\bigg\}\bigg]\nonumber\\
\label{eq:o2j}
O^{2,j}_{m,n}=&&
\Gamma_n(z_1)\bigg[\frac{1}{s_{2,j}}\bigg\{\frac{\spab m.|P_{(2,j)}|.j}{\spab m.|P_{(2,j)}|.1}\bigg(-\frac{\spaa j.|P_{(1,j)}.P_{(2,j)}|.m^3}{3\spab j.|P_{(2,j)}|.1^2}\frac{\spa m.j\spaa 2.|P_{(1,j)}.P_{(2,j)}|.m}{\spa j.2\spab m.|P_{(2,j)}|.1}\nonumber\\&&-
\frac{\spaa j.|P_{(1,j)}.P_{(2,j)}|.m^2}{2\spab j.|P_{(2,j)}|.1}\frac{\spa m.j^2\spaa 2.|P_{(1,j)}.P_{(2,j)}|.m^2}{\spa j.2^2\spab m.|P_{(2,j)}|.1^2}\bigg)+(j \rightarrow j+1,P_{(i,j)}\rightarrow P_{(i,j)})\nonumber\\&&
+\frac{\spab m.|P_{(2,j)}|.2}{\spab m.|P_{(2,j)}|.1}\bigg(-\frac{\spaa 2.|P_{(1,j)}.P_{(2,j)}|.m^3}{3\spab 2.|P_{(2,j)}|.1^2}A^{2j}_{m1}(z_1)
-\frac{\spaa 2.|P_{(1,j)}.P_{(2,j)}|.m^2}{2\spab 2.|P_{(2,j)}|.1}K^{2j}_{m1}(z_1)\bigg)\bigg\}\bigg],\nonumber\\
\label{eq:o2n}
O^{2,n}_{m,n}=&&\Gamma_n(z_1)\bigg[\frac{1}{s_{2,n}}\bigg\{\frac{\spab m.|P_{(2,n)}|.n}{\spab m.|P_{(2,n)}|.1}\bigg(-\frac{\spaa n.|P_{(1,n)}.P_{(2,n)}|.m^3}{3\spab n.|P_{(2,n)}|.1^2}\frac{\spa m.n\spaa 2.|P_{(1,n)}.P_{(2,n)}|.m}{\spa 2.n\spab m.|P_{(2,n)}|.1}\nonumber\\&&-
\frac{\spaa n.|P_{(1,n)}.P_{(2,n)}|.m^2}{2\spab n.|P_{(2,n)}|.1}\frac{\spa m.n^2\spaa 2.|P_{(1,n)}.P_{(2,n)}|.m^2}{\spa 2.n^2\spab m.|P_{(2,n)}|.1^2}\bigg)\nonumber\\&&+
\frac{\spab m.|P_{(2,n)}|.2}{\spab m.|P_{(2,n)}|.1}\bigg(-\frac{\spaa 2.|P_{(1,n)}.P_{(2,n)}|.m^3}{3\spab 2.|P_{(2,n)}|.1^2}\frac{\spa m.2\spaa n.|P_{(1,n)}.P_{(2,n)}|.m}{\spa 2.n\spab m.|P_{(2,n)}|.1}\nonumber\\&&
-\frac{\spaa 2.|P_{(1,n)}.P_{(2,n)}|.m^2}{2\spab 2.|P_{(2,n)}|.1}\frac{\spa m.2^2\spaa n.|P_{(1,n)}.P_{(2,n)}|.m^2}{\spa 2.n^2\spab m.|P_{(2,n)}|.1^2}
\bigg)\bigg\}\bigg].
\end{eqnarray}

\subsubsection{The overlap terms $O^{i,j}_{1,n}$, $O^{1,j}_{n}$ and $O^{i,1}_{n}$}
A similar set of overlap terms are generated in the $s_{i,j}$, $s_{1,j}$ and $s_{i,1}$ channels when
$p_1 \in P_{(i,j)}$.  We therefore use the shift $z_2=-s_{i,j}/\spab m.|P_{(i,j)}|.1$. Once again the tree factor, $\Gamma$ and the functions 
$A$ and $K$ must be evaluated under this shift; 
\begin{equation}
\Gamma_{n}(z_2)=-\frac{c_{\Gamma}N_P}{2\Pi_{\alpha=2}^{n-1}\spa \alpha.\ap}\frac{\spab m.|P_{(i,j)}|.1^2}
{\spaa m.|P_{(i,j)}.(p_1-P_{(i,j)})|.2\spaa n.|(p_1-P_{(i,j)}).P_{(i,j)}|.m},
\end{equation}
with,
\begin{eqnarray}
A^{ab}_{1m}(z_2)=\bigg(\frac{\spa b.m\spaa m.|P_{(i,j)}.(p_1-P_{(i,j)})|.a}{\spa a.b\spab m.|P_{(i,j)}|.1}-(b\rightarrow b+1)\bigg),\nonumber\\
K^{ab}_{1m}(z_2)=\bigg(\frac{\spa b.m^2\spaa m|.P_{(i,j)}.(p_1-P_{(i,j)})|.a^2}{\spa a.b^2\spab m.|P_{(i,j)}|.1^2}-(b\rightarrow b+1)\bigg).
\end{eqnarray}
Finally the prefactor multiplying the $A$ and $K$ functions is given by,
\begin{eqnarray}
\frac{\spaahl 1|.P_{(i,j)}.a|.m^n}{\spabh a.|P_{(i,j)}.{|a}^{n-1}}=(-1)^n\frac{\spab m.|P_{(i,j)}|.a}
{\spab m.|P_{(i,j)}|.1}\bigg(\frac{(P_{(i,j)}-p_1)^{2n}\spa a.m^n}{\spab a.|P_{(i,j)}|.1^{n-1}}\bigg).
\end{eqnarray}
The overlap contributions are given by,
\begin{eqnarray}
\label{eq:oij1}
O^{i,j}_{1,n}=&&\Gamma_n(z_2)\bigg[\frac{1}{s_{i,j}}\bigg\{\frac{\spab m.|P_{(i,j)}|.j}{\spab m.|P_{(i,j)}|.1}\bigg(-\frac{(P_{(i,j)}-p_1)^{6}\spa j.m^3}{3\spab j.|P_{(i,j)}|.1^{2}}A^{j(i-1)}_{1m}(z_2)\nonumber\\&&+\frac{(P_{(i,j)}-p_1)^{4}\spa j.m^2}{2\spab j.|P_{(i,j)}|.1}K^{j(i-1)}_{1m}(z_2)\bigg)+(j\rightarrow j+1,P_{(i,j)}\rightarrow P_{(i,j)})\nonumber\\&&
+\frac{\spab m.|P_{(i,j)}|.i}{\spab m.|P_{(i,j)}|.1}\bigg(\frac{(P_{(i,j)}-p_1)^{6}\spa i.m^3}{3\spab i.|P_{(i,j)}|.1^{2}}A^{ij}_{1m}(z_2)\nonumber\\&&-\frac{(P_{(i,j)}-p_1)^{4}\spa i.m^2}{2\spab i.|P_{(i,j)}|.1}K^{ij}_{1m}(z_2)\bigg)+(i\rightarrow i-1,{P_{(i,j)}\rightarrow P_{(i,j)}})\bigg\}\bigg]\nonumber\\
\label{eq:o1j}
O^{1,j}_n=&&\Gamma_n(z_2)\bigg[\frac{1}{s_{1,j}}\bigg\{\frac{\spab m.|P_{(1,j)}|.n}{3\spab m.|P_{(2,j)}|.1}\bigg(\frac{(P_{(2,j)})^{6}\spa n.m^3}{\spab n.|P_{(2,j)}|.1^{2}}A^{n j}_{1m}(z_2)-\frac{(P_{(2,j)})^{4}\spa n.m^2}{2\spab n.|P_{(2,j)}|.1}K^{n j}_{1m}(z_2)\bigg)\nonumber\\&&+\bigg(j\rightarrow j+1,{P_{a,j}\rightarrow P_{a,j}}\bigg)\nonumber\\&&+\bigg(-\frac{(P_{(2,j)})^{6}\spa j.m^3}{3\spab j.|P_{(2,j)}|.1^{2}}\bigg(\frac{\spa n.m\spaa m.|(p_1-P_{n,j}).P_{(1,j)}|.j}{\spa  j.n \spab m.|P_{(1,j)}|.1}+\spa 1.m\bigg)\nonumber\\&&+\frac{(P_{(2,j)})^{4}\spa j.m^2}{2\spab j.|P_{(2,j)}|.1}\bigg(\frac{\spa n.m^2\spaa m.|(p_1-P_{n,j}).P_{(1,j)}|.j^2}{\spa j.n^2 \spab m.|P_{(1,j)}|.1^2}-\spa 1.m^2\bigg)\bigg)\bigg\}\bigg]\nonumber\\
\label{eq:oi1}
O^{i,1}_n=&&\Gamma_n(z_2)\bigg[\frac{1}{s_{i,1}}\bigg\{-\frac{\spab m.|P_{(i,1)}|.i}{\spab m.|P_{(i,1)}|.1}\bigg(\frac{(P_{(i,n)})^{6}\spa i.m^3}{3\spab i.|P_{(i,n)}|.1^{2}}\bigg(\frac{\spa 2.m\spaa m.|P_{(i,n)}.P_{(i,1)}|.i}{\spa 2.i\spab m.|P_{(i,1)}|.1}+\spa 1.m\bigg)\nonumber\\&&+\frac{(P_{(i,n)})^{4}\spa i.m^2}{2\spab i.|P_{(i,n)}|.1}\bigg(-\frac{\spa 2.m^2\spaa m.|P_{(i,n)}.P_{(i,1)}|.i^2}{\spa 2.i^2\spab m.|P_{(i,1)}|.1^2}+\spa 1.m^2\bigg)\bigg)+(i\rightarrow i-1,{P_{i,a}\rightarrow P_{i,a}})\nonumber\\&&+\frac{\spab m.|P_{(i,1)}|.2}{\spab m.|P_{(i,1)}|.1}\bigg(-\frac{(P_{(i,n)})^{6}\spa 2.m^3}{3\spab 2.|P_{(i,n)}|.1^{2}}A^{2(i-1)}_{1m}(z_4)+\frac{(P_{(i,n)})^{4}\spa 2.m^2}{2\spab 2.|P_{(i,n)}|.1}K^{2(i-1)}_{1m}(z_4)\bigg)\bigg\}\bigg]\nonumber\\
\end{eqnarray}
At first glance there appear to be poles of order greater than one in $s_{12}$ and $s_{n1}$, however the presence of the 
factor $(P_{(i,j)}-p_1)^{2n}$ ensures that when $i=1,j=2$ or $i,j=1$ there are no issues with higher poles. 
As a result, poles in these channels are only generated by the multiplicative tree factor.

\subsubsection{The overlap terms $O^{1n}_{n}$ and $O^{12}_{n}$}

The final two overlap terms have a more subtle origin than the previous contributions and since they come from the tree factor the form of the cut-completion terms, 
can be 
compacted. We will however, need to have forms for $s_{i,j}$ when they acquire a $z$ dependence (with $z$ now in either the $s_{12}$ or $s_{n1}$ channel). 
We consider first the $s_{12}$ channel ($z_3=-\spa 1.2/\spa m.2$), we use the following form for $s_{i,j}$, when $p_m \in P_{(i,j)}$:
\begin{equation}
s_{i,j}(z_3)=\frac{\spaa m.|P_{(i,j)}.(P_{(i,j)}+p_1)|.2}{\spa m.2}.
\end{equation}
We will also require the tree factor $\Gamma$ and the functions $A$ and $K$:
\begin{eqnarray}
\Gamma_n(z_3)=-\frac{c_{\Gamma}N_{P}}{2\Pi_{\alpha=2}^{n-1}\spa \alpha.\ap}\frac{\spa 2.m}{\spa1.m\spa n.2\spa 1.2}\\
A^{ab}_{m1}(z_3)=\bigg(\frac{\spa b.2\spa1.m\spa m.a}{\spa 2.m\spa a.b}-(b\rightarrow b+1)\bigg)\\
K^{ab}_{m1}(z_3)=\bigg(\frac{\spa b.2^2\spa1.m^2\spa m.a^2}{\spa 2.m^2\spa a.b^2}-(b\rightarrow b+1)\bigg)
\end{eqnarray}
The overlap terms associated with this channel are defined by using the $(1\leftrightarrow m)$ symmetry of the cut-completion terms:
\begin{eqnarray}
O^{12}_n=O^{12}_{m,n}+O^{12}_{1,n}
\end{eqnarray} 
With $O^{12}_{m,n}$ defined by,
\begin{eqnarray}
O^{12}_{m,n}=&&\sum_{i=3}^{m}\sum_{j=m}^{n}\Gamma_n(z_3)\bigg[\bigg\{-\frac{\spaa m|.P_{(i,j)}.j|.2^3\spa1.m^3}{3\spaas m.|j(P_{(i,j-1)}+p_1)+P_{(i,j-1)}j|.2^2}A^{j(i-1)}_{m1}(z_3)\nonumber\\&&+\frac{\spaa m|.P_{(i,j)}.j|.2^2\spa1.m^2}{2\spaas m.|j(P_{(i,j-1)}+p_1)+P_{(i,j-1)}j|.2}K^{j(i-1)}_{m1}(z_3)\bigg\}\nonumber\\&&\times\bigg(\frac{1}{\spaa m.|P_{(i,j)}.(P_{(i,j)}+p_1)|.2}+\frac{1}{\spaa m.|P_{(i,j-1)}.(P_{(i,j-1)}+p_1)|.2}\bigg)\nonumber\\
&&+\sum_{i=3}^{m}\sum_{j=m}^{n-1}\bigg\{\frac{\spaa m.|P_{(i,j)}.i|.2^3\spa1.m^3}{3\spaas m.|i(P_{(i+1,j)}+p_1)+P_{(i+1,j)}i|.2^2}A^{ij}_{m1}(z_3)\nonumber\\&&-\frac{\spaa m.|P_{(i,j)}.i|.2^2\spa1.m^2}{2\spaas m.|i(P_{(i+1,j)}+p_1)+P_{(i+1,j)}i|.2}K^{ij}_{m1}(z_3)\bigg\}\nonumber\\&&\times\bigg(\frac{1}{\spaa m.|P_{(i,j)}.(P_{(i,j)}+p_1)|.2}+\frac{1}{\spaa m.|P_{(i+1,j)}.(P_{(i+1,j)}+p_1)|.2}\bigg)\nonumber\\
&&+\sum_{i=3}^{m}\bigg\{\frac{\spaa m.|P_{(i,n)}.i|.2^3\spa1.m^3}{3\spaas m.|iP_{(i+1,1)}+P_{(i+1,n)}i|.2^2}\frac{\spa m.i\spa n.2\spa1.m}{\spa i.n\spa2.m}\nonumber\\&&-\frac{\spaa
m.|P_{(i,n)}.i|.2^2\spa1.m^2}{2\spaas m.|iP_{(i+1,1)}+P_{(i+1,n)}i|.2}\frac{\spa m.i^2\spa n.2^2\spa1.m^2}{\spa i.n^2\spa2.m^2}\bigg\}\nonumber\\&&\times\bigg(\frac{1}{\spaa
m.|P_{(i,n)}.P_{(i,1)}|.2}+\frac{1}{\spaa m.|P_{(i+1,n)}.P_{(i+1,1)}|.2}\bigg)\bigg].\nonumber\\
\end{eqnarray}
For the second set of sums we will need to know $s_{i,j}$ with $p_1 \in P_{(i,j)}$,
\begin{equation}
s_{i,j}(z_3)=\frac{\spaa m.|P_{(i,j)}.(P_{(i,j)}-p_1)|.2}{\spa m.2}.
\end{equation}
We also require,
\begin{eqnarray}
A^{ab}_{1m}(z_3)=\bigg(\frac{\spa b.m\spa 1.m\spa 2.a}{\spa 2.m\spa a.b}-(b\rightarrow b+1)\bigg),\\
K^{ab}_{1m}(z_3)=\bigg(\frac{\spa b.m^2\spa 1.m^2\spa 2.a^2}{\spa 2.m^2\spa a.b^2}-(b\rightarrow b+1)\bigg),
\end{eqnarray}
so that,
\begin{eqnarray}
O^{12}_{1,n}=&&\Gamma_n(z_3)\bigg[\sum_{i=m+1}^{n}\sum_{j=2}^{m-1}\bigg\{-\frac{\spaa 2|.P_{(i,j)}.j|.m^3\spa1.m^3}{3\spaas m.|j(P_{(i,j-1)}-p_1)+P_{(i,j-1)}j|.2^2}A^{j(i-1)}_{1m}(z_3)\nonumber\\&&+\frac{\spaa 2|.P_{(i,j)}.j|.m^2\spa1.m^2}{2\spaas m.|j(P_{(i,j-1)}-p_1)+P_{(i,j-1)}j|.2}K^{j(i-1)}_{1m}(z_3)\bigg\}\nonumber\\&&\times\bigg(\frac{1}{\spaa m.|P_{(i,j)}.(P_{(i,j)}-p_1)|.2}+\frac{1}{\spaa m.|P_{(i,j-1)}.(P_{(i,j-1)}-p_1)|.2}\bigg)\nonumber\\&&+\sum_{j=2}^{m-1}\bigg\{-\frac{\spaa 2|.P_{(1,j)}.j|.m^3\spa1.m^3}{3\spaas m.|jP_{(2,j-1)}+P_{(1,j-1)}j|.2^2}\bigg(\frac{\spa2.j\spa n.m\spa 1.m}{\spa j.n\spa 2.m}+\spa 1.m\bigg)\nonumber\\&&
+\frac{\spaa 2|.P_{(1,j)}.j|.m^2\spa1.m^2}{2\spaas m.|jP_{(2,j-1)}+P_{(i,j-1)}j|.2}\bigg(\frac{\spa2.j^2\spa n.m^2\spa 1.m^2}{\spa j.n^2\spa 2.m^2}-\spa 1.m^2\bigg)\bigg\}\nonumber\\&&\times\bigg(\frac{1}{\spaa m.|P_{(1,j)}.P_{(3,j)}|.2}+\frac{1}{\spaa m.|P_{(2,j-1)}.P_{(3,j-1)}|.2}\bigg)\nonumber\\
&&+\sum_{i=m+1}^{n}\sum_{j=2}^{m-1}\bigg\{\frac{\spaa 2|.P_{(i,j)}.i|.m^3\spa1.m^3}{3\spaas m.|i(P_{(i+1,j)}-p_1)+P_{(i+1,j)}i|.2^2}A^{ij}_{1m}(z_3)\nonumber\\&&-\frac{\spaa
2|.P_{(i,j)}.i|.m^2\spa1.m^2}{2\spaas m.|i(P_{(i+1,j)}-p_1)+P_{(i+1,j)}i|.2}K^{ij}_{1m}(z_3)\bigg\}\nonumber\\&&\times\bigg(\frac{1}{\spaa
m.|P_{(i,j)}.(P_{(i,j)}-p_1)|.2}+\frac{1}{\spaa m.|P_{(i+1,j)}.(P_{(i+1,j)}-p_1)|.2}\bigg)\bigg].
\end{eqnarray}
In writing the above, we have used $A^{i1}_{m1}(z_3)=0$. 

The final overlap term is $O^{n1}_n$ and is calculated noting the $(n \leftrightarrow 2)$ symmetry in the shift. 
Once again we define the usual functions under the shift $(z_4)=-\spa1.n/\spa m.n$ , When  $p_m \in P_{(i,j)}$,
\begin{equation}
s_{i,j}(z_4)=\frac{\spaa m.|P_{(i,j)}.(P_{(i,j)}+p_1)|.n}{\spa m.n},
\end{equation}
together with,
\begin{eqnarray}
\Gamma_n(z_4)=-\frac{c_{\Gamma}N_{P}}{2\Pi_{\alpha=2}^{n-1}\spa \alpha.\ap}\frac{\spa n.m}{\spa1.m\spa n.1\spa 2.n},\\
A^{ab}_{m1}(z_4)=\bigg(\frac{\spa b.n\spa1.m\spa m.a}{\spa n.m\spa a.b}-(b\rightarrow b+1)\bigg),\\
K^{ab}_{m1}(z_4)=\bigg(\frac{\spa b.n^2\spa1.m^2\spa m.a^2}{\spa n.m^2\spa a.b^2}-(b\rightarrow b+1)\bigg).
\end{eqnarray}
The overlap terms in this channel are again split into two terms 
\begin{eqnarray}
O^{n1}_n=O^{n1}_{m,n}+O^{n1}_{1,n}, 
\end{eqnarray} 
with
\begin{eqnarray}
O^{n1}_{m,n}=&&\sum_{i=3}^{m}\sum_{j=m}^{n-1}\Gamma_n(z_4)\bigg[\bigg\{-\frac{\spaa m.|P_{(i,j)}.j|.n^3\spa1.m^3}{3\spaas m.|j(P_{(i,j-1)}+p_1)+P_{(i,j-1)}j|.n^2}A^{j(i-1)}_{m1}(z_4)\nonumber\\&&+\frac{\spaa m.|P_{(i,j)}.j|.n^2\spa1.m^2}{2\spaas m.|j(P_{(i,j-1)}+p_1)+P_{(i,j-1)}j|.n}K^{j(i-1)}_{m1}(z_4)\bigg\}\nonumber\\&&\times\bigg(\frac{1}{\spaa m.|P_{(i,j)}.(P_{(i,j)}+p_1)|.2}+\frac{1}{\spaa m.|P_{(i,j-1)}.(P_{(i,j-1)}+p_1)|.n}\bigg)\nonumber\\&&
+\sum_{j=m}^{n-1}\bigg\{\frac{\spaa m|.P_{(2,j)}.j|.n^3\spa1.m^3}{3\spaas m.|jP_{(1,j-1)}+P_{(2,j-1)}j|.n^2}\frac{\spa m.j\spa2.n\spa1.m}{\spa j.2\spa n.m}\nonumber\\&&-\frac{\spaa m|.P_{(2,j)}.j|.n^2\spa1.m^2}{2\spaas m.|jP_{(1,j-1)}+P_{(2,j-1)}j|.n}\frac{\spa m.j^2\spa2.n^2\spa1.m^2}{\spa j.2^2\spa n.m^2}\bigg\}\nonumber\\&&\times\bigg(\frac{1}{\spaa m.|P_{(2,j)}.P_{(1,j)}|.2}+\frac{1}{\spaa m.|P_{(2,j-1)}.P_{(1,j-1)}|.n}\bigg)\nonumber\\
&&+\sum_{i=2}^{m}\sum_{j=m}^{n-1}\bigg\{\frac{\spaa m.|P_{(i,j)}.i|.n^3\spa1.m^3}{3\spaas m.|i(P_{(i+1,j)}+p_1)+P_{(i+1,j)}i|.n^2}A^{ij}_{m1}(z_4)\nonumber\\&&-\frac{\spaa
m|.P_{(i,j)}.i|.n^2\spa1.m^2}{2\spaas m.|i(P_{(i+1,j)}+p_1)+P_{(i+1,j)}i|.n}K^{ij}_{m1}(z_3)\bigg\}\nonumber\\&&\times\bigg(\frac{1}{\spaa
m.|P_{(i,j)}.(P_{(i,j)}+p_1)|.n}+\frac{1}{\spaa m.|P_{(i+1,j)}.(P_{(i+1,j)}+p_1)|.n}\bigg)\bigg].\nonumber\\
\end{eqnarray}
For the second set of terms we need to evaluate $s_{i,j}$
when $p_1 \in P_{(i,j)}$, so that,
\begin{equation}
s_{i,j}(z_4)=\frac{\spaa m.|P_{(i,j)}.(P_{(i,j)}-p_1)|.n}{\spa m.n},
\end{equation}
and,
\begin{eqnarray}
A^{ab}_{1m}(z_4)=\bigg(\frac{\spa b.m\spa 1.m\spa n.a}{\spa n.m\spa a.b}-(b\rightarrow b+1)\bigg),\\
K^{ab}_{1m}(z_4)=\bigg(\frac{\spa b.m^2\spa 1.m^2\spa n.a^2}{\spa n.m^2\spa a.b^2}-(b\rightarrow b+1)\bigg).
\end{eqnarray}
We find that,
\begin{eqnarray}
O^{n1}_{1,n}=&&\Gamma_n(z_4)\bigg[\sum_{i=m+1}^{n}\sum_{j=2}^{m-1}\bigg\{-\frac{\spaa n|.P_{(i,j)}.j|.m^3\spa1.m^3}{3\spaas m.|j(P_{(i,j-1)}-p_1)+P_{(i,j-1)}j|.n^2}A^{j(i-1)}_{1m}(z_4)\nonumber\\&&+\frac{\spaa n|.P_{(i,j)}.j|.m^2\spa1.m^2}{2\spaas m.|j(P_{(i,j-1)}-p_1)+P_{(i,j-1)}j|.n}K^{j(i-1)}_{1m}(z_4)\bigg\}\nonumber\\&&\times\bigg(\frac{1}{\spaa m.|P_{(i,j)}.(P_{(i,j)}-p_1)|.n}+\frac{1}{\spaa m.|P_{(i,j-1)}.(P_{(i,j-1)}-p_1)|.n}\bigg)\nonumber\\
&&+\sum_{i=m+1}^{n}\sum_{j=2}^{m-1}\bigg\{\frac{\spaa n.|P_{(i,j)}.i|.m^3\spa1.m^3}{3\spaas m.|i(P_{(i+1,j)}-p_1)+P_{(i+1,j)}i|.n^2}A^{ij}_{1m}(z_4)\nonumber\\&&-\frac{\spaa n.|P_{(i,j)}.i|.m^2\spa1.m^2}{2\spaas m.|i(P_{(i+1,j)}-p_1)+P_{(i+1,j)}i|.n}K^{ij}_{1m}(z_4)\bigg\}\nonumber\\&&\times\bigg(\frac{1}{\spaa m.|P_{(i,j)}.(P_{(i,j)}-p_1)|.n}+\frac{1}{\spaa m.|P_{(i+1,j)}.(P_{(i+1,j)}-p_1)|.n}\bigg)\nonumber\\&&+\sum_{i=m+1}^{n}
\bigg\{-\frac{\spaa n.|P_{(i,1)}.i|.m^3\spa1.m^3}{3\spaas m.|iP_{(i+1,n)}+P_{(i+1,1)}i|.n^2}\bigg(\spa1.m+\frac{\spa n.i\spa 1.m\spa2.m}{\spa i.2\spa
n.m}\bigg)\nonumber\\&&-\frac{\spaa n.|P_{(i,1)}.i|.m^2\spa1.m^2}{2\spaas m.|iP_{(i+1,n)}+P_{(i+1,1)}i|.n}\bigg(\spa1.m^2-\frac{\spa n.i^2\spa 1.m^2\spa2.m^2}{\spa i.2^2\spa
n.m^2}\bigg)\bigg\}\nonumber\\&&\times\bigg(\frac{1}{\spaa m.|P_{(i,1)}.P_{(i,n)}|.n}+\frac{1}{\spaa m.|P_{(i+1,1)}.P_{(i+1,n)}|.n}\bigg)\bigg].
\end{eqnarray}

\section{The four point amplitude}
\label{sec:fourpoint}

The calculation of all Higgs plus four-gluon amplitudes at 
NLO in the heavy-top effective theory has been performed numerically in ~\cite{Campbell:2006xx}.
Here we provide an analytic form for 
the $A^{(1)}_4(H,1^-,2^+,3^-,4^+)$ to illustrate the use of 
our results for the $\phi$-MHV amplitude for general $n$. 

The cut-constructible part of the $\phi$-MHV four point amplitude is given by setting 
$n=4$ and $m=3$ in eq.~\eqref{eq:phimhv}, using the gluonic, fermionic and scalar
contributions given in eqs.~\eqref{eq:phiG}, \eqref{eq:phiFs} and \eqref{eq:phiSs} respectively,
\begin{eqnarray}
C_4(\phi,1^-,2^+,3^-,4^+)=A^{(0)}_4\bigg\{-
\frac{1}{2}\sum_{i=1}^{4}\Ftme(s_{i,i+3},s_{i+1,i+2};s_{i+1,i+3},s_{i,i+2})\nonumber\\-\frac{1}{2}\sum_{i=1}^{4}\Fom(s_{i,i+2};s_{i,i+1},s_{i+1,i+2})
+\sum_{i=1}^{4}(\Tri(s_{i,2+i})-\Tri(s_{i,3+i}))\nonumber\\
-4\bigg(1-\frac{\NF}{4N_c}\bigg)\bigg[\frac{1}{2}\frac{\trm(3241)\trm(3421)}{s^2_{24}s^2_{13}}\FFom(s_{234},s_{23},s_{34})\nonumber\\
-\frac{\trm(3241)\trm(3421)}{s_{24}s^2_{13}}L_1(s_{23},s_{234})
+(2\leftrightarrow 4)+(1\leftrightarrow3)+(1\leftrightarrow3,2\leftrightarrow4)\bigg]\nonumber\\
-2\bigg(1-\frac{\NF}{N_c}\bigg)\bigg[-\frac{1}{2}\frac{\trm(3241)^2\trm(3421)^2}{s^4_{24}s^4_{13}}\FFom(s_{234},s_{23},s_{34})\nonumber\\
-\frac{\trm(3241)\trm(3421)}{s^4_{13}}\bigg(\frac{\trm(3241)^2}{3s_{24}}L_3(s_{23},s_{234})\nonumber\\+\frac{\trm(3241)\trm(3421)}{2s^2_{24}}L_2(s_{23},s_{234})-\frac{\trm(3421)\trm(3241)}{s^3_{24}}L_1(s_{23},s_{234})\bigg)\nonumber\\
+(2\leftrightarrow 4)+(1\leftrightarrow3)+(1\leftrightarrow3,2\leftrightarrow4)\bigg]\bigg\}.
\end{eqnarray}
The cut completion terms are given by eq.~\eqref{eq:CR},
\begin{eqnarray}
\label{eq:CR4}
\lefteqn{CR_4(\phi,1^-,2^+,3^-,4^+)=\frac{N_P}{32\pi^2}\frac{1}{\spa1.2\spa2.3\spa3.4\spa4.1}}
\nonumber\\
&\times&
\bigg[\bigg(-\frac{\spaa 3|.2.4.{|1}^3}{3(s_{234}-s_{23})^2}\frac{\spa3.4\spa2.1}{\spa4.2}
-\frac{\spaa 3|.2.4.{|1}^2}{2(s_{234}-s_{23})}\frac{\spa3.4^2\spa2.1^2}{\spa4.2^2}\bigg)\bigg(\frac{1}{s_{23}}+\frac{1}{s_{234}}\bigg)\bigg] \nonumber \\
&&+(2\leftrightarrow 4)+(1\leftrightarrow3)+(1\leftrightarrow3,2\leftrightarrow4).
\end{eqnarray}

The remaining rational contributions are obtained by shifting the two negative helicity gluons,
\begin{equation}
\label{eq:z4shifts}
|\hat{1}\rangle=|1\rangle+z|3\rangle, \qquad |\hat{3}]=|3]-z|1].
\end{equation}
As discussed in subsection~\ref{subsec:infcr}, this shift generates a non-vanishing
contribution as $z \to \infty$ in the
cut completion term $CR_4$.   To compute this contribution, we use eq.~\eqref{eq:infC} with $m=3$ and $n=4$ to find,
\begin{eqnarray}
\mathrm{Inf}\,CR_4(\phi,1^-,2^+,3^-,4^+)=-\frac{N_{P}}{32\pi^2} 
\frac{\spa2.3\spa3.4\spb2.4^2}{\spa2.4^2\spb1.2\spb4.1}.
\end{eqnarray}

The direct rational contribution is generated by the recursion relation 
\eqref{eq:rats}, again with $m=3$ and $n=4$ and
is given by,
\begin{eqnarray}
R_4(\phi,1^-,2^+,3^-,4^+)&=&A^{(0)}(\phi,\hat{1}^-,\hat{P}^-_{234})\frac{1}{s_{234}}R(-\hat{P}^+_{234},2^+,\hat{3}^-,4^+)\nonumber\\
&&+R(4^+,\hat{1}^-,2^+,-\hat{P}^+_{412})\frac{1}{s_{412}}A^{(0)}(\phi,\hat{P}^-_{412},\hat{3}^-)\nonumber\\
&&+R(\phi,\hat{1}^-,2^+,-\hat{P}^+_{34},)\frac{1}{s_{34}}A^{(0)}(\hat{P}^-_{34},\hat{3}^-,4^+)\nonumber\\
&&+R(\phi,\hat{1}^-,4^+,-\hat{P}^+_{23})\frac{1}{s_{23}}A^{(0)}(\hat{P}^-_{23},2^+,\hat{3}^-)\nonumber\\
&&+A^{(0)}(\hat{1}^-,\hat{P}^+_{41},4^+)\frac{1}{s_{41}}R(\phi,-\hat{P}^-_{41},2^+,\hat{3}^-)\nonumber\\
&&+A^{(0)}(\hat{1}^-,\hat{P}^+_{12},2^+)\frac{1}{s_{12}}R(\phi,-\hat{P}^-_{12},\hat{3}^-,4^+),
\label{eq:rats4}
\end{eqnarray}
where we recycle the known lower point amplitudes.
For the four-point amplitude, we require the rational parts of the 
$\phi$ with one minus and two positive helicity gluons (\ref{eq:ncc3-mpp}),
the two and three-point  $\phi$-MHV amplitudes given in eqs.~\eqref{eq:ncc2-mm}, 
\eqref{eq:ncc3-mmp} and \eqref{eq:ncc3-mpm}, as well as the pure 
four-gluon QCD amplitude with a single negative helicity of eq.~\eqref{eq:ncc4-mppp}.

We find that 
\begin{eqnarray}
R^{234}_4=\frac{N_P}{96\pi^2}\frac{m^4_{H}}{s_{234}}\frac{\spa 2.4\spb2.4\spab 3.|P_{234}|.1^2}{\spab 4.|P_{234}|.1^2\spab2.|P_{234}|.1^2}.
\end{eqnarray}
Similarly, 
\begin{eqnarray}
&&R^{24}_4=-\frac{1}{8\pi^2}A^{(0)}(\phi^{\dagger},4^+,2^+,3^-,1^-)-\frac{N_P}{96\pi^2}s_{123}\frac{\spb 2.4\spb2.1}{\spb3.1\spb2.3}\frac{\spab4.|P_{123}|.2}{\spab
4.|P_{123}|.1^2},\nonumber\\
&&R^{34}_4=R^{23}_4 \quad (2\leftrightarrow4).
\end{eqnarray}
In the other channels,
\begin{eqnarray}
R^{41}_4=-\frac{1}{8\pi^2}A^{(0)}(\phi,1^-,3^-,2^+,4^+)\nonumber\\
R^{12}_4=R^{41}_4 \quad (4\leftrightarrow2),
\end{eqnarray}
and finally,
\begin{equation}
R^{412}_4=\frac{N_P}{96\pi^2}\frac{\spb2.4^3}{s_{412}}\frac{\spab 3.|P_{412}|.1^2}{\spa 2.4\spb1.2^2\spb4.1^2}.
\end{equation}

The overlap terms are given by,
\begin{equation}
O_4(\phi,1^-,2^+,3^-,4^+)=O^{234}_4+O^{23}_4+O^{34}_4+O^{41}_4+O^{12}_4+O^{412}_4.
\end{equation}
The first term is generated by eq.~(\ref{eq:o2n}) with $n=4$ and has the following form
\begin{eqnarray}
O^{234}_4=&&\frac{N_P}{32\pi^2s_{234}}\bigg(\frac{1}{3}\frac{\spaa3|.P_{234}.P_{1234}|.2^2\spb4.2}{ \spa2.4\spab2.|P_{234}|.1^2}
\nonumber\\&&+
 \frac{1}{2}\frac{\spa3.2\spaa3.|P_{234}.P_{1234}|.2\spaa3.|P_{234}.P_{1234}|.4\spb4.2}{\spa2.4^2\spab2.|P_{234}|.1\spab3.|P_{234}|.1}
+(2\leftrightarrow4)\bigg).
\end{eqnarray}
The overlap pieces in the $23$ and $34$ channels are given by eq.~\eqref{eq:o2j} and eq.~\eqref{eq:oin} (with $i=j=3$),
\begin{eqnarray}
O^{23}_4=&&-\frac{N_P}{32\pi^2s_{23}}\bigg(-\frac{\spa3.2^2\spab4.|P_{123}|.2^2\spb2.4}{3\spab4.|P_{123}|.1^2\spa4.2}
\nonumber\\&&+\frac{\spa3.2\spa3.4\spb2.4\spab2.|P_{123}|.2\spab4.|P_{123}|.2}{2\spb1.2\spa4.2^2\spab 4.|P_{123}|.1}\bigg),\\
O^{34}_4=&&O^{23}_4 \quad (4\leftrightarrow2).
\end{eqnarray}
$O^{41}$ and $O^{12}$ both vanish,
while eq.~(\ref{eq:oij1}) with $i=4, j=2$ leads to,
\begin{eqnarray}
O^{412}_4=-\frac{N_P}{32\pi^2s_{412}}\bigg(\frac{1}{2}\frac{\spa2.3\spa4.3\spab3.|P_{412}|.4\spb4.2^2}{\spa2.4\spab3.|P_{412}|.1\spb4.1}
- \frac{1}{3}\frac{\spa2.3^2\spb4.2^3}{\spa2.4\spb4.1^2} 
+ (2\leftrightarrow4)\bigg).
\end{eqnarray}

Combining contributions, the full four-point amplitude is given by,
\begin{eqnarray}
A^{(1)}_4(\phi,1^-,2^+,3^-,4^+)&=&C_4(\phi,1^-,2^+,3^-,4^+)+CR_4(\phi,1^-,2^+,3^-,4^+)\nonumber\\
&+&\hat{R}_4(\phi,1^-,2^+,3^-,4^+),
\label{eq:phiTot}
\end{eqnarray}
with
\begin{eqnarray}
\hat{R}(\phi,1^-,2^+,3^-,4^+)&=&O_4(\phi,1^-,2^+,3^-,4^+)+R_4(\phi,1^-,2^+,3^-,4^+)\nonumber\\&-&\mathrm{Inf}CR_4(\phi,1^-,2^+,3^-,4^+).
\end{eqnarray}
After some algebra, the combination of overlapping and recursive terms can be written
in the following form,  free of spurious singularities \footnote{Which we have checked with the aid of the package S@M ~\cite{Maitre:2007jq}},
\begin{eqnarray}
\lefteqn{\hat{R}_4(\phi,1^-,2^+,3^-,4^+)=-\frac{1}{8\pi^2}A^{(0)}(A,1^-,2^+,3^-,4^+)}\nonumber\\
&+&\frac{N_P}{192\pi^2}\frac{\spb2.4^4}{\spb1.2\spb2.3\spb3.4\spb4.1}
\bigg(-\frac{s_{23}s_{34}}{s_{24}s_{412}}+3\frac{s_{23}s_{34}}{s_{24}^2}-\frac{s_{12}s_{41}}{s_{24}s_{234}}+3\frac{s_{12}s_{41}}{s_{24}^2}\bigg),\nonumber\\
\label{eq:phiTotRat}
\end{eqnarray}
where $A^{(0)}(A,1^-,2^+,3^-,4^+)$ is the difference of $\phi$ and $\phi^{\dagger}$ amplitudes. 
Finally the full Higgs amplitude is given by the sum of $\phi$ and $\phi^{\dagger}$ amplitudes
\begin{eqnarray}
A^{(1)}_4(H,1^-,2^+,3^-,4^+)&=&A^{(1)}_4(\phi,1^-,2^+,3^-,4^+)+A^{(1)}_4(\phi^{\dagger},1^-,2^+,3^-,4^+),
\end{eqnarray}
with,
\begin{eqnarray}
A^{(1)}_4(\phi^{\dagger},1^-,2^+,3^-,4^+)&=&
A^{(1)}_4(\phi,2^-,3^+,4^-,1^+)_{\spa i.j \leftrightarrow \spb i.j}.
\end{eqnarray}
We note that the rational terms not proportional to $N_P$ in 
eq.~\eqref{eq:phiTotRat}
cancel when forming the Higgs amplitude, just as for the 
$A^{(1)}_4(H,1^-,2^-,3^+,4^+)$ amplitude of ref.~\cite{Badger:2007si}.

The one-loop amplitudes in this paper are computed in the four-dimensional
helicity scheme and are not renormalised.   To perform an $\overline{MS}$
renormalisation, one should subtract an $\overline{MS}$ counterterm from
$A^{(1)}_n$,
\begin{equation}
A^{(1)}_n \to A^{(1)}_n - c_\Gamma \frac{n}{2}\frac{\beta_0}{\epsilon}
A^{(0)}_n.
\end{equation}
The Wilson coefficient \eqref{eq:C} produces an additional finite contribution,
\begin{equation}
A^{(1)}_n \to A^{(1)}_n +\frac{11}{2}~A^{(0)}_n.
\end{equation}


\section{Cross Checks and Limits}
\label{sec:checks}

\subsection{Infrared poles}

The infrared pole structure of a one-loop $\phi$-amplitude has the following form,
\begin{equation}
A^{(1)}_n=-\frac{c_{\Gamma}}{\epsilon^2}A^{(0)}\sum_{i=1}^{n}\bigg(\frac{\mu^2}{-s_{i,i+1}}\bigg)^{\epsilon}+\mathcal{O}(\epsilon^0).
\end{equation}
Since only $A^{\phi,G}_{n;1}(m,n)$ contributes at $\mathcal{O}(\epsilon^{-2})$ the IR pole structure of the 
general $\phi$-MHV amplitude is identical to that of the adjacent minus  case (apart from the trivial change in the tree amplitude). 
This combination was shown to have the correct IR behaviour in~\cite{Badger:2007si}.

\subsection{Collinear limits}

The general behaviour of a one-loop amplitude when gluons $i$ and $j$ become collinear, such that
$p_i\rightarrow zK$ and $p_{i+1}\rightarrow (1-z)K$,
is well known,
\begin{eqnarray}
&&A^{(1)}_n(\dots,i^{\lambda_i},i+1^{\lambda_{i+1}},\dots)^{i\underrightarrow{\parallel i+}1}\nonumber\\
\sum_{h=\pm}&\big [&
A_{n-1}^{(1)}(\dots,i-1^{\lambda_{i-1}},K^h,i+2^{\lambda_{i+2}},\dots)\mathrm{Split}^{(0)}(-K^{-h};i^{\lambda_i},i+1^{\lambda_{i+1}})\nonumber\\
&&+
A_{n-1}^{(0)}(\dots,i-1^{\lambda_{i-1}},K^h,i+2^{\lambda_{i+2}},\dots)\mathrm{Split}^{(1)}(-K^{-h};i^{\lambda_i},i+1^{\lambda_{i+1}})\big ].
\end{eqnarray}
The universal splitting functions are given by~\cite{BDDK:uni1,BDDK:uni2,Bern:allorder},
\begin{eqnarray}
&&\mathrm{Split}^{(0)}(-K^{+};1^{-},2^{+})=\frac{z^2}{\sqrt{z(1-z)}\spa1.2},\\
&&\mathrm{Split}^{(0)}(-K^{+};1^{+},2^{-})=\frac{(1-z)^2}{\sqrt{z(1-z)}\spa1.2},\\
&&\mathrm{Split}^{(0)}(-K^{-};1^{+},2^{+})=\frac{1}{\sqrt{z(1-z)}\spa1.2},\\
&&\mathrm{Split}^{(0)}(-K^{-};1^{-},2^{-})=0.
\end{eqnarray}
The one-loop splitting function can be written in terms of cut-constructible and rational components,
\begin{equation}
\mathrm{Split}^{(1)}(-K^{-h},1^{\lambda_1},2^{\lambda_2})=\mathrm{Split}^{(1),C}(-K^{-h},1^{\lambda_1},2^{\lambda_2})
+\mathrm{Split}^{(1),R}(-K^{-h},1^{\lambda_1},2^{\lambda_2})
\end{equation}
where
\begin{eqnarray}
&&\mathrm{Split}^{(1),C}(-K^{\pm},1^{-},2^{+})=\mathrm{Split}^{(0)}(-K^{\pm},1^{-},2^{+})\frac{c_{\Gamma}}{\epsilon^2}\times\nonumber\\
&&\bigg(\frac{\mu^2}{-s_{12}}\bigg)^{\epsilon}\bigg(1-\FF{1,-\e;1-\e; \frac{z}{z-1}}\bigg)-\FF{1,-\e;1-\e; \frac{z}{z-1}}\bigg)\bigg),\\
&&\mathrm{Split}^{(1),C}(-K^{+},1^{-},2^{-})=\mathrm{Split}^{(0)}(-K^{+},1^{-},2^{-})\frac{c_{\Gamma}}{\epsilon^2}\times\nonumber\\
&&\bigg(\frac{\mu^2}{-s_{12}}\bigg)^{\epsilon}\bigg(1-\FF{1,-\e;1-\e; \frac{z}{z-1}}\bigg)-\FF{1,-\e;1-\e; \frac{z}{z-1}}\bigg)\bigg),\\
&&\mathrm{Split}^{(1),C}(-K^{-},1^{-},2^{-})=0,\\
&&\mathrm{Split}^{(1),R}(-K^{\pm},1^{-},2^{+})=0,\\
&&\mathrm{Split}^{(1),R}(-K^{+},1^{-},2^{-})=\frac{N_P}{96\pi^2}\frac{\sqrt{z(1-z)}}{\spb1.2},\\
&&\mathrm{Split}^{(1),R}(-K^{-},1^{-},2^{-})=\frac{N_P}{96\pi^2}\frac{\sqrt{z(1-z)}\spa1.2}{\spb1.2^2}.
\end{eqnarray}
Explicitly, the cut-constructible parts should satisfy, 
\begin{eqnarray}
\lefteqn{C_n(\dots,i^{\lambda_i},i+1^{\lambda_{i+1}},\dots)^{i\underrightarrow{\parallel i+}1}\sum_{h=\pm}}\nonumber\\
&&
\phantom{+}C_{n-1} (\dots,i-1^{\lambda_{i-1}},K^h,i+2^{\lambda_{i+2}},\dots)~\mathrm{Split}^{(0)}(-K^{-h};i^{\lambda_i},i+1^{\lambda_{i+1}})\nonumber\\
&&+
A_{n-1}^{(0)}(\dots,i-1^{\lambda_{i-1}},K^h,i+2^{\lambda_{i+2}},\dots)~\mathrm{Split}^{(1),C}(-K^{-h};i^{\lambda_i},i+1^{\lambda_{i+1}}),
\end{eqnarray}
while the rational pieces obey,
\begin{eqnarray}
\lefteqn{R_n(\dots,i^{\lambda_i},i+1^{\lambda_{i+1}},\dots)^{i\underrightarrow{\parallel i+}1}\sum_{h=\pm}}\nonumber\\
&&
\phantom{+}R_{n-1} (\dots,i-1^{\lambda_{i-1}},K^h,i+2^{\lambda_{i+2}},\dots)~\mathrm{Split}^{(0)}(-K^{-h};i^{\lambda_i},i+1^{\lambda_{i+1}})\nonumber\\
&&+
A_{n-1}^{(0)}(\dots,i-1^{\lambda_{i-1}},K^h,i+2^{\lambda_{i+2}},\dots)~\mathrm{Split}^{(1),R}(-K^{-h};i^{\lambda_i},i+1^{\lambda_{i+1}}).
\end{eqnarray}

\subsection{Collinear factorisation of the cut-constructible contributions}

In Ref.~\cite{Badger:2007si}, it was demonstrated that the helicity independent cut-constructible 
gluonic contribution obeys,
\begin{eqnarray}
\lefteqn{C^{\phi\{G\}}_n(\dots,i^{\lambda_i},i+1^{\lambda_{i+1}},\dots)^{i\underrightarrow{\parallel i+}1}\sum_{h=\pm}}\nonumber\\
&&
\phantom{+}C^{\phi\{G\}}_{n-1} (\dots,i-1^{\lambda_{i-1}},K^h,i+2^{\lambda_{i+2}},\dots)~\mathrm{Split}^{(0)}(-K^{-h};i^{\lambda_i},i+1^{\lambda_{i+1}})\nonumber\\
&&+
A_{n-1}^{(0)}(\dots,i-1^{\lambda_{i-1}},K^h,i+2^{\lambda_{i+2}},\dots)~\mathrm{Split}^{(1),C}(-K^{-h};i^{\lambda_i},i+1^{\lambda_{i+1}}).
\end{eqnarray}
Therefore to check the collinear behaviour of the general $\phi$-MHV amplitude, we simply need to check  that 
the fermionic and scalar contributions satisfy the following relation,
\begin{eqnarray}
&&C^{\phi\{F,S\}}_n(\dots,i^{\lambda_i},i+1^{\lambda_{i+1}},\dots)^{i\underrightarrow{\parallel i+}1}\nonumber\\
&&\sum_{h=\pm}
C_{n-1}^{\phi\{F,S\}}(\dots,i-1^{\lambda_{i-1}},K^h,i+2^{\lambda_{i+2}},\dots)~\mathrm{Split}^{(0)}(-K^{-h};i^{\lambda_i},i+1^{\lambda_{i+1}}).
\end{eqnarray}
In other words,  the $F$ and $S$ contributions should factorise onto the tree-level splitting amplitude for the helicity of the gluons considered. 
According to the definition of $C_n$ in eq.~\eqref{eq:phimhv}, there is an overall factor $A^{(0)}_{n}$, which in the collinear limit
produces the correct tree-level splitting function.
It therefore remains to show that,
\begin{equation}
A_{n;1}^{\phi F,\phi S}  \to A_{n-1;1}^{\phi F,\phi S} 
\end{equation}
in the collinear limit 
with $A^{\phi F}_{n;1}(m,n)$ and $A^{\phi S}_{n;1}(m,n)$ given in eqs.~\eqref{eq:phiFs} and \eqref{eq:phiSs} respectively.

\subsubsection{Collinear behaviour of mixed helicity gluons}

We first consider the limit where two adjacent gluons become collinear, one of which has negative helicity. For definiteness, 
we take the limit $\mm \parallel m$. 

The coefficient of the box function $b_{m1}^{ij}$ 
enters both $A^{\phi S}$ and $A^{\phi F}$.  In this limit,
\begin{equation}
b^{ij}_{m1}\,^{\underrightarrow{m-1\parallel m}}\,\frac{\trm(K,i,j,1)\trm(K,j,i,1)}{s^2_{ij}s^2_{1K}} \equiv b^{ij}_{K1}.
\end{equation}
For the special cases, $i=m-1$ and $j=m-1$, we have,
\begin{equation}
b^{m-1,j}_{m1}=b^{i,m-1}_{m1}=0
\end{equation}
so that the box contribution correctly factorises onto the lower point amplitude. 

The remaining terms in the sub-amplitudes are proportional to one of 
the auxiliary functions $\mathcal{F}^{ij}_{m1}$ with $\mathcal{F} = \mathcal{A},\mathcal{K}$ and $\mathcal{I}$
and which are defined in eqs.~\eqref{eq:Adef},~\eqref{eq:Kdef} and \eqref{eq:Idef}. 
We shall see that these too 
have the correct factorisation properties.
Let us first consider the ranges $2 \le i \le m-1$ and $m \le j \le n$. 
When $i \le m-2$, the momentum $P_{(i,j)}$ always contains both $m-1$ and $m$,
while   $P_{(j,i)}$ never includes either $m-1$ or $m$, and we find relations 
such as,
\begin{eqnarray}
\frac{\trm(m,P_{(i,j)},i,1)}{s^2_{1m}} \mathcal{A}^{ij}_{m1} &\phantom{}
 \,^{\underrightarrow{m-1\parallel m}}\,&
\frac{\trm(K,P_{(i,j)},i,1)}{s^2_{1K}} \mathcal{A}^{ij}_{K1},\nonumber \\
\frac{\trm(1,P_{(j,i)},i,m)}{s^2_{1m}} \mathcal{A}^{i(j-1)}_{1m} &\phantom{}
 \,^{\underrightarrow{m-1\parallel m}}\,&
\frac{\trm(1,P_{(j,i)},i,K)}{s^2_{1K}} \mathcal{A}^{i(j-1)}_{1K}.
\end{eqnarray}
We note that for the special case $i=m-1$,
\begin{eqnarray}
\mathcal{A}^{m-1,j}_{m1}=\frac{\trm(m,j,m-1,1)}{s_{m-1,j}}-\frac{\trm(m,j,m,1)}{s_{m,j}} 
&& \phantom{~} \,^{\underrightarrow{m-1\parallel m}}\, 0,\nonumber \\
\mathcal{A}^{m-1,j}_{1m} && \phantom{~} \,^{\underrightarrow{m-1\parallel m}}\, 0,\nonumber \\
\mathcal{A}^{i,m-1}_{m1} && \phantom{~} \,^{\underrightarrow{m-1\parallel m}}\, 0.
\end{eqnarray}
Similar relations hold for the terms involving $\mathcal{K}$ and $\mathcal{I}$.
Therefore, all terms in the $n$-gluon version of 
$A_{n;1}^{\phi F}$ and $A_{n;1}^{\phi S}$ therefore either collapse onto similar terms,
or vanish in such a way that the reduced summation precisely matches onto 
the corresponding $A_{n-1;1}^{\phi F}$ and $A_{n-1;1}^{\phi S}$.

\subsubsection{Two positive collinear limit}

Next we consider the limit when two positive helicity gluons become collinear.  We focus on
the specific example where
$\ell-1\parallel \ell$ with $3\le \ell \le m-1$.
As in the previous subsection, let first consider the ranges $2 \le i \le m-1$ and $m \le j \le n$.
We note that,
\begin{eqnarray}
b^{\ell-1 j}_{1m} &\,^{\underrightarrow{\ell-1 \parallel \ell}}& \,b^{K j}_{1m},\nonumber \\
b^{\ell j}_{1m} &\,^{\underrightarrow{\ell-1 \parallel \ell}}& \,b^{K j}_{1m}. 
\end{eqnarray}
The collinear factorisation of box functions has been well 
studied~\cite{BDDK:uni1,BDDK:uni2,Bern:allorder} and in this case, the relation,
\begin{eqnarray}
\left(b^{\ell-1 j}_{1m}\right)^n \Fftme(s_{\ell-1,j},s_{\ell,j-1};s_{\ell,j},s_{\ell-1,j-1})
+\left(b^{\ell j}_{1m}\right)^n\Fftme(s_{\ell,j},s_{\ell+1,j-1};s_{\ell+1,j},s_{\ell,j-1})
\nonumber\\ ^{\underrightarrow{\ell-1 \parallel \ell}} 
\left(b^{K j}_{1m}\right)^n
\Fftme(s_{K,j},s_{\ell+1,j-1};s_{K,j},s_{\ell+1,j-1})\nonumber \\
\end{eqnarray}
ensures the box terms correctly factorise onto the lower point amplitude.

The next set of functions we consider are the 
triangle functions which have $j$ as the second index, these functions possess the general form:
\begin{equation}
\sum_{i=\ell-1}^{\ell}
\trm(m,P_{(i,j)},j,1)^n\mathcal{F}^{j(i-1)}_{m1} L_{n}(P_{(i,j-1)},P_{(i,j)}).
\end{equation}
There is no contribution when $i=\ell$, because $\mathcal{F}^{j(\ell-1)}_{m1}=\mathcal{F}^{j(\ell-1)}_{1m}=0$, while
the remaining $i=\ell-1$ contribution collapses onto the correct term,
\begin{equation}
\trm(m,P_{(K,j)},\ell-1,1)^n\mathcal{F}^{j(K-1)}_{m1} L_{n}(P_{(K,j-1)},P_{(K,j)}).
\end{equation}
Similarly, when we consider
\begin{equation}
\sum_{i=\ell-1}^{\ell} \trm(m,P_{(j,i)},j,1)^n\mathcal{F}^{ji}_{1m} L_{n}(P_{(j+1,i)},P_{(j,i)}),
\end{equation}
there is no contribution when $i=\ell-1$, while for $i=\ell$, we recover the correct contribution.

The remaining types of triangle function are of the form 
\begin{equation}
\sum_{i=\ell-1}^{\ell} \trm(m,P_{(i,j)},i,1)^n\mathcal{F}^{ij}_{m1} L_{n}(P_{(i+1,j)},P_{(i,j)}).
\end{equation}
Since $\mathcal{F}^{\ell j}_{m1}=\mathcal{F}^{(\ell-1)j}_{m1}$ we have contributions from both terms,
however, it is straightforward to show that,
\begin{eqnarray}
\trm(m,P_{(\ell-1,j)},\ell-1,1)^nL_n(P_{(\ell,j)},P_{(\ell-1,j)})+\trm(m,P_{(\ell+1,j)},\ell,1)^nL_n(P_{(\ell+1,j)},P_{(\ell,j)})\nonumber\\
^{\ell\underrightarrow{-1\parallel}\ell} \trm(m,P_{(\ell+1,j)},K,1)^nL_n(P_{(\ell+1,j)},P_{(K,j)}).\nonumber \\
\end{eqnarray}
Similar considerations apply to
\begin{equation}
\sum_{i=\ell-1}^{\ell} \trm(1,P_{(j,i)},i,m)^n\mathcal{F}^{i(j-1)}_{1m} L_{n}(P_{(j,i-1)},P_{(j,i)}),
\end{equation}
thus ensuring the correct collinear factorisation. 

\subsection{The cancellation of unphysical singularities}

The cut constructible terms eq.~(\ref{eq:phiFs}) - (\ref{eq:phiSs}) contain poles in $\spa i.j$. For the most part, $i$ and $j$ are non-adjacent 
gluons and as such there should be no singularity as these become collinear. 
In the following section we prove that this is indeed the case. 
To be explicit, we consider the collinear limit $i \parallel j$ with,
\begin{eqnarray}
i\rightarrow zK,\nonumber\\
j\rightarrow (1-z)K.
\end{eqnarray}

Let us consider the cut-constructible pieces associated with the fermionic loop contribution,
$A^{\phi F}_{n;1}(m,n)$ given in eq.~(\ref{eq:phiFs}). 
There are ten terms containing an explicit pole in $s_{ij}$ which are given by,
\begin{eqnarray}
&&\phantom{+} b^{ij}_{1m}\Fftme(s_{i,j},s_{i+1,j-1};s_{i+1,j},s_{i,j-1})\nonumber\\
&&+b^{ij}_{1m}\Fftme(s_{j,i},s_{j+1,i-1};s_{j+1,i},s_{j,i-1})\nonumber\\
&&-\frac{\trm(m,P_{(i+1,j)},i,1)}{s_{1m}^2}\frac{\trm{(m,i,j,1)}}{s_{ij}}  L_1(P_{(i+1,j)},P_{(i,j)})\nonumber\\
&&+\frac{\trm(m,P_{(i+1,j-1)},i,1)}{s_{1m}^2}\frac{\trm{(m,i,j,1)}}{s_{ij}}L_1(P_{(i+1,j-1)},P_{(i,j-1)})\nonumber\\
&&-\frac{\trm(1,P_{(j,i-1)},i,m)}{s_{1m}^2}\frac{\trm{(1,i,j,m)}}{s_{ij}}  L_1(P_{(j,i-1)},P_{(j,i)})\nonumber\\
&&+\frac{\trm(1,P_{(j+1,i-1)},i,m)}{s_{1m}^2}\frac{\trm{(1,i,j,m)}}{s_{ij}}L_1(P_{(j+1,i-1)},P_{(j+1,i)})\nonumber\\
&&-\frac{\trm(m,P_{(i,j-1)},j,1)}{s_{1m}^2}\frac{\trm{(m,j,i,1)}}{s_{ij}}  L_1(P_{(i,j-1)},P_{(i,j)})\nonumber\\
&&+\frac{\trm(m,P_{(i+1,j-1)},j,1)}{s_{1m}^2}\frac{\trm{(m,j,i,1)}}{s_{ij}}L_1(P_{(i+1,j-1)},P_{(i+1,j)})\nonumber\\
&&-\frac{\trm(1,P_{(j+1,i)},j,m)}{s_{1m}^2}\frac{\trm{(1,j,i,m)}}{s_{ij}}  L_1(P_{(j+1,i)},P_{(j,i)})\nonumber\\
&&+\frac{\trm(1,P_{(j+1,i-1)},j,m)}{s_{1m}^2}\frac{\trm{(1,j,i,m)}}{s_{ij}}L_1(P_{(j+1,i-1)},P_{(j,i-1)}).
\end{eqnarray}
Using $P_{(i+1,j)}=P_{(i+1,j-1)}+p_j$, $P_{(j,i-1)}=P_{(j+1,i-1)}+p_j$,
$P_{(i,j-1)}=P_{(i+1,j-1)}+p_i$ and $P_{(j+1,i)}=P_{(j+1,i-1)}+p_i$, as well as
$\trm{(1,j,i,m)}=-\trm{(1,i,j,m)} + {\cal O}(s_{ij})$ etc, we can rewrite these terms as 
\begin{eqnarray}
&&\phantom{+} b^{ij}_{1m}
\left(\Fftme(s_{i,j},s_{i+1,j-1};s_{i+1,j},s_{i,j-1})
-s_{ij}L_1(P_{(i+1,j)},P_{(i,j)})
-s_{ij}L_1(P_{(i,j-1)},P_{(i,j)})\right)\nonumber\\
&&+b^{ij}_{1m}
\left(
\Fftme(s_{j,i},s_{j+1,i-1};s_{j+1,i},s_{j,i-1})
 -s_{ij} L_1(P_{(j,i-1)},P_{(j,i)}) 
 -s_{ij} L_1(P_{(j+1,i)},P_{(j,i)})\right)\nonumber\\
&&-\frac{\trm(m,P_{(i+1,j-1)},i,1)}{s_{1m}^2}\frac{\trm{(m,i,j,1)}}{s_{ij}}\times \left(L_1(P_{(i+1,j)},P_{(i,j)})-L_1(P_{(i+1,j-1)},P_{(i,j-1)})\right)\nonumber\\
&&+\frac{\trm(m,P_{(i+1,j-1)},j,1)}{s_{1m}^2}\frac{\trm{(m,i,j,1)}}{s_{ij}}\times \left(L_1(P_{(i,j-1)},P_{(i,j)})-L_1(P_{(i+1,j-1)},P_{(i+1,j)})\right)\nonumber\\
&&-\frac{\trm(1,P_{(j+1,i-1)},i,m)}{s_{1m}^2}\frac{\trm{(1,i,j,m)}}{s_{ij}}\times \left(L_1(P_{(j,i-1)},P_{(j,i)})-L_1(P_{(j+1,i-1)},P_{(j+1,i)})\right)\nonumber\\
&&+\frac{\trm(1,P_{(j+1,i-1)},j,m)}{s_{1m}^2}\frac{\trm{(1,i,j,m)}}{s_{ij}}\times \left( L_1(P_{(j+1,i)},P_{(j,i)})-L_1(P_{(j+1,i-1)},P_{(j,i-1)})\right).\nonumber \\
\end{eqnarray}
Finally, in the $i \parallel j$ collinear limit,
\begin{eqnarray}
&&\trm(m,P_{(i+1,j-1)},i,1) \left(L_1(P_{(i+1,j)},P_{(i,j)})-L_1(P_{(i+1,j-1)},P_{(i,j-1)})\right)\nonumber \\
&\to&
\trm(m,P_{(i+1,j-1)},j,1) \left(L_1(P_{(i,j-1)},P_{(i,j)})-L_1(P_{(i+1,j-1)},P_{(i+1,j)})\right)
\end{eqnarray}
and noting that the combination,
\begin{equation}
\Fftme(s_{i,j},s_{i+1,j-1};s_{i+1,j},s_{i,j-1})
-s_{ij}L_1(P_{(i+1,j)},P_{(i,j)})
-s_{ij}L_1(P_{(i,j-1)},P_{(i,j)}) \to {\cal O}(s_{ij}^2),
\end{equation}
we see that all singularities cancel.
The same arguments apply to the cut-constructible pieces associated with the scalar pieces. 

\subsection{Collinear factorisation of the rational pieces}

This section is devoted to the collinear factorisation of the rational pieces of the four point amplitude. Since there is a
$(1\leftrightarrow 3)$ and $(2\leftrightarrow4)$ symmetry there are two independent limits $1\parallel2$ and $2\parallel3$. We
first consider the collinear limit $2\parallel3$. It is straightforward to see that the amplitude correctly factorises onto:
\begin{eqnarray}
\hat{R}_4(\phi,1^-,2^+,3^-,4^+)+CR_4(\phi,1^-,2^+,3^-,4^+)^{2\underrightarrow{\parallel}3} R_3(\phi,1^-,K^+,4^+)\mathrm{Split}^{(0)}(-K^-,2^+,3^-)\nonumber\\+
R_3(\phi,1^-,K^-,4^+)\mathrm{Split}^{(0)}(-K^+,2^+,3^-).\nonumber\\
\end{eqnarray}
In a similar fashion the remaining non-trivial collinear limit takes the form,
\begin{eqnarray}
\hat{R}_4(\phi,1^-,2^+,3^-,4^+)+CR_4(\phi,1^-,2^+,3^-,4^+)^{1\underrightarrow{\parallel}2} R_3(\phi,K^+,3^-,4^+)\mathrm{Split}^{(0)}(-K^-,1^-,2^+)\nonumber\\+
R_3(\phi,K^-,3^-,4^+)\mathrm{Split}^{(0)}(-K^+,1^-,2^+).\nonumber\\
\end{eqnarray}

\subsection{Soft limit of $A^{(1)}_4(\phi,1^-,2^+,3^-,4^+)$}

The final test is to take the limit as the $\phi$ momentum becomes soft.  Our
naive expectation is that in this limit, the $\phi$ field is essentially
constant so that
\begin{equation}
C \phi \mathrm{tr}G_{SD\,\mu\nu}G^{\mu,\nu}_{SD} \to
\mathrm{tr}G_{SD\,\mu\nu}G^{\mu,\nu}_{SD}.
\end{equation}
In other words, the amplitude should collapse onto the gluon-only amplitude.
Following~\cite{Berger:higgsrecfinite}, we expect that,
\begin{equation}
\label{eq:naivesoft}
A^{(1)}_n(\phi,n_- g^-, n_+ g^+)  \overset{p_\phi \to 0}{\to} n_-\, A^{(1)}_n(n_- g^-, n_+ g^+),
\end{equation}
while
\begin{equation}
A^{(1)}_n(\phi^\dagger,n_- g^-, n_+ g^+)  \overset{p_\phi^\dagger \to 0}{\to} n_+\, A^{(1)}_n(n_- g^-, n_+ g^+).
\end{equation}

We first consider the cut constructible contributions. These factorise onto the four gluon
amplitude in rather trivial manner since in our construction we separated gluon-only like
diagrams and those which require a non-vanishing $\phi$-momentum.  
In the soft limit, the one and two mass easy box and triangle functions have smooth limits so
that,
\begin{eqnarray}
\bigg(\frac{\mu^2}{-m_{\phi}^2}\bigg)^{\epsilon}\,  \overset{p_ \phi \to 0}{\to}\ 0,\\
\bigg(\frac{\mu^2}{-s_{\phi i}}\bigg)^{\epsilon}\ \overset{p_ \phi \to 0}{\to}0.\\
\end{eqnarray}
Furthermore, in the soft limit the $L_k$ functions 
become the massless $T_i$ functions defined in eq.~\eqref{eq:Ti}, 
\begin{eqnarray}
L_k(s_{234},s_{23}) =\frac{{\rm Bub}(s_{234})-{\rm Bub}(s_{23})}{(s_{234}-s_{23})^k} 
\overset{p_ \phi \to 0}{\to} \frac{(-1)^k}{s_{23}^k\epsilon(1-2\epsilon)}\bigg(\frac{\mu^2}{-s_{23}}\bigg)^{\epsilon}.
\end{eqnarray}
Altogether, we find that
\begin{eqnarray}
C_4(\phi,1^-,2^+,3^-,4^+) \overset{p_ \phi \to 0}{\to}2C_4(1^-,2^+,3^-,4^+),
\end{eqnarray}
where $C_4(1^-,2^+,3^-,4^+)$ is given by eq.~\eqref{eq:gluMHV} with $n=4$. This confirms that 
the cut-constructible terms of the amplitude do follow the naive factorisation of 
eq.~\eqref{eq:naivesoft}

The rational terms of eqs.~\eqref{eq:phiTotRat} and \eqref{eq:CR4}, 
are each apparently singular in this limit.  However, careful 
combination reveals the soft behaviour,
\begin{eqnarray}
\hat{R}_4(\phi,1^-,2^+,3^-,4^+)+CR_4(\phi,1^-,2^+,3^-,4^+) \overset{p_ \phi \to 0}{\to} 
\frac{N_Pc_{\Gamma}}{3}A^{(0)}(1^-,2^+,3^-,4^+).
\end{eqnarray}
This is similar to the soft limit found in ref.~\cite{Badger:2007si,Risager:2008yz} for the 
MHV  amplitudes with adjacent negative helicities, but, as anticipated in
ref.~\cite{Berger:higgsrecfinite}, is not consistent with the naive limit of
eq.~\eqref{eq:naivesoft}.

\newpage
\section{Conclusions}
\label{sec:conclusions}

 Previous
analytic calculations of  
$\phi$-amplitudes at one-loop with arbitrary numbers of gluons
are the adjacent minus
$\phi$-MHV~\cite{Badger:2007si}, the all minus~\cite{Badger:1lhiggsallm}, 
and the finite all plus and single plus~\cite{Berger:higgsrecfinite} 
$\phi$-amplitudes.   Higgs amplitudes produced by the effective
interaction between Higgs and gluons induced by a heavy top quark loop,
may be constructed from the sum of a $\phi$-amplitude and its parity
conjugate $\phi^{\dagger}$. 
In this paper, we have extended the calculation of one-loop MHV
$\phi$-amplitudes to include the general MHV configuration.

One-loop amplitudes naturally divide into cut-containing,  $C_n$, and rational, $R_n$,
parts. As in Ref.~\cite{Badger:2007si}, we used the  double-cut unitarity approach of
ref.~\cite{Brandhuber:n4} to apply the one-loop MHV rules to derive all the multiplicity
results for  the cut-constructible contribution $C_n$. 
In this paper we also used the 
spinor integration technique of ref.~\cite{Britto:sqcd,Britto:ccqcd} 
to determine $C_n$, 
finding complete agreement between the two methods. We found that the
cut-constructible terms had a natural decomposition in terms of the pure glue MHV
amplitude, we discovered that the new diagrams which arose as a result of the $\phi$
interaction could be easily described by the basis functions used in the construction of
the pure glue result. An explicit formula for the cut-constructible part of the
$\phi$-amplitude are given in eq.~\eqref{eq:phimhv},  with the gluonic, fermionic and
scalar contributions given in eqs.~\eqref{eq:phiG},~\eqref{eq:phiFs} and
\eqref{eq:phiSs}. 

The rational terms have several
sources - first the cut-completion term $CR_n$ which eliminates the
unphysical poles present in $C_n$, second the direct on-shell recursion 
contribution $R_n^D$, third the overlap term $O_n$ and finally from the large $z$ limit of the
cut completion terms $\mathrm{Inf}\,CR_n$.
Explicit formulae for each of these contributions are given in
eqs.~\eqref{eq:CR}, \eqref{eq:rats}, \eqref{eq:otot} and \eqref{eq:infC} respectively.

The four gluon case is worked through in detail, and an explicit solution
for the $\phi$-amplitude with split helicities, $A^{(1)}_4(\phi,1^-,2^+,3^-,4^+)$, together
with instructions for how to assemble the Higgs amplitude 
$A^{(1)}_4(H,1^-,2^+,3^-,4^+)$ are given in section~\ref{sec:fourpoint}. 
Numerical results for this amplitude have previously been obtained in Ref.~\cite{Campbell:2006xx}.
We have
checked our analytic expressions in the limit where two of the gluons are collinear, 
in the limit where the $\phi$ becomes soft and against previously known 
results for up to four gluons.

\acknowledgments 

Part of this work was carried out while two of the authors were attending the programme
``Advancing Collider Physics: From Twistors to Monte Carlos" of the Galileo Galilei Institute
for Theoretical Physics (GGI) in Florence. We thank the GGI for its hospitality and
the Istituto Nazionale di Fisica Nucleare (INFN) for partial support.
CW acknowledges
the award of an STFC studentship.


\newpage

\appendix
\section{Evaluation of the $\widehat{\cal G}$, $\widehat{\cal F}$ and $\widehat{\cal S}$ functions.}
\label{app:X}

\subsection{ $\widehat{\cal G}(i,i+1,j,j+1)$}

The function $\widehat{\cal G}$ is defined in eq.~\eqref{eq:gdef}.
Using the Schouten identity, we can rewrite it as,
\begin{equation}
\widehat{\cal G} (i,i+1,j,j+1)   =
{\cal G}(i,j)+{\cal G}(i+1,j+1)-{\cal G}(i+1,j)-{\cal G}(i,j+1),
\end{equation}
where the function ${\cal G}(i,j)$ is given by,
\begin{equation}
\label{eq:GGij}
\mathcal{G}(i,j)=\frac{\spa i.\lt\spa j.\lo}{\spa i.\lo\spa j.\lt}=\frac{T(i,\lt,j,\lo)}{2\lo.p_i 2\lt.p_j}.
\end{equation}
Clearly, 
\begin{equation}
\mathcal{G}(i,i)=1.
\end{equation}
If $i\neq j$ then
\begin{equation}
\mathcal{G}(i,j)=1+\frac{P.p_i}{2\lo.p_i}-\frac{P.p_j}{2\lt.p_j}+\frac{N(P,p_i,p_j)}{2\lo.p_i 2\lt.p_j},
\label{eq:gTr}
\end{equation}
where 
\begin{equation}
N(P,i,j) = P^2 p_i\cdot p_j - 2 P\cdot p_i P\cdot p_j.
\end{equation}
${\cal G}$ is now written in terms of
scalar integrals so we can directly use the results of van Neerven~\cite{vanNeerven:dimreg} to perform the phase space integration:
\begin{align}
	 \int d^D{\rm LIPS}&(-l_1,l_2,P)
		\frac{N(P,p_1,p_2)}{(l_1+p_1)^2(l_2+p_2)^2}
		=\nonumber\\
		&\frac{c_\Gamma}{(4\pi)^2\e^2}2i\sin(\pi\e) 
		\mu^{2\e}|P^2|^{-\e}\FF{1,-\e;1-\e; \frac{ p_1\cdot p_2 P^2 }{ N(P,p_1,p_2) } } \\
	 \int d^D{\rm LIPS}&(-l_1,l_2,P)
		\frac{2(P\cdot p_1)}{(l_1+p_1)^2}
		=-\frac{c_\Gamma}{(4\pi)^2\e^2}2i\sin(\pi\e) 
		\mu^{2\e}|P^2|^{-\e} \\
	 \int d^D{\rm LIPS}&(-l_1,l_2,P)\phantom{\frac{2(P\cdot p_1)}{(l_1+p_1)^2}}
		=
		-\frac{c_\Gamma}{(4\pi)^2\e(1-2\e)}2i\sin(\pi\e) 
		\mu^{2\e}|P^2|^{-\e}
	\label{eq:dlips}
\end{align}
where the factor $c_\Gamma$ is given by,
\begin{equation}
	c_\Gamma = (4\pi)^{\e-2}\frac{\Gamma(1+\e)\Gamma^2(1-\e)}{\Gamma(1-2\e)}.
\end{equation}
The final integration is over the $z$ variable. However, the only dependence on $z$ appears through the
quantity $\wh{P}_{1,n}$\footnote{Through a suitable choice of 
$\eta$, one can always ensure that $N(P,p_1,p_2)$ is independent of $z$~\cite{Brandhuber:n4}}
 so it is convenient to make a change of variables,
\begin{equation}
	\frac{dz}{z} = \frac{d(\wh{P})^2}{\wh{P}^2-{P}^2}
\end{equation}
to produce a dispersion integral that will re-construct the parts of
the cut-constructible amplitude proportional to $(s_{1,n})^{-\e}$,
\begin{equation}
	\int \frac{d(\wh{P})^2}{P^2-\wh{P}^2} 2i\sin(\pi\e)|\wh{P}^2|^{-\e} = 2\pi i (-P^2)^{\e}.
	\label{eq:dispersion}
\end{equation}
We define the function $G(i,j)$ to be the reconstructed contribution after
integration over phase space, and after performing the dispersion integration,
\begin{equation}
G(i,j) = \int \frac{dz}{z} ~\int d^D{\rm LIPS}(-l_1,l_2,P)  ~{\cal G}(i,j).
\end{equation}
Explicitly, we find that in the $P^2$ channel
\begin{equation}
 G(i,j)=\frac{c_{\Gamma}}{\epsilon^2}\bigg(\frac{\mu^2}{-P^2}\bigg)^{\epsilon}\bigg(1-\FF{1,-\e;1-\e; \frac{ p_i\cdot p_j P^2 }{ N(P,i,j) }} +\frac{\epsilon}{1-2\epsilon}\bigg)
\end{equation}
The terms associated with triangle and bubble contributions will always cancel in the summation of the $G(i,j)$ leaving only the contributions from the hypergeometric function as one would expect. 

\subsubsection{Spinorial Integration}

\def\spab#1.#2.#3{\left\langle#1|\,#2\,|#3\right]}
\def\spba#1.#2.#3{\left[#1|\,#2\,|#3\right\rangle}
\def\spaa#1.#2.#3.#4{\left\langle#1|\,#2\,#3\,|#4\right\rangle}

Let us show how the function $\widehat{\cal G}$ can be
computed {\it via} spinorial integration.
It is convenient to rearrange the integrand by applying 
different Schouten identities from the ones used above, 
so that
\bea
{\hat{{\cal G}}}(i,i+1,j,j+1) = 
- \Gamma(i,j) + \Gamma(i+1,j) - \Gamma(i,j+1) + \Gamma(i+1,j+1)
\eea
where
\bea
\Gamma(i,j) = 
 {
\spa{i}.{j}
\spa{\ell_1}.{\ell_2} 
\over
\spa{i}.{\ell_1}
\spa{j}.{\ell_2}
}.
\eea
By using momentum conservation, $l_2 = P + l_1$,
one can rewrite $\Gamma(i,j) $ in terms of $l_1$,
\bea
\Gamma(i,j) &=&
P^2 {\spa{i}.{j} \over \spa{i}.{l_1} \spab j.P.{l_1}} \ .
\eea
Then, one uses the rescaling in eq.~(\ref{pm:rescaling}), so that,
\bea
\Gamma(i,j) &=&
{1 \over t} \ P^2 {\spa{i}.{j} \over \spa{i}.{\ell} \spab j.P.{\ell}} \ .
\eea
The above expression is the integrand of the double-cut integration,
defined as,
\bea
G^\prime(i,j) = \int d{\rm LIPS}^{(4)} \ \Gamma(i,j) \ .
\eea
By substituting the parametrization of $d{\rm LIPS}^{(4)}$ given in 
eq.~(\ref{pm:phi4massless}), one has
\bea
{(2\pi)^4} \ G^\prime(i,j) &=&
\int {\dea \deb \over \spab \ell.P.\ell}
\int t \ dt \ 
\delta \Bigg(t - {P^2 \over \spab \ell.P.\ell}\Bigg)
\ 
{1 \over t} \ P^2 {\spa{i}.{j} \over \spa{i}.{\ell} \spab j.P.{\ell}}
\nonumber\\
&=&
\int \dea \deb \ P^2 
{
\spa{i}.{j}
\over
\spa{i}.{\ell}
\spab \ell.P.\ell
\spab j.P.\ell
} \nonumber \\
&=&
\int \dea \deb \ P^2 
{
\spa{i}.{j}
\spb{\ell}.i
\over
\spab j.P.\ell
\spab \ell.{i}.{\ell}
\spab \ell.P.\ell
}
\eea
where the $t$-integration has been performed trivially.
Before carrying through the spinor integration, we
introduce a Feynman parameter to combine the two denominators
depending on $|\ell\rangle$
\bea
G^\prime(i,j) &=&
\frac{1}{(2\pi)^4}
\int_0^1 dx
\int \dea \deb \ P^2 
{
\spa{i}.{j}
\spb{\ell}.i
\over
\spab c.P.\ell
} \
{1 
\over
\spab \ell.R.\ell^2
}
\eea
where
\bea
\s{R} = x \s{k}_i + (1-x) \s{P} \ .
\eea
Integrating-by-parts in $|\ell\rangle$,
using the idenity,
\bea
{\dea \over \spab \ell.R.\ell^2}
 = \dedea {\spa j.\ell \over \spab j.R.\ell \spab \ell.R.\ell } 
\eea
we obtain,
\bea
G^\prime(i,j) &=&
\frac{1}{(2\pi)^4}
\int_0^1 dx
\int \dedea \deb \ P^2 
{
\spa{i}.{j}
\spb{\ell}.i
\spa j.\ell
\over
\spab j.P.\ell
\spab j.R.\ell 
\spab \ell.R.\ell
} \ .
\eea
The integration on $|\ell]$ can be performed by
Cauchy's residues theorem, by taking
the residues at the two poles,
$|\ell] = \s{P}|j\rangle$ and 
$|\ell] = \s{R}|j\rangle$,
\bea
G^\prime(i,j) &=&
\frac{2 \pi i}{(2\pi)^4}
\int_0^1 dx
\bigg\{
{ 
P^2
\spa i.j
\spab j.R.i
\over
R^2
\spaa j.P.R.j
}
-
{ 
P^2
\spa i.j
\spab j.P.i
\spab j.P.j
\over
\spaa j.P.R.j
\spab j.{P \ R \ P}.j
}
\bigg\}
\eea
Inserting the definition of $\s{R}$ in terms of $x$
(paying attention to $R^2$ that is quadratic in $x$),
we can perform the parametric integration, and by
using some spinor identities, find that
\bea
G^\prime(i,j) &=&
\int d{\rm LIPS}^{(4)} \ \Gamma(i,j) =
\frac{2\pi i}{(2\pi)^4}
 \ln \left(
1 - P^2 { \spab i.j.i \over \spab i.P.i \spab j.P.j }
\right) \\
&=&
\frac{2 \pi i}{(2\pi)^4}
 \ln \left(
1 -
P^2 { 
 (2 p_i \cdot p_j) 
\over
 (2 P \cdot p_i) 
 (2 P \cdot p_j) }
\right) \ ,
\label{pm:eq:GprimeResult}
\eea
which corresponds to the (discontinuity of)
the double-cut of (the finite part of) the one-loop box function.

\subsection{ $\widehat{\cal F}(i,i+1,j,j+1)$}
The function $\widehat{\cal F}$ is defined in eq.~\eqref{eq:fdef}.
Again we define 
\begin{equation}
\hat{\mathcal{F}}(i,i+1,j,j+1)=\mathcal{F}(i,j)+\mathcal{F}(i+1,j+1)-\mathcal{F}(i+1,j)-\mathcal{F}(i,j+1)
\end{equation}
with,
\begin{equation}
\label{eq:Fij}
\mathcal{F}(i,j)=\frac{\spa i.m\spa j.m\spa 1.\lt\spa 1.\lo}{\spa i.\lo\spa j.\lt\spa 1.m^2}
\end{equation}
Then after using the Schouten Identity twice this can be written as, 
\begin{equation}
\mathcal{F}(i,j)=\frac{\spa i.1\spa i.m\spa j.m\spa 1.\lo}{\spa i.j\spa i.\lo\spa 1.m^2}
+\frac{\spa i.m\spa j.m\spa 1.j\spa 1.\lt}{\spa i.j\spa j.\lt\spa 1.m^2} 
+\frac{\spa i.1\spa i.m\spa j.1\spa j.m\spa \lo.\lt}{\spa i.j\spa i.\lo\spa j.\lt\spa 1.m^2}
\end{equation}
Promoting to traces
\begin{eqnarray}
\mathcal{F}(i,j)=\frac{\trm(1,i,j,m)\trm(1,\lo,i,m)}{s_{ij}s^2_{1m}(2\lo.p_i)}+\frac{\trm(1,j,i,m)\trm(1,\lt,j,m)}{s_{ij}s^2_{1m}(2\lt.p_j)}\nonumber\\-\frac{\trm(1,i,j,m)\trm(1,j,i,m)\trm(j,i,\lo,\lt)}{s^2_{ij}s^2_{1m}(2\lt.p_j)(2\lo.p_i)}
\end{eqnarray}
Which we recognise as two linear triangles and a box function similar to those in $\mathcal{G}$. If we commute $\lt$ and $j$ in the final term we can get something which looks like eq.~(\ref{eq:gTr}).
\begin{equation}
\frac{\trm(j,i,\lo,\lt)}{(2\lo.p_i)(2\lt.p_j)}=1-\frac{\trm(j,\lo,i,\lt)}{(2\lo.p_i)(2\lt.p_j)}
\end{equation}
The first term will cancel the bubbles which arise in the calculation and the remaining terms are triangles and boxes. However, since the coefficients of $\mathcal{F}$ depend on $i$ and $j$ there 
will no longer be a cancellation between the four terms. This is important in controlling the IR divergences of the amplitude, the triangle pieces are needed to cancel off the IR poles coming from the box functions. After performing Passarino-Veltman reduction on the tensor integrals and performing the dispersion integrals we find. 
\begin{eqnarray}
F(i,j)=\frac{c_{\Gamma}}{\epsilon^2}\bigg(\frac{\mu^2}{-P^2}\bigg)^{\epsilon}\bigg[\frac{\trm(1,i,j,m)\trm(1,j,i,m)}{s^2_{ij}s^2_{1m}}
\bigg(-1+\FF{1,-\e;1-\e; \frac{ p_i\cdot p_j P^2 }{ N(P,i,j) }}\bigg)\nonumber\\
+\bigg(\frac{\trm(1,i,j,m)\trm(1,P,i,m)}{s_{ij}s^2_{1m}}\frac{1}{2(P.p_i)}
+\frac{\trm(1,j,i,m)\trm(1,P,j,m)}{s_{ij}s^2_{1m}}\frac{1}{2(P.p_j)}\bigg)\frac{\epsilon}{1-2\epsilon}\bigg].\nonumber\\
\end{eqnarray}

\subsubsection{Spinorial Integration}

\def\spab#1.#2.#3{\left\langle#1|\,#2\,|#3\right]}
\def\spba#1.#2.#3{\left[#1|\,#2\,|#3\right\rangle}
\def\spaa#1.#2.#3.#4{\left\langle#1|\,#2\,#3\,|#4\right\rangle}

Alternatively the function $\widehat{\cal F}$ can be
computed {\it via} spinorial integration.
Using momentum conservation, $l_2 = P + l_1$,
and the rescaling in eq.~(\ref{pm:rescaling}),
one can rewrite $\mathcal{F}(i,j)$ of eq.~\eqref{eq:Fij} 
in terms of $\ell$, and $t$,
\bea
\mathcal{F}(i,j) &=&
-{
\spa i.m
\spa j.m
\spa \ell.1
\spab 1.P.\ell
\over
\spa i.\ell
\spa m.1^2
\spab j.P.\ell
}
\eea
The above expression, which turns out to be independent of $t$, 
is the integrand of the double-cut integration,
defined as,
\bea
F^\prime(i,j) = \int d{\rm LIPS}^{(4)} \ \mathcal{F}(i,j) \ .
\eea
By substituting the parametrization of $d{\rm LIPS}^{(4)}$ given in 
eq.~(\ref{pm:phi4massless}), and performing the phase-space integration
with spinor-variables, one finds,
\bea
F^\prime(i,j) &=&
- {
\spab{m}.{i\ j \ 1}.{m}
\spab{m}.{j\ i \ 1}.{m}
\over
s_{ij}^2 \ 
s_{1m}^2
} 
G^\prime(i,j) 
+ 
\nn \\ &&
-
\frac{2 \pi i}{(2\pi)^4} \ 
\left\{
{
\spab{m}.{j\ i \ 1}.{m}
\spab{m}.{i\ P \ 1}.{m}
\over
s_{ij} \ 
s_{1m}^2
\spab i.P.i
} 
+ (i \leftrightarrow j) 
\right\} \ ,
\label{pm:eq:FprimeResult}
\eea
where $G^\prime(i,j)$ was given in eq.~(\ref{pm:eq:GprimeResult}).
We remark that the term proportional to $G^\prime(i,j)$ 
corresponds to the (discontinuity of)
the double-cut of (the finite part of) the one-loop box function;
while the rational part of eq.~(\ref{pm:eq:FprimeResult}) corresponds
to the discontinuity of logarithmic functions associated with 
a combination of 2-point and (1m- and 2m-) 3-point functions.

\subsection{ $\widehat{\cal S}(i,i+1,j,j+1)$}
The final pieces of the amplitude, associated with the 
propagation of scalar particles around the loop, are the 
most complicated.  
The function $\widehat{\cal S}$ is defined in eq.~\eqref{eq:sdef}.
In a similar fashion to the gluonic and fermionic pieces we define,
\begin{equation}
\hat{\mathcal{S}}(i,i+1,j,j+1)=\mathcal{S}(i,j)+\mathcal{S}(i+1,j+1)-\mathcal{S}(i+1,j)-\mathcal{S}(i,j+1)
\end{equation}
with 
\begin{equation}
\label{eq:Sij}
\mathcal{S}(i,j)=\frac{\spa1.\lo^2\spa1.\lt^2\spa m.\lo\spa m.\lt \spa i.m \spa j.m}{\spa1.m^4\spa\lo.\lt^2\spa i.\lo\spa j.\lt}
\end{equation}
After using the Schouten Identity the above can be reduced to 
a scalar box and
third rank triangles which can be solved via Passarino-Veltman reduction generating,
\begin{eqnarray}
S(i,j)=\frac{c_{\Gamma}}{\epsilon^2}\bigg(\frac{\mu^2}{-P^2}\bigg)^{\epsilon}\bigg[\frac{\trm(1,i,j,m)^2\trm(1,j,i,m)^2}{s^4_{ij}s^4_{1m}}
\bigg(1-\FF{1,-\e;1-\e; \frac{ p_i\cdot p_j P^2 }{ N(P,i,j) }}\bigg)\nonumber\\
+\bigg\{\frac{1}{3}\bigg(\frac{\trm(1,i,j,m)\trm(1,P,i,m)^3}{s_{ij}s^4_{1m}}\frac{1}{(2p_i.P)^3}+(i\leftrightarrow j)\bigg)\nonumber\\
-\frac{1}{2}\bigg(\frac{\trm(1,i,j,m)^2\trm(1,P,i,m)^2}{s_{ij}^2s^4_{1m}}\frac{1}{(2p_i.P)^2}+(i\leftrightarrow j)\bigg)\nonumber\\
-\bigg(\frac{\trm(1,i,j,m)^2\trm(1,P,i,m)\trm(1,j,i,m)}{s_{ij}^3
s^{4}_{1m}}\frac{1}{(2p_i.P)}+(i\leftrightarrow
j)\bigg)\bigg\}\frac{\epsilon}{1-2\epsilon}\bigg].\nonumber\\
\end{eqnarray}

\subsubsection{Spinorial Integration}

\def\spab#1.#2.#3{\left\langle#1|\,#2\,|#3\right]}
\def\spba#1.#2.#3{\left[#1|\,#2\,|#3\right\rangle}
\def\spaa#1.#2.#3.#4{\left\langle#1|\,#2\,#3\,|#4\right\rangle}

Alternatively the function $\widehat{\cal S}$ can be
computed {\it via} spinorial integration.
By using momentum conservation, $l_2 = P + l_1$,
and the rescaling in eq.~(\ref{pm:rescaling}),
one can rewrite eq.~\eqref{eq:Sij}
in terms of $\ell$ and $t$,
\bea
\mathcal{S}(i,j) &=&
t^2 \ {
\spa i.m
\spa j.m
\spa m.\ell
\spa \ell.1^2
\spab m.P.\ell
\spab 1.P.\ell^2
\over
P^4
\spa i.\ell
\spa m.1^4
\spab j.P.\ell
}.
\eea
By substituting the parametrization of $d{\rm LIPS}^{(4)}$ given in 
eq.~(\ref{pm:phi4massless}), and performing the phase-space integration
with spinor-variables, one obtains,
\bea
S^\prime(i,j) &=&\int d{\rm LIPS}^{(4)} \ \mathcal{S}(i,j)\nn\\
&=&
{
\spab{m}.{i\ j \ 1}.{m}^2 \ 
\spab{m}.{j\ i \ 1}.{m}^2 \ 
\over
s_{ij}^4 \ 
s_{1m}^4
} 
G^\prime(i,j)
+
\frac{2 \pi i}{(2\pi)^4} 
\Bigg\{
{
\spab{m}.{1 \ i \ j}.{m} \ 
\spab{m}.{i\ P \ 1}.{m} 
\over
s_{ij} \ 
s_{1m}^4
\spab i.P.i
} \times  \nn \\
&&
\quad
\Bigg(
-{
\spab{m}.{i\ P \ 1}.{m}^2
\over
3 \ 
\spab i.P.i^2
} 
+ 
{
\spab{m}.{1 \ i \ j}.{m} \ 
\spab{m}.{i\ P \ 1}.{m} \ 
\over
2 \ 
s_{ij} \ 
\spab i.P.i
} 
\nn \\ && \qquad 
+
{
\spab{m}.{1 \ i \ j}.{m} \ 
\spab{m}.{i \ j \ 1}.{m} \ 
\over 
s_{ij}^2 \ 
} 
\Bigg)
+
(i \leftrightarrow j)
\Bigg\} \ ,
\label{pm:eq:SprimeResult}
\eea
where $G^\prime(i,j)$ was given in eq.~(\ref{pm:eq:GprimeResult}).
We remark that the term proportional to $G^\prime(i,j)$ 
corresponds to the (discontinuity of)
the double-cut of (the finite part of) the one-loop box function;
while the rational part of eq.~(\ref{pm:eq:SprimeResult}) corresponds
to the discontinuity of logarithmic functions associated with 
a combination of 2-point and (1m- and 2m-) 3-point functions.

\section{Scalar integrals}
\label{app:scalarintegrals}

The one-loop functions that appear in the all-orders cut-constructible contribution
${C}_n$ given in section 3 are defined by,
\begin{align}
	F^{0m}_4(s,t) = \frac{2}{\e^2}
	\bigg[
	&\left (\frac{\mu^2}{-s}\right )^{\e}\FF{1,-\e;1-\e;-\frac{u}{t}} \nonumber\\
	+&\left (\frac{\mu^2}{-t}\right )^{\e}\FF{1,-\e;1-\e;-\frac{u}{s}} \bigg]
	\label{eq:f0m},\\
	F^{1m}_4(P^2;s,t) = \frac{2}{\e^2}
	\bigg[
	&\left (\frac{\mu^2}{-s}\right )^{\e}\FF{1,-\e;1-\e;-\frac{u}{t}} \nonumber\\
	+&\left (\frac{\mu^2}{-t}\right )^{\e}\FF{1,-\e;1-\e;-\frac{u}{s}} \nonumber\\
	-&\left (\frac{\mu^2}{-P^2}\right )^{\e}\FF{1,-\e;1-\e;-\frac{uP^2}{st}} 
	\bigg],
	\label{eq:f1m}\\
	F^{2me}_4(P^2,Q^2;s,t)  = \frac{2}{\e^2}
	\bigg[
	&\left (\frac{\mu^2}{-s}\right )^{\e}\FF{1,-\e;1-\e;
	\frac{
	us
	}{
	P^2Q^2-st
	}} \nonumber\\
	+&\left (\frac{\mu^2}{-t}\right )^{\e}\FF{1,-\e;1-\e;
	\frac{
	ut
	}{
	P^2Q^2-st
	}} \nonumber\\
	-&\left (\frac{\mu^2}{-P^2}\right )^{\e}\FF{1,-\e;1-\e;
	\frac{
	uP^2
	}{
	P^2Q^2-st
	}} \nonumber\\ 
	-&\left (\frac{\mu^2}{-Q^2}\right )^{\e}\FF{1,-\e;1-\e;
	\frac{
	uQ^2
	}{
	P^2Q^2-st
	}}
	\bigg],
	\label{eq:f2me}
\end{align}
and
\begin{align}	
	F^{1m}_3(s) = \frac{1}{\e^2}&\left (\frac{\mu^2}{-s}\right )^{\e},
	\label{eq:tri}\\
	{\rm Bub}(s) = \frac{1}{\e(1-2\e)}&\left (\frac{\mu^2}{-s}\right )^{\e}.
\end{align}

\bibliographystyle{JHEP-2}
\providecommand{\href}[2]{#2}\begingroup\raggedright\endgroup
\end{document}